\documentclass[onecolumn, prd, aps, tightenlines, preprintnumbers, showpacs, nofootinbib, superscriptaddress, notitlepage]{revtex4-2}
\pdfoutput=1
\usepackage{dsfont}
\usepackage{float}
\usepackage{amsmath}
\usepackage{color}
\usepackage{graphicx}
\usepackage[dvipsnames]{xcolor}
\usepackage{url}
\usepackage{epsfig}
\usepackage[T1]{fontenc}
\usepackage{multirow}
\usepackage{physics} % use \bra{...} and \ket{...}, \abs{x}, \norm{x}, \eval{x}_0^\infty, \order{x^2}, comm{A,B}, acomm{A, B}, \vdot, \grad \div \curl, \dd, \pdv{x}, \pdv{f}{x} \pdv{f}{x}{y}, \braket{a}{b}, \expval{A}, \dyad{a}{b}
\usepackage{booktabs} % for much better looking tables
\usepackage{array} % for better arrays (eg matrices) in maths
\usepackage{paralist} % very flexible & customisable lists (eg. enumerate/itemize, etc.)
\usepackage{grffile}
\usepackage{verbatim} % adds environment for commenting out blocks of text & for better verbatim
\usepackage{subfig} % make it possible to include more than one captioned figure/table in a single float
\usepackage{amsmath,amsthm,amssymb,bm,amsfonts}
\usepackage{slashed}
\usepackage[utf8]{inputenc}
\usepackage{hyperref}
\allowdisplaybreaks[1]

\usepackage{color}
\usepackage[normalem]{ulem}
\usepackage{multirow}
\usepackage[title]{appendix}

\def\be{\begin{equation}}
\def\ee{\end{equation}}
\def\ba{\begin{eqnarray}}
\def\ea{\end{eqnarray}}

\newcommand{\slv}{\raise.15ex\hbox{$/$}\kern-.53em\hbox{$v$}}
\newcommand{\slnbar}{\raise.15ex\hbox{$/$}\kern-.53em\hbox{$\bar{n}$}}
\newcommand{\slF}{\raise.15ex\hbox{$/$}\kern-.53em\hbox{$F$}}
\newcommand{\sllbar}{\raise.15ex\hbox{$/$}\kern-.40em\hbox{$\bar{l}$}}
\newcommand{\slh}{\raise.15ex\hbox{$/$}\kern-.40em\hbox{$h$}}
\newcommand{\slP}{\raise.15ex\hbox{$/$}\kern-.53em\hbox{$P$}}
\newcommand{\slR}{\raise.15ex\hbox{$/$}\kern-.53em\hbox{$R$}}
\newcommand{\slz}{\raise.15ex\hbox{$/$}\kern-.53em\hbox{$Z$}}
\newcommand{\slzbar}{\raise.15ex\hbox{$/$}\kern-.53em\hbox{$\bar{Z}$}}
\newcommand{\slQ}{\raise.15ex\hbox{$/$}\kern-.53em\hbox{$Q$}}
\newcommand{\slK}{\raise.15ex\hbox{$/$}\kern-.53em\hbox{$K$}}
\newcommand{\slkbar}{\raise.15ex\hbox{$/$}\kern-.53em\hbox{$\bar{k}$}}
\newcommand{\slkone}{\raise.15ex\hbox{$/$}\kern-.53em\hbox{$k_1$}}
\newcommand{\slpone}{\raise.15ex\hbox{$/$}\kern-.53em\hbox{$p_1$}}
\newcommand{\slpbarone}{\raise.15ex\hbox{$/$}\kern-.53em\hbox{$\bar{p}_1$}}
\newcommand{\slptwo}{\raise.15ex\hbox{$/$}\kern-.53em\hbox{$p_2$}}
\newcommand{\slpbartwo}{\raise.15ex\hbox{$/$}\kern-.53em\hbox{$\bar{p}_2$}}
\newcommand{\slqone}{\raise.15ex\hbox{$/$}\kern-.53em\hbox{$q_1$}}
\newcommand{\slD}{\raise.15ex\hbox{$/$}\kern-.53em\hbox{$\!D$}}
\newcommand{\slC}{\raise.15ex\hbox{$/$}\kern-.53em\hbox{$C$}}
\newcommand{\slA}{\raise.15ex\hbox{$/$}\kern-.73em\hbox{$A$}}
\newcommand{\slSigma}{\raise.15ex\hbox{$/$}\kern-.53em\hbox{$\Sigma$}}
\newcommand{\slpartial}{\raise.15ex\hbox{$/$}\kern-.53em\hbox{$\partial$}}
\newcommand{\slcalP}{\raise.15ex\hbox{$/$}\kern-.63em\hbox{$\cal P$}}
\newcommand{\sleps}{\raise.15ex\hbox{$/$}\kern-.53em\hbox{$\epsilon$}}
\newcommand{\slepsbar}{\raise.15ex\hbox{$/$}\kern-.53em\hbox{$\overline{\epsilon}$}}
\newcommand{\slepsstar}{\raise.15ex\hbox{$/$}\kern-.53em\hbox{$\epsilon$}^\star}
\newcommand{\slS}{\raise.15ex\hbox{$/$}\kern-.73em\hbox{$S$}}

\newcommand{\slq}{\slashed{q}}
\newcommand{\slk}{\slashed{k}}
\newcommand{\sll}{\slashed{l}}
\newcommand{\sln}{\slashed{n}}

\newcommand{\nn}{\nonumber\\}

\newcommand{\td}{\text{d}}

\newcommand{\bk}{\mathbf{k}}

\newcommand{\bq}{\mathbf{q}}
\newcommand{\bx}{\mathbf{x}}
\newcommand{\by}{\mathbf{y}}
\newcommand{\bbperp}{\mathbf{b}}
\newcommand{\bg}{\mathbf{\gamma}}
\newcommand{\bl}{\mathbf{l}}

\def\T{{\cal T}}
\def\td{\textrm d}
\def\tx{{\textrm x}}
\def\tz{{\textrm z}}
 
\def\l{{\mathbf l}}
\def\b{{\mathbf b}}

\def\k{{\mathbf k}}
  
\def\y{{\mathbf y}}

\begin{document}

\title{From target to projectile: CSS evolution of quark TMD in different light-cone gauges }

\author{Tolga Altinoluk}
\affiliation{Theoretical Physics Division, National Centre for Nuclear Research, Pasteura 7, Warsaw, 02-093, Poland}
 
\author{Guillaume Beuf}
\affiliation{Theoretical Physics Division, National Centre for Nuclear Research, Pasteura 7, Warsaw, 02-093, Poland}

\author{Jamal Jalilian-Marian}
\affiliation{Department of Natural Sciences, Baruch College, CUNY, 17 Lexington Avenue, New York, NY 10010, USA}
\affiliation{The City University of New York Graduate Center, 365 Fifth Avenue, New York, NY 10016, USA}

\author{Mirja Tevio}
\affiliation{Theoretical Physics Division, National Centre for Nuclear Research, Pasteura 7, Warsaw, 02-093, Poland}
\affiliation{
Department of Physics, University of Jyväskylä,  P.O. Box 35, 40014 University of Jyväskylä, Finland
}

%\date{March 2025}
\begin{abstract}
We calculate the one-loop corrections to the quark TMD in the projectile light-cone gauge using the background field formalism, with the Mandelstam-Leibbrandt (ML) prescription for the extra singularity present in the light-cone gauge propagator. We use the pure rapidity regulator for rapidity divergences. The Collins-Soper-Sterman (CSS) evolution equations are obtained after one-loop renormalization of the quark TMD in this gauge. We discuss how the structure of the rapidity divergences and the double-logarithmic contributions to the CSS resummation compares with the analogous calculation performed in the target light-cone gauge, and discuss the implications for the connection between the TMD factorization and Color Glass Condensate frameworks.
\end{abstract}

\maketitle

\section{Introduction}
Understanding the relation between transverse-momentum-dependent (TMD) factorization \cite{Bomhof:2006dp,Bacchetta:2006tn,Collins:2011zzd,Angeles-Martinez:2015sea,Boussarie:2023izj} and the Color Glass Condensate (CGC) \cite{Gelis:2010nm,Albacete:2014fwa,Blaizot:2016qgz} remains an important open problem in high-energy QCD, particularly in their overlapping regime of validity. While TMD evolution is governed by the Collins–Soper–Sterman (CSS) equations \cite{Collins:1981uk, Collins:1981uw, Collins:1984kg} (see also \cite{Collins:2011zzd,Boussarie:2023izj} for detailed discussions) and is most naturally formulated in covariant or target light-cone gauges, the CGC is conventionally formulated in the  projectile light-cone gauge $A^+=0$ (in the case of right-moving projectile and left-moving target), where observables are expressed in terms of correlations of Wilson lines. Bridging these two descriptions therefore requires understanding how standard TMD evolution equations emerges in a gauge and formalism that are structurally aligned with CGC computations. However, this is a nontrivial task. In projectile light-cone gauge the operator structure of TMDs becomes more involved, the gluon propagator contains additional singular terms, and the Mandelstam–Leibbrandt (ML) prescription \cite{Mandelstam:1982cb,Leibbrandt:1983pj,Leibbrandt:1983zd,Bassetto:1984dq} introduces nonstandard pole structures that can qualitatively modify intermediate contributions. It is, therefore, necessary to understand the emergence of the CSS equations within this setup, and how the familiar logarithmic structure is reorganized at the level of individual diagrams. In this work, we address this question by computing the one-loop renormalization of the quark TMD in projectile light-cone gauge within the background field formalism and demonstrating how the CSS equations emerge in this framework.

At high energies (small $x$), hadronic structure is dominated by gluonic degrees of freedom, and the CGC provides an effective field theory description of QCD in this regime. Rapidity evolution of observables in the CGC formalism is governed by the nonlinear BK/JIMWLK equations~\cite{Balitsky:1995ub,Kovchegov:1999yj,Kovchegov:1999ua,Jalilian-Marian:1996mkd,Jalilian-Marian:1997qno,Jalilian-Marian:1997jhx,Jalilian-Marian:1997ubg,Kovner:2000pt,Weigert:2000gi,Iancu:2000hn,Iancu:2001ad,Ferreiro:2001qy}. In parallel, substantial progress has been made in relating the CGC and TMD frameworks, including the extraction of gluon TMDs in the correlation limit~\cite{Dominguez:2011wm}, higher-order analyses of dijet production~\cite{Marquet:2017xwy,Altinoluk:2018uax,Altinoluk:2018byz,Altinoluk:2020qet,Taels:2022tza,Caucal:2023fsf,Caucal:2024nsb,Taels:2023czt}, and the development of interpolating iTMD frameworks~\cite{Kotko:2015ura,vanHameren:2016ftb,Bury:2020ndc,Fujii:2020bkl,Altinoluk:2021ygv,Boussarie:2020vzf,Boussarie:2021ybe}. Moreover, the correlation limit of dijet production has been also studied beyond the eikonal approximation \cite{Altinoluk:2023qfr, Altinoluk:2024zom, Altinoluk:2024tyx,Mukherjee:2026cte} to further study the connection between TMD and CGC approaches. More recently, direct derivations of CSS evolution within the CGC have been pursued~\cite{Duan:2024nlr,Duan:2024qck,Duan:2024qev,Caucal:2024bae}, further motivating a unified understanding of these two descriptions.

A particularly useful approach to connecting these frameworks is the background field formalism, in which the gauge field is decomposed into a classical background and perturbative quantum fluctuations. This method is widely used in small-$x$ physics and provides a natural setting for connecting CGC and TMD factorization. In earlier work, we applied this formalism to compute one-loop corrections to collinear PDFs~\cite{Altinoluk:2023dww}, and it has since been extended to the study of gluon and quark TMDs in different gauges~\cite{Mukherjee:2023snp,Mukherjee:2025aiw,Altinoluk:2025ewj}. Related applications of the background field method within TMD factorization have also been developed in Refs.~\cite{Scimemi:2019gge,Rein:2022odl,Vladimirov:2021hdn}.

In a previous study~\cite{Altinoluk:2025ewj}, we demonstrated that the CSS evolution equations for quark TMDs are correctly reproduced at one loop in the target light-cone gauge $A^-=0$  (for left-moving target) within the background field formalism, using different prescriptions for the light-cone singularity and a pure rapidity regulator~\cite{Ebert:2018gsn}. A central result of that work was the role of the ML prescription~\cite{Mandelstam:1982cb,Leibbrandt:1983pj,Leibbrandt:1983zd,Bassetto:1984dq}, where the double-logarithmic contribution to the CSS kernel was traced back to a specific zero-mode structure associated with radiation of a gluon by the quark (antiquark) field where the radiated gluon is connected to the transverse Wilson line at infinity.

The target light-cone gauge $A^-=0$ is naturally adapted to a highly boosted target and simplifies the operator structure of TMDs since longitudinal Wilson lines are unity. By contrast, the projectile light-cone gauge $A^+=0$, which is the standard choice in CGC calculations, leads to qualitatively different contributions from the Wilson-line structure and a more involved gluon propagator. As a result, while the CSS evolution equations are gauge invariant, their emergence from the diagrammatic structure in different gauges is nontrivial and the organization of the underlying logarithmic contributions at the level of individual diagrams is not immediately transparent.

In this work, we perform the one-loop renormalization of the quark TMD in projectile light-cone gauge using the background field formalism, together with the ML prescription and a pure rapidity regulator. We show that the CSS evolution equations are reproduced in this framework and analyze in detail how they arise compared to the target light-cone gauge calculation. This provides a nontrivial consistency check of the formalism and clarifies how identical evolution equations emerge from distinct intermediate diagrammatic structures. Our results constitute a step toward a more complete understanding of TMD evolution within a CGC-compatible framework and set the stage for future studies of gluon TMDs and beyond-eikonal corrections.

The paper is organized as follows. In Sec.~\ref{sec:setup}, we introduce the operator definition of quark TMD in projectile light-cone gauge. In Sec.~\ref{sec:NLO_diagrams}, we compute the one-loop diagrams using the background field formalism. In Sec.~\ref{sec:CSS}, we discuss the renormalization of the fields and extract the CSS evolution equations. In Sec.~\ref{sec:comparison}, we compare the projectile and target light-cone gauge results in detail. Finally, in Sec.~\ref{sec:conc}, we present our conclusions and outlook. Technical details are collected in the appendices.

%================================================================================
\section{Setup and operator definition of the  Quark TMD}
\label{sec:setup}
%================================================================================

In this section, we closely follow the notation and arguments introduced in Sec.~II of Ref.~\cite{Altinoluk:2025ewj}, adapting them to the projectile light-cone gauge $A^+=0$. We refer the reader to that reference for further details on the general setup and conventions. 

Throughout this work, ultraviolet (UV) divergences are regulated using dimensional regularization in $D=4-2\epsilon$ dimensions. Since dimensional regularization alone does not regulate rapidity divergences, we employ the pure rapidity regulator introduced in  \cite{Ebert:2018gsn} to regularize them, which amounts to including a factor 
\begin{align}
 w^2\bigg[
\frac{{|k^-|}}{|k^+|} \, \frac{\nu^+}{\nu^-}
\bigg]^{\eta/2}
\label{def:pure_rap_reg}
\end{align}
in the integrand when integrating over the gluon momentum $k^\mu$. Here, $\eta$ is a dimensionless parameter analogous to $\epsilon$ in dimensional regularization while $\nu^+$ and $\nu^-$ are the $+$ and $-$ components of a reference momentum scale $\nu$. In order to ensure that the factor \eqref{def:pure_rap_reg} regulates only the rapidity divergences, one must always take the limit $\eta\to0$ at a finite non-zero $\epsilon$. Finally, $ w^2$ is a bookkeeping parameter \cite{Ebert:2018gsn,Chiu:2011qc,Chiu:2012ir} that satisfies
\begin{align}
\lim_{\eta\to 0} w^2&=1 \\
\frac{\nu^+}{\nu^-}\frac{\partial}{\partial\big(\frac{\nu^+}{\nu^-}\big)} w^2&=-\frac{\eta}{2}
\end{align}
so that $w^2$ reduces to unity in the absence of rapidity divergences, while the full expression in \eqref{def:pure_rap_reg} remains independent of the ratio $\nu^+/\nu^-$.

The operator definition of the unpolarized quark TMD is given as 
\begin{align}
\label{def:q_op_def_fixed_order}
q^{\textrm{n.r.}}_{\textrm{unsub.}}(\tx, \mathbf{b};\mu^2) = & \lim_{Y^+\rightarrow +\infty}
\int 
\frac{\td b^{ +}}{2 \pi}    \, 
e^{-i\tx P^- b^+}
\big\langle P\big|
{\overline\Psi}(b^+, \mathbf{b}, 0^-) \, \frac{\gamma^-}{2} \, 
U(Y^+ , \mathbf{b}, 0^-;b^+, \mathbf{b}, 0^-)^{\dag}
\nonumber \\
& \times \, 
U (Y^+, \mathbf{b}, 0^- ; Y^+,0_\perp  , 0^-) 
\, 
U (Y^+, 0_\perp , 0^- ; 0)   
\Psi(0) 
\big|P\big\rangle
\, 
\end{align}
up to UV and rapidity renormalizations. Here, the (left-moving) target state is defined via its momentum $P^\mu\simeq (0^+, P^-,0_\perp)$ (up to target mass corrections). The rapidity divergences in the unpolarized quark TMD given in Eq. \eqref{def:q_op_def_fixed_order} are regularized using \eqref{def:pure_rap_reg} but not yet subtracted which is indicated by the label ``unsub.''. Moreover, the quark TMD given in Eq. \eqref{def:q_op_def_fixed_order} is written in terms of renormalized fields and couplings, the UV divergences that arise from its perturbative expansion have not yet been removed which is indicated by the label ``n.r.''.

Different choices of gauge-link geometry give rise to distinct quark TMDs, each associated with a particular class of hard scattering processes.
In this work, we consider the quark TMD with a future-pointing gauge staple which appears in the TMD factorization of semi-inclusive deep inelastic scattering (SIDIS). In this case, two of the Wilson lines appearing in the operator definition in Eq. \eqref{def:q_op_def_fixed_order} are light-like and are defined as
\begin{align}
\label{def:Wilson_lightlike}
 U(Y^+ , \mathbf{b}, 0^-;b^+, \mathbf{b}, 0^-) 
 =&\,
  \Bigg[
 {\cal{P}} \exp\left\{- i \mu^\epsilon g \int_{b^+}^{Y^+} \td x^+
 t^a A_a^- (x^+, \mathbf{b} , 0^-)\right\}
 \Bigg]
 \, ,
\end{align}
where 
${\cal{P}}$ indicates path ordering. In addition, there is also a space-like Wilson line located in the far future
\begin{align}
\label{def:Wilson_para}
U (Y^+, \bx + \mathbf{b}, 0^- ; Y^+,  \bx , 0^-) 
 &\,=
 {\cal{P}} \exp\left\{- i \mu^\epsilon g \int_0^1 d \tau\,  
\mathbf{b}^i\, t^a\, A_i^a (Y^+, \bx +\tau \mathbf{b} , 0^-)\right\}
\, ,
\end{align}
whose contribution is non-trivial and physically significant in light-cone gauges \cite{Ji:2002aa,Belitsky:2002sm,Cherednikov:2009wk,Idilbi:2010im,Garcia-Echevarria:2011ewi}. 

The operator definition of the unpolarized quark TMD, Eq.~\eqref{def:q_op_def_fixed_order}, contains contributions from gluon exchanges occurring entirely within the gauge-link structure. These Wilson-line self-energy terms do not belong to the physical TMD and must therefore be removed. This is conventionally achieved \cite{Collins:2011zzd,Boussarie:2023izj} by normalizing the TMD operator with the square root of a soft factor $S(\b)$, defined as the vacuum expectation value of a closed Wilson loop composed of light-like and space-like Wilson lines separated by a transverse distance $\b$: 
\begin{align}
S(\mathbf{b}) = \frac{1}{N_C}\tr_F\bigg[
\Big\langle 0\Big|
\mathcal{T}\,\mathcal{P}\exp\!\left(-ig\mu^\epsilon\oint_\ell dx^\mu\,t^a A^a_\mu(x)\right)
\Big|0\Big\rangle\bigg]\,,
\label{def:soft_factor}
\end{align}
where the contour $\ell$ forms a closed path in the $(x^+,x^-)$ plane at fixed transverse separation $\mathbf{b}$, with its light-like segments extending to $Y^\pm\to+\infty$. By construction, the soft factor contains precisely the same Wilson line self energy contributions as those appearing in the TMD operator. Dividing $\sqrt{S(\mathbf{b})}$ therefore removes these unphysical contributions while preserving gauge invariance. The fully renormalized and rapidity subtracted quark TMD can therefore be defined schematically as 
\begin{align}
q(x, \mathbf{b}; \mu^2, \zeta)
\propto \frac{1}{\sqrt{S(\mathbf{b})}}\,
q^{\textrm{n.r.}}_{\textrm{unsub.}}(x, \mathbf{b}; \mu^2)\,,
\label{def:q_subtracted}
\end{align}
up to UV and  rapidity renormalization factors that will be discussed and specified in Sec.~\ref{sec:CSS}. Using the pure rapidity regulator \eqref{def:pure_rap_reg}, rapidity divergences cancel in the soft factor. In that case, the soft factor is not directly related to the subtraction or renormalization of rapidity divergences, by contrast to the situation with various other rapidity regulators~\cite{Boussarie:2023izj}.

The perturbative expansion of \eqref{def:q_op_def_fixed_order} can be simplified by equivalently rewriting it as a time-ordered product which reads 
\begin{align}
\label{def:q_op_def_T_ord}
q^{\textrm{n.r.}}_{\textrm{unsub.}}(\tx, \mathbf{b};\mu^2) = & \lim_{Y^+\rightarrow +\infty}
\int 
\frac{\td b^{ +}}{2 \pi}    \, 
e^{-i\tx P^- b^+}
\big\langle P\big| \T \Big[ 
{\overline\Psi}(b^+, \mathbf{b}, 0^-) \, \frac{\gamma^-}{2} \, 
U(Y^+ , \mathbf{b}, 0^-;b^+, \mathbf{b}, 0^-)^{\dag}
\nonumber \\
& \times\, 
U (Y^+, \mathbf{b}, 0^- ; Y^+,0_\perp  , 0^-) 
\, 
U (Y^+, 0_\perp , 0^- ; 0)   
\Psi(0) \Big]
\big|P\big\rangle_c
\, .
\end{align}
where only the diagrams connected to the target are included. The perturbative expansion of the time-ordered operator given in Eq. \eqref{def:q_op_def_T_ord} involves only Feynman propagators which simplifies the calculation considerably.  

As a remark, one can perform a spacetime translation of the nonlocal operator in Eqs.~\eqref{def:q_op_def_fixed_order} or \eqref{def:q_op_def_T_ord}, and obtain, for example,
\begin{align}
\label{def:q_op_def_T_ord_transl}
q^{\textrm{n.r.}}_{\textrm{unsub.}}(\tx, \mathbf{b};\mu^2) = & \lim_{Y^+\rightarrow +\infty}
\int 
\frac{\td b^{ +}}{2 \pi}    \, 
e^{-i\tx P^- b^+}
\big\langle P\big| \T \Big[ 
{\overline\Psi}(0) \, \frac{\gamma^-}{2} \, 
U(Y^+ , 0_\perp, 0^-;0)^{\dag}
\nonumber \\
& \times\, 
U (Y^+,0_\perp, 0^- ; Y^+,  -\mathbf{b}, 0^-) 
\, 
U (Y^+, -\mathbf{b} , 0^- ; -b^+, 
-\mathbf{b},0^-)   
\Psi(-b^+, -\mathbf{b},0^-) \Big]
\big|P\big\rangle_c
\, ,
\end{align}
without an extra phase factor since the target states on the left and on the right have the same momentum.

In this study, we choose to work in the light-cone gauge $A^+=0$ which is known as the projectile light-cone gauge for a left-moving target. This gauge choice is the standard one employed in the CGC framework, and therefore working in this gauge is a natural step toward establishing a closer connection between the TMD and the CGC frameworks. We emphasize that this represents a key structural difference compared to our previous analysis in \cite{Altinoluk:2025ewj}, which was performed in the target light-cone gauge $A^-=0$. Although the operator definition of the quark TMD retains the same form in both gauges, the role played by the gauge links is fundamentally modified by the gauge choice. In the projectile light-cone gauge, the longitudinal Wilson lines along $x^-$ direction no longer reduce to unity. As a consequence,  additional diagrams involving fluctuation fields contribute at one loop order compared to the target light-cone gauge computation.

In the light-cone gauge $n_{\mu} A^{\mu}=0$ (with $n^2=0$), the free Feynman propagator for a gluon in momentum space is
\begin{align}
{\tilde G}_{0,F}^{\mu \nu}(k)=&\,
\frac{i}{\left(k^2+i0\right)}\,
\left\{
-g^{\mu \nu}
+\frac{(k^{\mu}n^{\nu}+n^{\mu}k^{\nu})}{[n\!\cdot\!k]}
\right\}
\label{eq:Feyn-prop_def}
\, ,
\end{align} 
which contains, apart from the usual scalar denominator, a second denominator $n\!\cdot\!k$, for which we employ the ML prescription \cite{Mandelstam:1982cb,Leibbrandt:1983pj} defined as 

\begin{align}
\frac{1}{[n\!\cdot\!k]}
\equiv&\,
\frac{(\bar n\!\cdot\!k)}{(n\!\cdot\!k)(\bar n\!\cdot\!k)+i0} 
= \frac{\theta(\bar n\!\cdot\!k)}{\left((n\!\cdot\!k)+i0\right)}\,  +  \frac{\theta(-\bar n\!\cdot\!k)}{\left((n\!\cdot\!k)-i0\right)}\, 
\label{eq:ML_prescription}
\, 
\end{align} 
involving a second light-like vector $\bar n^{\mu}$. In the projectile light-cone gauge, $n^{\mu}=g^{\mu+}$ and $\bar n^{\mu}=g^{\mu-}$, which is opposite to the assignment used in the target light-cone gauge calculation in \cite{Altinoluk:2025ewj}. 
As a remark,
the Feynman propagator \eqref{eq:Feyn-prop_def} with the ML prescription can equivalently be written as 
\begin{align}
{\tilde G}_{0,F}^{\mu \nu}(k)
=&\,
\frac{i}{\left(k^2+i0^+\right)}\,
\left\{
-g^{\mu \nu}
+\frac{2(\bar n\!\cdot\!k)}{\k^2}
(k^{\mu}n^{\nu}+n^{\mu}k^{\nu})
\right\}
%\nonumber\\
%&\,
-i\bigg[
\frac{\theta(\bar n\!\cdot\!k)}{\left((n\!\cdot\!k)+i0\right)}\,  +  \frac{\theta(-\bar n\!\cdot\!k)}{\left((n\!\cdot\!k)-i0\right)}
\bigg]
\frac{(k^{\mu}n^{\nu}+n^{\mu}k^{\nu})}{\k^2}
\label{eq:Feyn-prop_2}
\end{align} 
in which the second part can be interpreted as a zero-mode $n\!\cdot\!k=0$ ghost, which results from the residual gauge freedom in the light-cone gauge~\cite{Bassetto:1984dq}.

An important practical difference with respect to the target light-cone gauge calculation concerns the order of integration in the loop integrals. In the projectile light-cone gauge $A^+=0$, the ML prescription \eqref{eq:ML_prescription} requires that pole integration over $n\cdot k=k^+$ be performed first. After this integration, rapidity divergences appear in the subsequent $\bar n \cdot k=k^-$, in the regime $k^-\to0$. This is in contrast to the target light-cone gauge calculation performed in \cite{Altinoluk:2025ewj}, where rapidity divergences appeared in the regime $k^+\to \infty$. As a consequence, in the present study the factor included in the integrand to regularize rapidity divergences is the on-shell analog of the form given in Eq. \eqref{def:pure_rap_reg} which reads 
\begin{align}
 w^2\bigg[
\frac{2 |k^-|^2}{{\k^2}} \, \frac{\nu^+}{\nu^-}
\bigg]^{\eta/2}
\label{def:pure_rap_reg_OS}
\, .
\end{align}
This form ensures that the rapidity divergences in the $k^-$ integration are always regulated without interfering with the pole integration over $k^+$, which is performed first. This is one of the key technical distinctions of the present calculation compared to \cite{Altinoluk:2025ewj}. 

%

%%%%%%%%%%%%%%%%%%%%%%%%%%%%%%%%%%%%%%%%%%%%%%%%%%%%%%%%%%%%%%%%%%%%%%%%%%%%%%%%%%%%%%
\begin{figure}[h!]
\subfloat[Diagram of gluon emission from the quark to the lower part of the gauge link \label{Fig:diag6}]{%
       \includegraphics[width=0.40\textwidth]{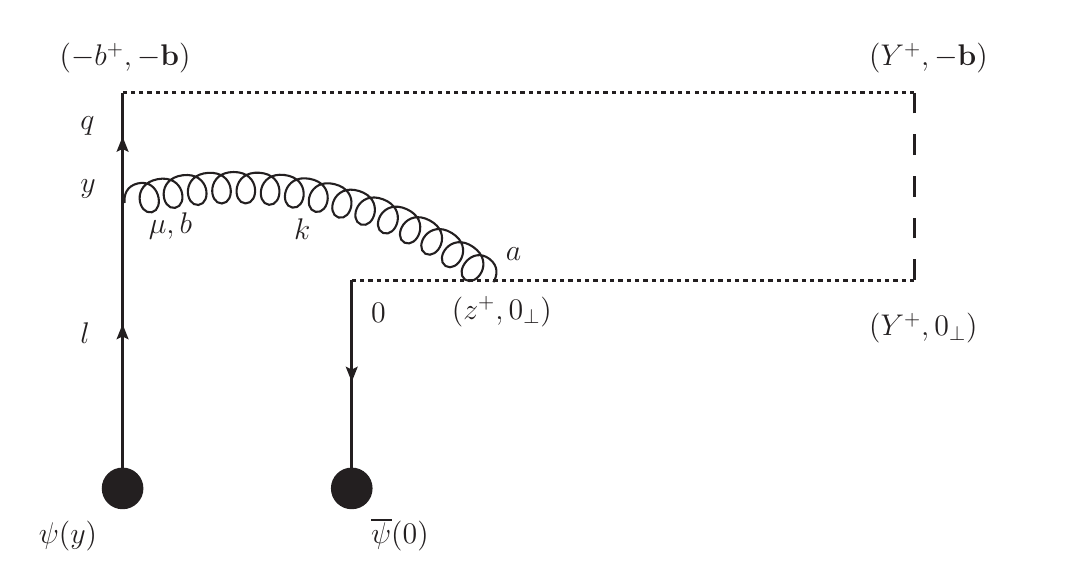}
     }
\hfill     
\subfloat[Diagram of gluon emission from the quark to the upper part of the gauge link \label{Fig:diag4}]{%
       \includegraphics[width=0.40\textwidth]{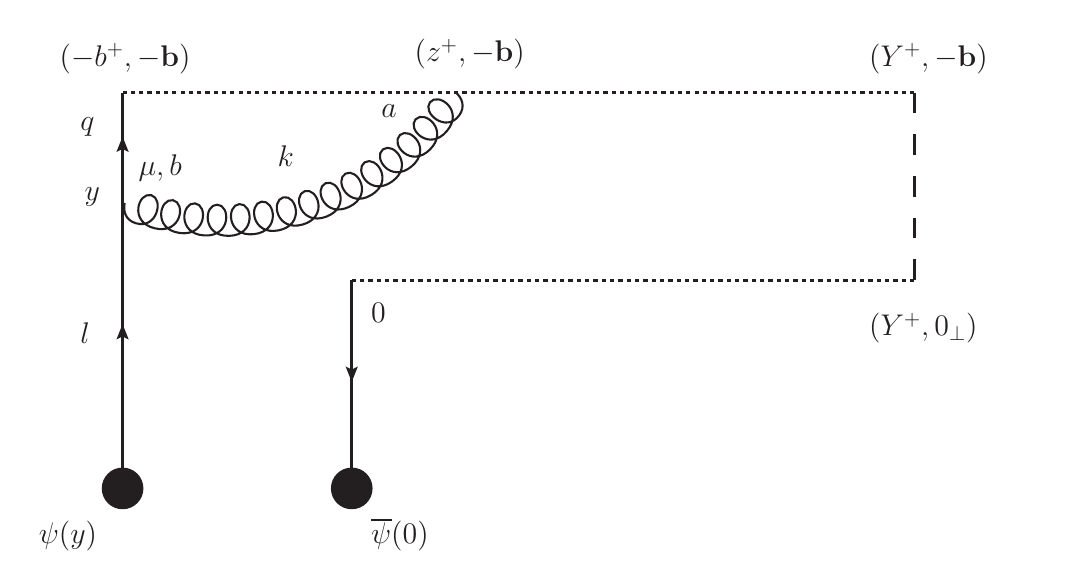}
     }

\subfloat[Diagram of gluon emission from the quark to infinity \label{Fig:diag5}]{%
       \includegraphics[width=0.40\textwidth]{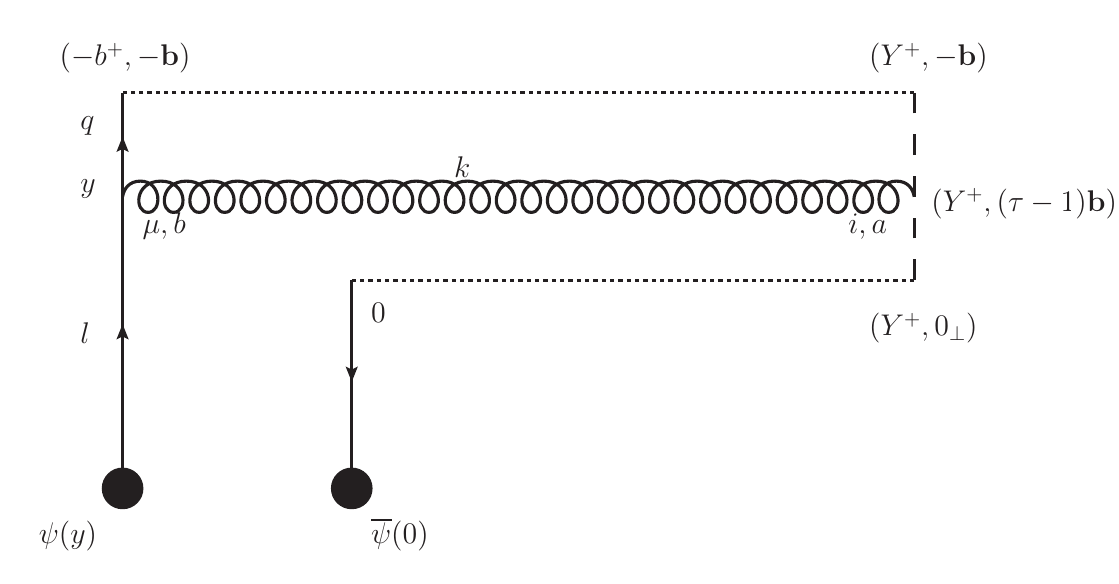}
     }
\hfill
\subfloat[Quark-to-quark ladder diagram \label{Fig:diag1}]{%
       \includegraphics[width=0.40\textwidth]{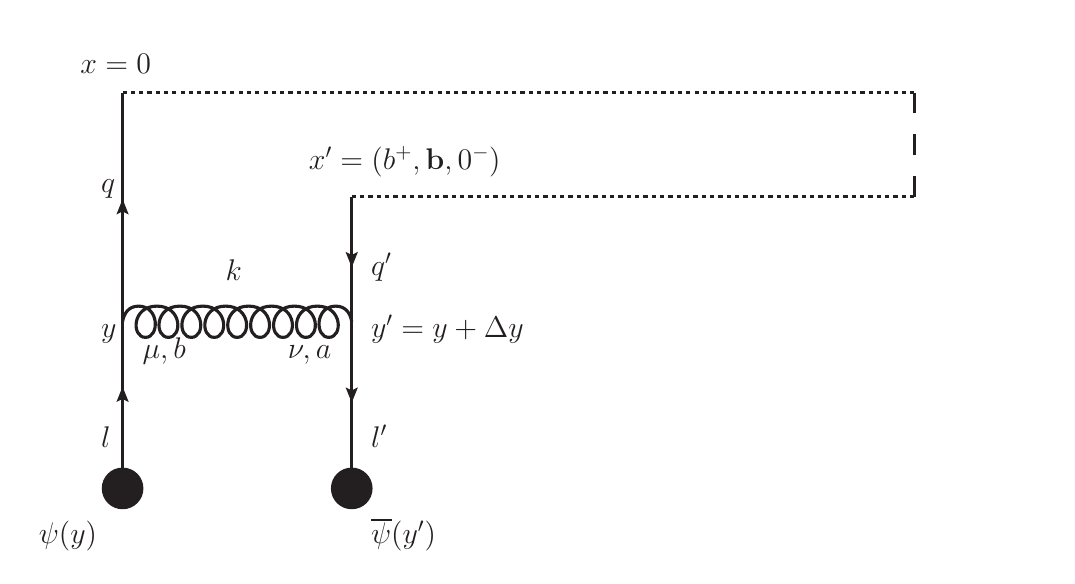}
     }

\subfloat[Wilson line self-energy at infinity \label{Fig:diag11}]{%
       \includegraphics[width=0.40\textwidth]{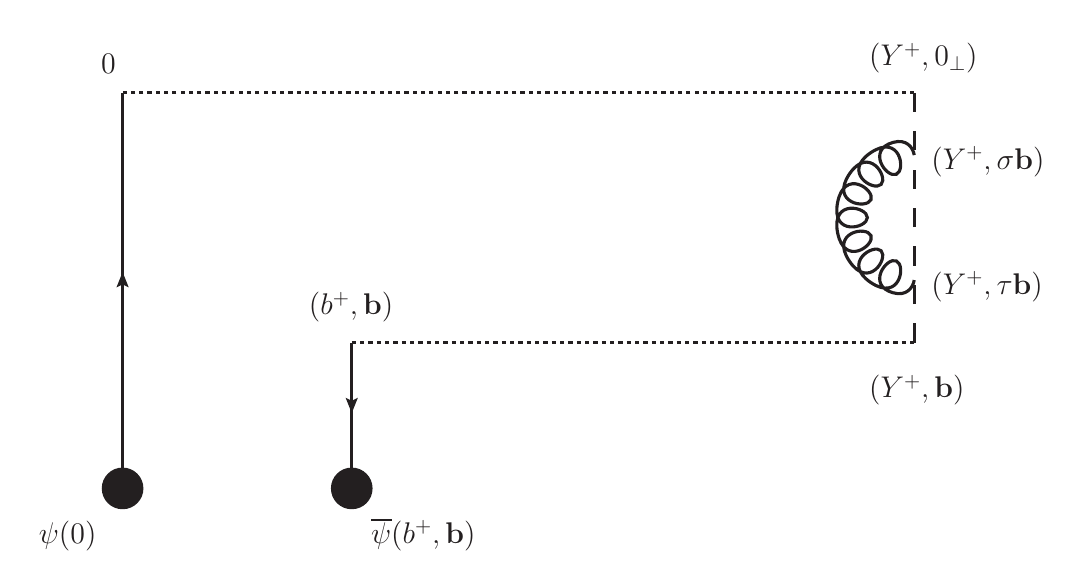}
     } 
\hfill
\subfloat[Wilson line self-energy diagram in the lower part of the gauge link
\label{Fig:diag12}]{%
       \includegraphics[width=0.40\textwidth]{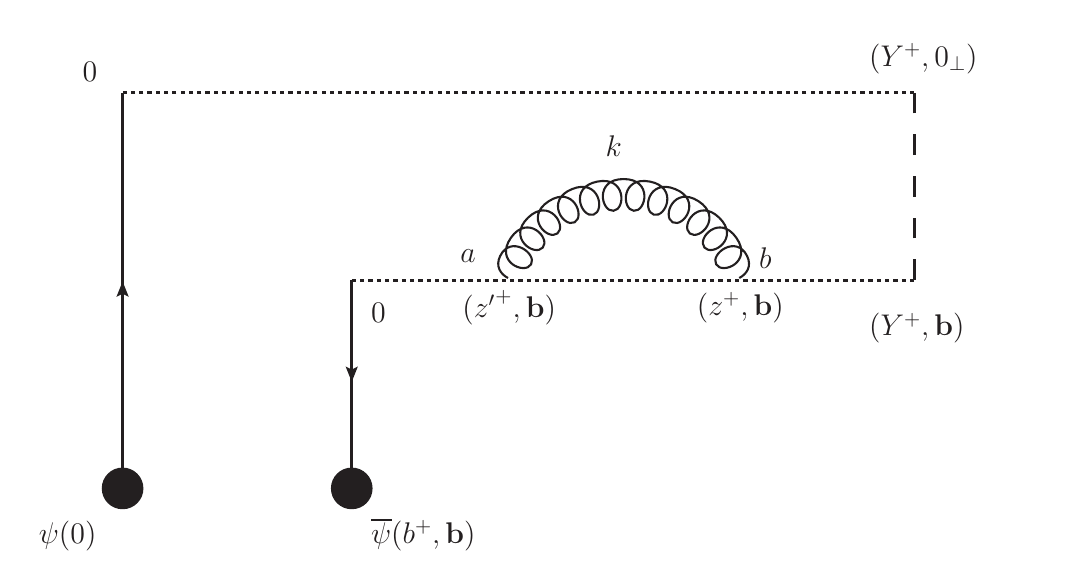}
     }  

\subfloat[Emission of gluon from the upper part of the gauge link to infinity \label{Fig:diag13}]{%
       \includegraphics[width=0.40\textwidth]{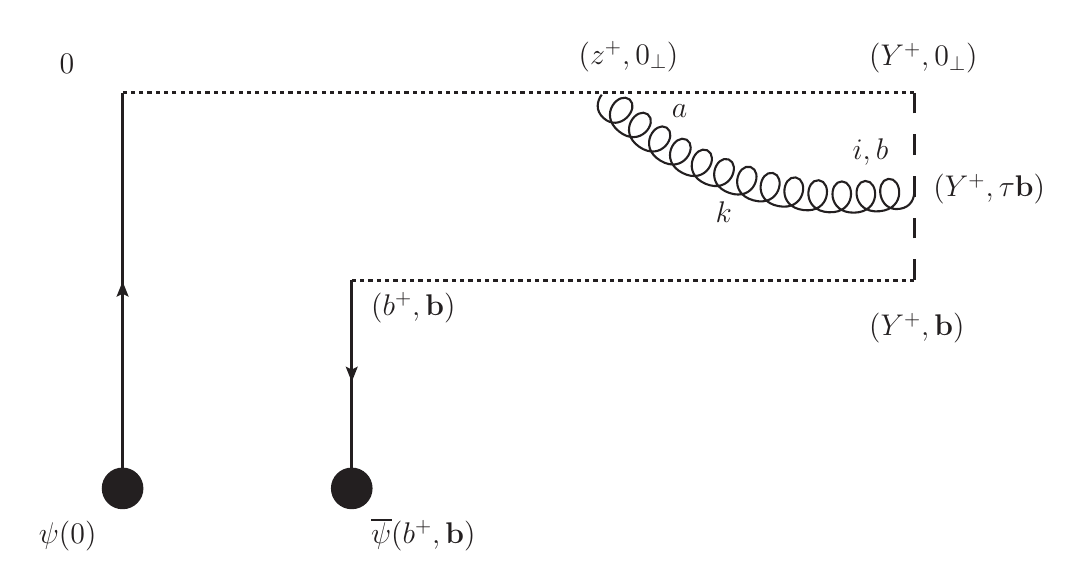}
     } 
\hfill
\subfloat[Gluon exchange between two light-like Wilson lines in the gauge link \label{Fig:diag14}]{%
       \includegraphics[width=0.40\textwidth]{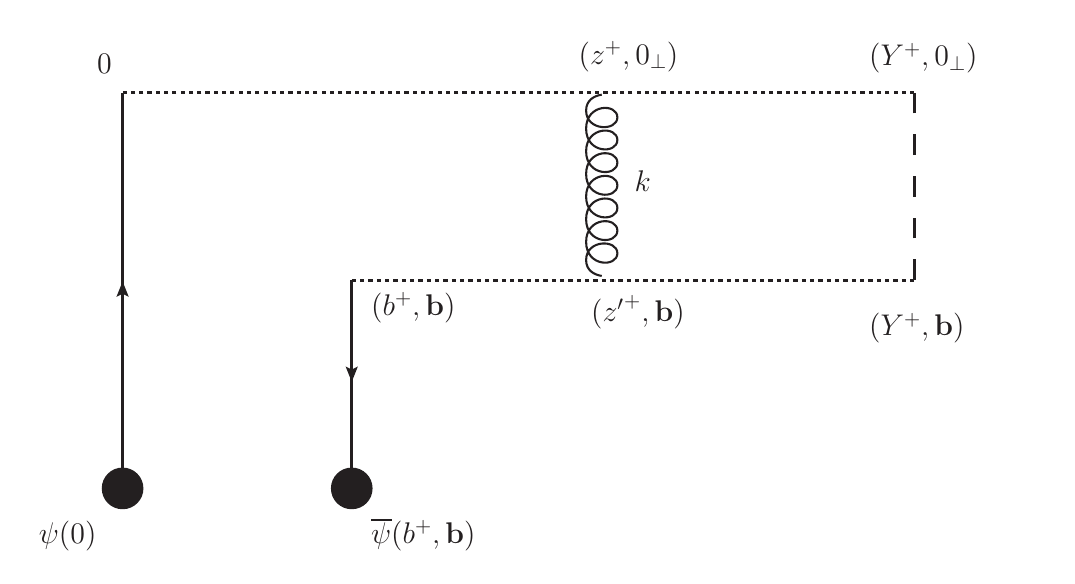}
     }

\caption{\label{Fig:real_diags 1} NLO diagrams in the expansion of the quark TMD around its background contribution in the projectile light-cone gauge. The symmetric counterparts to the diagrams~\ref{Fig:diag6}, \ref{Fig:diag4}, \ref{Fig:diag5}, \ref{Fig:diag12}, and \ref{Fig:diag13} are presented in Appendix~\ref{appendix_symmetric_diagrams}.}
\end{figure}
%%%%%%%%%%%%%%%%%%%%%%%%%%%%%%%%%%%%%%%%%%%%%%%%%%%%%%%%%%%%%%%%%%%%%%%%%%%%%%%%%%%%%%
%%%%%%%%%%%%%%%%%%%%%%%%%%%%%%%%%%%%%%%%%%%%%%%%%%%%%%%%%%%%%%%%%%%%%%%%%%%%%%%%%%%%%%

%%%%%%%%%%%%%%%%%%%%%%%%%%%%%%%%%%%%%%%%%%%%%%%
%%%%%%%%%%%%%%%%%%%%%%%%%%%%%%%%%%%%%%%%%%%%%%%
%%%%%%%%%%%%%%%%%%%%%%%%%%%%%%%%%%%%%%%%%%%%%%%
%%%%%%%%%%%%%%%%%%%%%%%%%%%%%%%%%%%%%%%%%%%%%%%
%%%%%%%%%%%%%%%%%%%%%%%%%%%%%%%%%%%%%%%%%%

%======================================================================================
\section{Calculation of NLO diagrams from  the operator definition}
\label{sec:NLO_diagrams}
%======================================================================================

Let us outline the perturbative expansion of the quark TMD operator \eqref{def:q_op_def_T_ord} within the background field formalism, following the same strategy introduced in Ref.~\cite{Altinoluk:2025ewj}. The quark and gluon fields are decomposed into a classical background component and a perturbative quantum fluctuation as 
\begin{align}
\Psi(x)=&\, \psi(x)+\delta\Psi(x)\,,
\label{eq:BFM_split_Psi}
\\
A^\mu(x)=& \, {\cal A}^{\mu}(x)+\delta A^{\mu}(x) \,.
\label{eq:BFM_split_A_mu}
\end{align}
We perform the calculation in the projectile light-cone gauge. More precisely, we choose the projectile light-cone gauge for the fluctuation field, $\delta A^+ = 0$, and we do not need to choose a gauge for the background field ${\cal A}^{\mu}$.  When the fluctuation fields are switched off entirely, and in the dilute approximation for the target, the quark TMD operator \eqref{def:q_op_def_T_ord} reduces to the so-called background TMD which is defined as 
\begin{align}
\label{def:q_Bckgd}
q_{\textrm{Bckgd}}(\mathbf{x}, \mathbf{b};\mu^2) = &
\int
\frac{d b^{ +}}{2 \pi}    \,
e^{-i\mathbf{x} P^- b^+}
\big\langle P\big| \mathcal{T} \Big[
{\overline\psi}(b^+, \mathbf{b}, 0^-) \, \frac{\gamma^-}{2} \,
\psi(0) \Big]
\big|P\big\rangle_c
\, .
\end{align}
Indeed, in the dilute regime, the gluon background field would contribute only at higher orders in $\alpha_s$. 
 The one-point functions of the fluctuation fields are set to zero by requiring that the background fields satisfy their equations of motion, which incorporate both classical contributions and quantum corrections generated by the fluctuations. The leading perturbative corrections to the background TMD given in Eq. \eqref{def:q_Bckgd} within the operator definition \eqref{def:q_op_def_T_ord} arises from the terms that are quadratic in the fluctuation fields. In what follows we restrict ourselves to those contributions and systematically neglect corrections involving three or more fluctuation fields. 

We emphasize that the projectile light-cone gauge, $A^+=0$, exhibits an important structural difference compared to the target light-cone gauge, $A^-=0$. In the latter case, the longitudinal Wilson lines reduce to unity, and only the transverse Wilson line at infinity contributes non-trivially. By contrast, in the projectile light-cone gauge the longitudinal Wilson lines along $x^-$ remain nontrivial, generating additional one-loop diagrams involving $\delta A^-$ fluctuations. Expanding the full operator to quadratic order in the fluctuation fields, and restricting ourselves to the dilute target limit where each background field insertion is perturbatively suppressed, one finds 
\begin{align} 
& 
\big\langle P\big| \T \Big[ 
{\overline\Psi}(b^+, \mathbf{b}, 0^-) \, \frac{\gamma^-}{2} \, 
U(Y^+ , \mathbf{b}, 0^-;b^+, \mathbf{b}, 0^-)^{\dag}
U (Y^+, \mathbf{b}, 0^- ; Y^+,0_\perp  , 0^-) 
\, 
U (Y^+, 0_\perp , 0^- ; 0)   
\Psi(0) \Big]
\big|P\big\rangle_c
\nonumber \\
&-
\big\langle P\big| \T \Big[ 
{\overline\psi}(b^+, \mathbf{b}, 0^-) \, \frac{\gamma^-}{2} 
\psi(0) \Big]
\big|P\big\rangle_c
\nonumber
\\
=&\,
\big\langle P\big| \T \Big[ 
{\overline{\delta\Psi}}(b^+, \mathbf{b}, 0^-) \, \frac{\gamma^-}{2} \, 
\, 
\delta\Psi(0) \Big]
\big|P\big\rangle_c
\nonumber\\
&\,
+
\big\langle P\big| \T \Big[ 
{\overline\psi}(b^+, \mathbf{b}, 0^-) \, \frac{\gamma^-}{2} \, 
\left\{- i \mu^\epsilon g \int_0^1 d \tau\,  
\mathbf{b}^i\, t^a\, \delta A_i^a (Y^+, \tau \mathbf{b} , 0^-)\right\} 
\, 
\delta\Psi(0) \Big]
\big|P\big\rangle_c
\nonumber\\
&\,
+
\big\langle P\big| \T \Big[ 
{\overline{\psi}}(b^+, \mathbf{b}, 0^-) \, \frac{\gamma^-}{2} \, 
\frac{1}{2} {\cal{P}}\left\{- i \mu^\epsilon g \int_0^1 d \tau\,  
\mathbf{b}^i\, t^a\, \delta A_i^a (Y^+, \tau \mathbf{b} , 0^-)\right\}^2 
\, 
\psi(0) \Big]
\big|P\big\rangle_c
\nonumber\\
&\,
+
\big\langle P\big| \T \Big[ 
{\overline\psi}(b^+, \mathbf{b}, 0^-) \, \frac{\gamma^-}{2} \, 
\left\{+ i \mu^\epsilon g \int_{b^+}^{Y^+} \td z^+
 t^a \delta A_a^- (z^+, \mathbf{b} , 0^-)\right\}
\, 
\delta\Psi(0) \Big]
\big|P\big\rangle_c
\nonumber\\
&\,
+
\big\langle P\big| \T \Big[ 
{\overline{\psi}}(b^+, \mathbf{b}, 0^-) \, \frac{\gamma^-}{2} \, 
\frac{1}{2} {\cal{P}}\left\{+ i \mu^\epsilon g \int_{b^+}^{Y^+} \td z^+
 t^a \delta A_a^- (z^+, \mathbf{b} , 0^-)\right\}^2 
\, 
\psi(0) \Big]
\big|P\big\rangle_c
\nonumber\\
&\,
+
\big\langle P\big| \T \Big[ 
{\overline\psi}(b^+, \mathbf{b}, 0^-) \, \frac{\gamma^-}{2} \, 
\left\{- i \mu^\epsilon g \int_{0}^{Y^+} \td z^+
 t^a \delta A_a^- (z^+, 0_\perp , 0^-)\right\}
\, 
\delta\Psi(0) \Big]
\big|P\big\rangle_c
\nonumber\\
&\,
+
\big\langle P\big| \T \Big[ 
{\overline\psi}(b^+, \mathbf{b}, 0^-) \, \frac{\gamma^-}{2} \, 
\left\{- i \mu^\epsilon g \int_0^1 d \tau\,  
\mathbf{b}^i\, t^a\, \delta A_i^a (Y^+, \tau \mathbf{b} , 0^-)\right\}
\left\{- i \mu^\epsilon g \int_{0}^{Y^+} \td z^+
 t^b \delta A_b^- (z^+, 0_\perp , 0^-)\right\}
\, 
\psi(0) \Big]
\big|P\big\rangle_c
\nonumber\\
&\,
+
\big\langle P\big| \T \Big[ 
{\overline\psi}(b^+, \mathbf{b}, 0^-) \, \frac{\gamma^-}{2} \, 
\left\{+ i \mu^\epsilon g \int_{b^+}^{Y^+} \td z^+
 t^a \delta A_a^- (z^+, \bbperp , 0^-)\right\}
\left\{- i \mu^\epsilon g \int_{0}^{Y^+} \td z^+
 t^b \delta A_b^- (z^+, 0_\perp , 0^-)\right\}
\, 
\psi(0) \Big]
\big|P\big\rangle_c\nonumber
\\
&+\dots\,.
\label{eq: operator expansion in background field}
\end{align}
The first line on the right-hand side corresponds to the ladder diagrams, while the second and third lines arise from the transverse Wilson line at infinity and its self-energy, as in the target light-cone gauge. The forth, fifth, and sixth lines are new compared to Ref.~\cite{Altinoluk:2025ewj} and arise from the longitudinal Wilson lines along $x^-$, which are nontrivial in the projectile light-cone gauge. The seventh and eighth lines correspond to mixed diagrams involving both transverse and longitudinal parts of the gauge link. 
The remaining one-loop corrections, which are symmetric to the terms above, are listed in the Appendix~\ref{appendix_symmetric_diagrams}.
The corresponding Feynman diagrams are presented in Fig.~\ref{Fig:real_diags 1}. 

In the remainder of this section, we compute these diagrams.
%%%%%%%%%%%%%%%%%%%%%%%%%%%%%%%%%%%%%%%%%%%%%%%
%%%%%%%%%%%%%%%%%%%%%%%%%%%%%%%%%%%%%%%%%%%%%%%
%%%%%%%%%%%%%%%%%%%%%%%%%%%%%%%%%%%%%%%%%%%%%%%

%====================================================================================
\subsection{Diagrams~\ref{Fig:diag6} and \ref{Fig:diag4}: gluon emissions from the quark to the lower and upper part of the gauge link}
\label{sec: diag 6 and 4}
%====================================================================================

We begin our analysis with the diagrams in which a gluon is radiated by the background quark field and subsequently absorbed by one of the longitudinal Wilson lines in the gauge staple. Diagram~\ref{Fig:diag6} corresponds to absorption by the lower part of the staple (the Wilson line running from $0$ to $Y^+$ along $x^+$ direction at $0_\perp$), while diagram~\ref{Fig:diag4} corresponds to absorption by the upper part (the Wilson line running from $-b^+$ to $Y^+$ along $x^+$ direction  at $ -\mathbf{b}$). These diagrams are genuinely new compared to the target light-cone gauge calculation of Ref.~\cite{Altinoluk:2025ewj}, where the longitudinal Wilson lines reduce to unity and do not contribute to perturbative corrections. 

%====================================
\subsubsection{Diagram~\ref{Fig:diag6}}
%====================================

\begin{figure}[h!]
{%
       \includegraphics[width=0.40\textwidth]{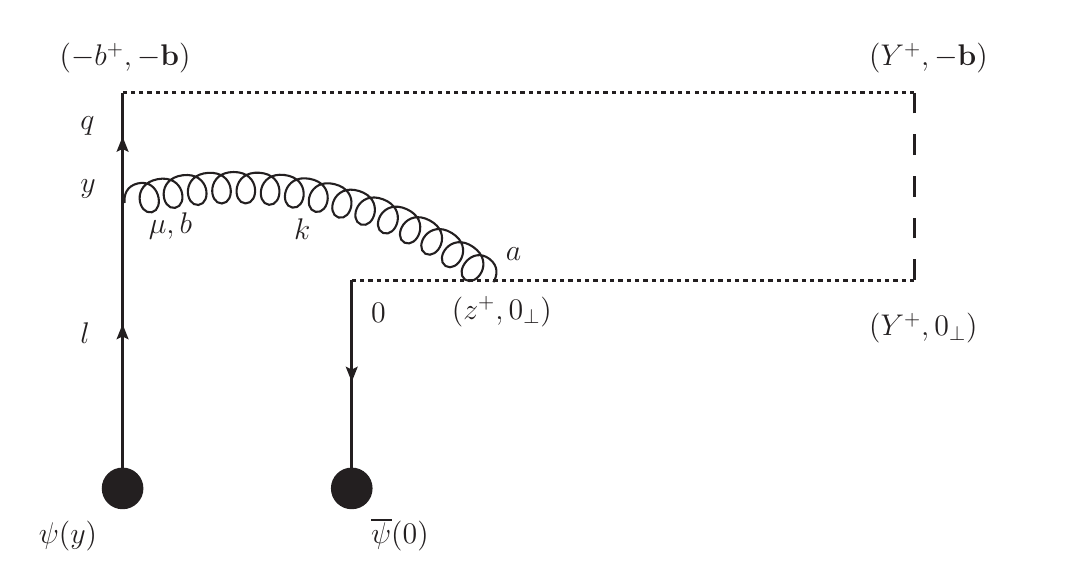}
     }
\caption{Gluon emission from the quark to the lower part of the gauge link - diagram~\ref{Fig:diag6}.
}
\end{figure}

After performing the spacetime translation introduced in Eq. \eqref{def:q_op_def_T_ord_transl}, the contribution of diagram~\ref{Fig:diag6} to the quark TMD reads  
\begin{align}
\label{def:q_dist-1-loop-rad-to-lower-ww}
 q^{\textrm{n.r.}}_{\textrm{unsub.}}&(\tx, \mathbf{b};\mu^2)\Big|_{\ref{Fig:diag6}}  =
%\nonumber \\
%= & 
\lim_{Y^+\rightarrow +\infty}
\int \frac{\td b^+}{(2\pi)}\, 
e^{-i\tx P^-b^+}
\nonumber \\
& \hspace{3cm}\times 
\big\langle P\big| \T \Big[ 
{\overline\psi} (0)  \, \frac{\gamma^-}{2} \, 
(+ i) g \mu^\epsilon \int_0^{Y^+} d z^+ \,  
\delta A^-_a(z^+, 0_\perp , 0^-) \, t_a
%\nn
%& \hspace{8cm}
%\,
%\times
\delta \Psi ( - b^+, - \mathbf{b} , 0^-) \Big]
\big|P\big\rangle_c
\, .
\end{align}
Expressing the two-point correlator of the fluctuation fields in terms of free Feynman propagators and background field $\psi$ insertion vertex, one finds
\begin{align}
q^{\textrm{n.r.}}_{\textrm{unsub.}}(\tx, \mathbf{b};\mu^2)\Big|_{\ref{Fig:diag6}} & =  
\lim_{Y^+\rightarrow +\infty}
\int \frac{\td b^+}{(2\pi)}\, 
e^{-i\tx P^-b^+}
\big\langle P\big| \T \Big[ 
{\overline\psi} (0)  \, \frac{\gamma^-}{2}
\bigg(+ i g \mu^\epsilon  
\int_0^{Y^+} d z^+ \,  
\delta A^-_a (z^+, 0_\perp , 0^-) \, t_a 
\nonumber \\
& \hspace{2cm}
\times \int \td^{4 - 2 \epsilon} y \, 
S_{0,F}(- b^+ , - \mathbf{b} , 0^- ; y)
\left[- i g \mu^\epsilon \, t_b \gamma_\mu\right] 
\delta A^\mu_b (y) \bigg)
\psi (y)
\big|P\big\rangle_c
\nonumber \\
&
= g^2 \mu^{2\epsilon} C_F \, 
\lim_{Y^+\rightarrow +\infty} 
\int \frac{\td b^+}{(2\pi)}\, 
e^{-i\tx P^-b^+}
\int \td^{4 - 2 \epsilon} y \, 
\big\langle P\big|  {\overline\psi}_\alpha (0) \,  \psi_\beta (y)\big|P\big\rangle_c
\nonumber \\
& \hspace{2cm}
\times\int_0^{Y^+} d z^+ 
\bigg\{
 \frac{\gamma^-}{2} S_{0,F}(- b^+ , - \mathbf{b} , 0^- ; y) \gamma_\mu
 \bigg\}_{\alpha\beta}
G_{0,F}^{- \mu}(z^+, 0_\perp, 0^- ; y) \,,
\label{eq:Diag_1a_inter_1}
\end{align}
where $\alpha,\beta$ denote Dirac spinor indices of the quark fields. Using the explicit forms of the free quark and gluon propagators in momentum space, Eq. \eqref{eq:Diag_1a_inter_1} can be written as 
\begin{align}
 q^{\textrm{n.r.}}_{\textrm{unsub.}}&(\tx, \mathbf{b};\mu^2)\Big|_{\ref{Fig:diag6}}   
  =
g^2 \mu^{2\epsilon} C_F \, 
\lim_{Y^+\rightarrow +\infty} \int \frac{\td b^+}{(2\pi)}\, 
e^{-i\tx P^-b^+} \int \td^{4 - 2 \epsilon} y \, 
\big\langle P\big|  {\overline\psi}_\alpha (0) \,  \psi_\beta (y)\big|P\big\rangle_c
\nonumber \\
 & \hspace{1.8cm}\times
\int_0^{Y^+} d z^+ 
\int \frac{\td^{4-2\epsilon} q}{(2\pi)^{4-2\epsilon}} 
\int \frac{\td^{4-2\epsilon} k}{(2\pi)^{4-2\epsilon}} \, 
\bigg\{
 \frac{\gamma^-}{2} \frac{i \slq}{q^2 + i0} \gamma_\mu
 \bigg\}_{\alpha\beta}\, \frac{i}{k^2 + i0}
\nonumber \\
&
  \hspace{1.8cm} \times 
 \bigg\{
 - g^{-\mu} + \frac{n^- k^\mu + n^\mu k^-}{n\cdot k}
 \bigg\} \; 
 e^{i [q^- (y^+ + b^+) + q^+ y^- - \bq\cdot(\by + \bbperp)]}
  e^{i [k^- (y^+ - z^+) + k^+ y^- - \bk\cdot\by]} \,.
\end{align}
Upon integrating over $b^+$ and then over $q^-$ which sets $q^-=\tx\, P^-$, the contribution from diagram~\ref{Fig:diag6} takes the form    
\begin{align}
q^{\textrm{n.r.}}_{\textrm{unsub.}}(\tx, \mathbf{b};\mu^2)\Big|_{\ref{Fig:diag6}}  = &
\, \alpha_s \, \mu^{2\epsilon} C_F 
\lim_{Y^+\rightarrow +\infty} 
\int \td^{4 - 2 \epsilon} y \,
\big\langle P\big|\T\, {\overline\psi}_\alpha (0)  \, \psi_\beta (y) \big|P\big\rangle_c \, 
\int_0^{Y^+} \td z^+ \, 
\int \frac{\td^{4-2\epsilon} k}{(2\pi)^{4-2\epsilon}} 
e^{i (\tx P^- + k^-) y^+}\nonumber \\
&
\times
\int \frac{\td^{2-2\epsilon} \bl}{(2\pi)^{2-2\epsilon}} 
\int \frac{\td l^+}{(2 \pi)} 
 e^{i (l^+ y^- - \bl\cdot\by)}  
e^{-i (\bl - \bk)\cdot\bbperp} e^{-i k^- z^+} \,
\frac{i}{k^2 + i0} \frac{i}{q^2 + i0}\bigg|_{q^- = \tx P^-}
\nonumber \\
&
\times
\bigg\{
 \gamma^- \left[\tx P^- \gamma^+ - (\bl - \bk)^i\bg^i\right] \gamma_\mu
 \bigg\}_{\alpha\beta}
 \left[
 - g^{-\mu} + \frac{n^- k^\mu + n^\mu k^-}{n\cdot k}
 \right] \,,
 \label{eq:Diag_1a_inter_2}
\end{align}
where we have performed a change of variables from $q$ to $l$ such that 
\begin{align}
\label{eq:cov-q-to-l}
   q^+ \mapsto &\,  l^+ \equiv q^+ + k^+ \, ,  
    \nonumber \\
   \bq \mapsto &\, \bl \equiv \bq + \bk \, . 
\end{align}
It is straightforward to show that the tensor structure of the gluon propagator contracted with $\gamma_\mu$ can be evaluated as
\begin{align}
\gamma_\mu \,
 \left[
 - g^{-\mu} + \frac{n^- k^\mu + n^\mu k^-}{n\cdot k}
 \right] =
 \left[- \gamma^- + \frac{\slk + \sln k^-}{n\cdot k} \right] \, . 
 \label{eq:g_prop_contracted_gamma_mu}
\end{align}
Substituting Eq. \eqref{eq:g_prop_contracted_gamma_mu} into Eq. \eqref{eq:Diag_1a_inter_2}, the contribution of diagram~\ref{Fig:diag6} can be written as  
\begin{align}
q^{\textrm{n.r.}}_{\textrm{unsub.}}(\tx, \mathbf{b};\mu^2)\Big|_{\ref{Fig:diag6}}  = &
\alpha_s \mu^{2\epsilon} C_F 
\lim_{Y^+\rightarrow +\infty} 
\int \td^{4 - 2 \epsilon} y \,
\big\langle P\big|\T\, {\overline\psi}_\alpha (0)\,\psi_\beta (y) \big|P\big\rangle_c \, 
e^{i \tx P^- y^+}\, 
\int_0^{Y^+} \td z^+ \, 
\int \frac{\td^{4-2\epsilon} k}{(2\pi)^{4-2\epsilon}} 
e^{i k^- (y^+ - z^+)}\nonumber \\
&
\times
\int \frac{\td^{2-2\epsilon} \bl}{(2\pi)^{2-2\epsilon}} 
\int \frac{\td l^+}{(2 \pi)} 
 e^{i (l^+ y^- - \bl\cdot\by)} \,  
e^{-i (\bl - \bk)\cdot\bbperp} 
\frac{i}{k^2 + i0} \frac{i}{q^2 + i0}_{q^- = \tx P^-}
\frac{1}{[k^+]}
\nonumber \\
&
\times\bigg\{
 \gamma^- \left[\tx P^- \gamma^+ - (\bl - \bk)^i\bg^i\right]
 \left[2 k^- \gamma^+ - \bk^j\bg^j\right] 
 \bigg\}_{\alpha\beta} \,.
 \label{eq:Diag_1a_inter_3}
\end{align}
The $[k^+]$ denominator in the second line of  Eq.~\eqref{eq:Diag_1a_inter_3} is the extra denominator originating  from the gluon propagator in light-cone gauge given in Eq. \eqref{eq:Feyn-prop_def} and as mentioned earlier it is treated within the ML prescription \eqref{eq:ML_prescription}, such that 
\begin{align}
\frac{1}{[k^+]}
\equiv \frac{\theta(k^-)}{(k^+\!+\!i0)} \, 
+\frac{\theta(-k^-)}{(k^+\!-\!i0)}\, . 
\label{eq:ml-prescription_lc_var} 
\end{align}
The contour integration over $k^+$ can now be performed by using Eq. \eqref{eq:ml-prescription_lc_var} which reads, for $\tx>0$, 
\begin{align}
 \int \frac{\td k^+}{(2 \pi)} 
\left\{
\frac{\theta(k^-)}{k^+ + i0} + \frac{\theta(-k^-)}{k^+ - i0}
\right\} \, &
\frac{i}{2 k^- \left[k^+ - \frac{\bk^2 - i0}{2 k^-}\right]}
\frac{i}{\left[2 \tx P^- (l^+ - k^+)  - (\bl - \bk)^2 + i0\right]}
\nonumber \\
& \hspace{1cm}
= \frac{i \theta (k^-) \tx P^-}
{\left[2 \tx P^- l^+ - (\bl - \bk)^2\right] 
\left[k^- \left(2 \tx P^- l^+ - (\bl - \bk)^2\right) - \tx P^- \bk^2\right]}
\nonumber \\
&  \hspace{1cm}
=\frac{i \theta (k^-)}{\bk^2} \left[ -\frac{1}{\left[2 \tx P^- l^+ - (\bl - \bk)^2\right]} +\frac{k^-}{k^-\left[2 \tx P^- l^+ - (\bl - \bk)^2\right]-\tx P^- \bk^2}\right]\,,
\end{align}
where the contributions from negative $k^-$ vanish since all poles lie on the same side of the real axis in the complex plane in that case. The integration over $z^+$ can also be performed in a trivial manner. After all, the contribution from diagram~\ref{Fig:diag6} reads  
\begin{align}
q^{\textrm{n.r.}}_{\textrm{unsub.}}(\tx, \mathbf{b};\mu^2)\Big|_{\ref{Fig:diag6}} & =  
- \alpha_s \mu^{2\epsilon} C_F \, 
\lim_{Y^+\rightarrow +\infty} 
\int \td^{4 - 2 \epsilon} y \,
\big\langle P\big|\T\, {\overline\psi}_\alpha (0)\,\psi_\beta (y) \big|P\big\rangle_c \, 
e^{i \tx P^- y^+}
\int \frac{\td^{2-2\epsilon} \bl}{(2\pi)^{2-2\epsilon}} 
\int \frac{\td l^+}{(2 \pi)} \, e^{i (l^+ y^- - \bl\cdot(\by+\bbperp))} 
\nonumber \\
& 
\times\int \frac{\td^{2-2\epsilon} \bk}{(2\pi)^{2-2\epsilon}}
\,\frac{e^{i \bk\cdot\bbperp}}{\bk^2\left[2 \tx P^- l^+ - (\bl - \bk)^2\right]}
\int_0^\infty \frac{\td k^-}{2 \pi} \,\frac{1}{k^-}\,
\left[e^{-i k^- (Y^+ - y^+)} - e^{i k^- y^+}\right] \nonumber \\
&
\times\left[ -1+\frac{k^-}{k^--\frac{\tx P^- \bk^2}{\left[2 \tx P^- l^+ - (\bl - \bk)^2\right]}}\right]\bigg\{
 \gamma^- \left[\tx P^- \gamma^+ - (\bl - \bk)^i\bg^i\right]
 \left[2 \gamma^+k^- - \bk^j\bg^j\right] 
 \bigg\}_{\alpha\beta}\,.
 \label{eq:Diag_1a_inter_4}
\end{align}
We now focus on the $k^-$ integral. We note that, for finite $Y^+$, the integral is free of divergences in both $k^- \rightarrow 0$ and $k^- \rightarrow \infty$ limits. However, upon taking the limit $Y^+ \rightarrow \infty$, the first term in the square brackets in the second line of \eqref{eq:Diag_1a_inter_4} vanishes due to rapidly oscillating phase factor. By contrast, the second term develops a rapidity singularity as $k^-\to 0$, which needs to be regulated. We implement pure rapidity regulator to regulate this singularity which corresponds to multiplying the integrand by the factor given in Eq. \eqref{def:pure_rap_reg_OS}. Then, the contribution from diagram~\ref{Fig:diag6} can be written as 
\begin{align}
q^{\textrm{n.r.}}_{\textrm{unsub.}}(\tx, \mathbf{b};\mu^2)\Big|_{\ref{Fig:diag6}} & =  
- \alpha_s \mu^{2\epsilon} C_F \, 
\lim_{Y^+\rightarrow +\infty} 
\int \td^{4 - 2 \epsilon} y \,
\big\langle P\big|\T\, {\overline\psi}_\alpha (0)\,\psi_\beta (y) \big|P\big\rangle_c \, 
e^{i \tx P^- y^+}
\int \frac{\td^{2-2\epsilon} \bl}{(2\pi)^{2-2\epsilon}} 
\int \frac{\td l^+}{(2 \pi)} \, e^{i (l^+ y^- - \bl\cdot(\by+\bbperp))} 
\nonumber \\
& 
\times\int \frac{\td^{2-2\epsilon} \bk}{(2\pi)^{2-2\epsilon}}
\,\frac{e^{i \bk\cdot\bbperp}}{\bk^2\left[2 \tx P^- l^+ - (\bl - \bk)^2\right]} \, 
 w^2\left[\frac{2 \nu^+}{\bk^2 \nu^-}\right]^{\frac{\eta}{2}}
\int_0^\infty \frac{\td k^-}{2 \pi} \,
\left[k^-\right]^{\eta-1}\,
%\int_0^\infty \frac{\td k^-}{2 \pi} \,\frac{1}{k^-}\,
\left[e^{-i k^- (Y^+ - y^+)} - e^{i k^- y^+}\right] \nonumber \\
&
\times\left[ -1+\frac{k^-}{k^--\frac{\tx P^- \bk^2}{\left[2 \tx P^- l^+ - (\bl - \bk)^2\right]}}\right]\bigg\{
 \gamma^- \left[\tx P^- \gamma^+ - (\bl - \bk)^i\bg^i\right]
 \left[2 \gamma^+k^- - \bk^j\bg^j\right] 
 \bigg\}_{\alpha\beta} \,.
 \label{eq:Diag_1a_inter_5}
\end{align}
Now that the rapidity regulator is introduced in Eq. \eqref{eq:Diag_1a_inter_5}, the limit $Y^+\to \infty$ can be taken safely. Thanks to the pure rapidity regulator, the first phase factor now vanishes. After introducing a new change of variable 
\begin{align}
\label{cov:k_minus_2_z}
k^-\mapsto z \equiv \frac{\tx P^-}{k^- + \tx P^-}\,,
\end{align}
Eq. \eqref{eq:Diag_1a_inter_5} can be rewritten as 
\begin{align}
q^{\textrm{n.r.}}_{\textrm{unsub.}}(\tx, \mathbf{b};\mu^2)\Big|_{\ref{Fig:diag6}} & =  
- \alpha_s \mu^{2\epsilon} C_F \, 
%\lim_{Y^+\rightarrow +\infty} 
\int \td^{4 - 2 \epsilon} y \,
\big\langle P\big|\T\, {\overline\psi}_\alpha (0)\,\psi_\beta (y) \big|P\big\rangle_c \, 
e^{i \tx P^- y^+}
\int \frac{\td^{2-2\epsilon} \bl}{(2\pi)^{2-2\epsilon}} 
\int \frac{\td l^+}{(2 \pi)} \, e^{i (l^+ y^- - \bl\cdot(\by+\bbperp))} 
\nonumber \\
& 
\times\int \frac{\td^{2-2\epsilon} \bk}{(2\pi)^{2-2\epsilon}}
\frac{e^{i \bk\cdot\bbperp}}{\bk^2\left[2 \tx P^- l^+ - (\bl - \bk)^2\right]} \, 
 w^2\left[\frac{2 \nu^+}{\bk^2 \nu^-}\right]^{\frac{\eta}{2}}
\frac{[\tx P^-]^\eta}{2\pi} e^{- i \tx P^- y^+} 
\int_0^1 \td z 
[1 - z]^{\eta -1} [z]^{- \eta} e^{i  \tx P^- y^+ \frac{1}{z} } 
\nonumber \\
&
\times
\left[ 1-\frac{1-z}{1-z-\frac{z \bk^2}{\left[2 \tx P^- l^+ - (\bl - \bk)^2\right]}}\right]
\bigg\{
 \gamma^- \left[\tx P^- \gamma^+ - (\bl - \bk)^i\bg^i\right]
\left[2 \gamma^+\tx P^-\frac{1-z}{z} - \bk^i\bg^i\right]
 \bigg\}_{\alpha\beta} \,,
 \label{eq:Diag_1a_inter_6}
\end{align}
where the rapidity divergence appears as a pole in $\eta$ as $\eta\to0$ at $z=1$. We note that the terms proportional to $[1-z]^\eta$ (with $\eta > 0$) are finite as $z\to 1$. 
.% 
To isolate the rapidity divergent part of Eq. \eqref{eq:Diag_1a_inter_6}, we focus on the term proportional to $[1 - z]^{\eta -1}$ and employ the same procedure as in Ref.~\cite{Altinoluk:2025ewj} to extract the rapidity pole which takes the form 
\begin{align}
\int_0^1 \td \tz\, (1\!-\! \tz)^{\eta-1}\, f(\tz, \eta) 
=&\,
f(1, \eta)\int_0^1 \td \tz\, (1\!-\!\tz)^{\eta-1}\, 
+\int_0^1 \td \tz\, (1\!-\!\tz)^{\eta-1}
\Big[
f(\tz, \eta) -f(1, \eta)
\Big]
\nonumber\\
=&\,
\frac{f(1, \eta)}{ \eta}
+\int_0^1 \td \tz\, \frac{\big[f(\tz, 0)\!-\!f(1, 0)\big]}{(1\!-\!\tz)}
+O(\eta)\,,
\label{eq:eta_expand}
\end{align}
with
\begin{align}
\label{def:fzeta}
    f(z, \eta)\equiv 
 w^2\left[\frac{2 \nu^+}{\bk^2 \nu^-}\right]^{\frac{\eta}{2}}\frac{[\tx P^-]^\eta}{2\pi} e^{- i \tx P^- y^+}  [z]^{- \eta} e^{i  \tx P^- y^+ \frac{1}{z} }(-\bk^i\bg^i)\,.
\end{align}
Using Eq.\eqref{eq:eta_expand} together with Eq. \eqref{def:fzeta}, the rapidity divergent part of the contribution from diagram~\ref{Fig:diag6} takes the following form: 
\begin{align}
\label{eq: diag 6 ML intermediate result}
q^{\textrm{n.r.}}_{\textrm{unsub.}}&(\tx, \mathbf{b};\mu^2,\zeta)\Big|_{\ref{Fig:diag6}}^{1/\eta}  \nonumber\\
& =  
- \frac{\alpha_s}{2 \pi} \mu^{2\epsilon} C_F \, [\tx P^-]^\eta \,
 w^2\left[\frac{2 \nu^+}{\nu^-}\right]^{\frac{\eta}{2}} \,
\int \td^{4 - 2 \epsilon} y \,
\big\langle P\big|\T\, {\overline\psi}_\alpha (0)\,\psi_\beta (y) \big|P\big\rangle_c \, 
e^{i \tx P^- y^+}\, 
\int \frac{\td^{2-2\epsilon} \bl}{(2\pi)^{2-2\epsilon}} 
\int \frac{\td l^+}{(2 \pi)} \, e^{i (l^+ y^- - \bl\cdot\by)} 
\nonumber \\
& 
\times\int \frac{\td^{2-2\epsilon} \bk}{(2\pi)^{2-2\epsilon}}
\frac{e^{-i (\bl - \bk)\cdot\bbperp}}
{\left[2 \tx P^- l^+ - (\bl - \bk)^2\right]}
\frac{1}{[\bk^2]^{\frac{\eta}{2} + 1}} 
\left\{
\gamma^- \left[\tx P^- \gamma^+ - (\bl - \bk)^i\bg^i\right] (- \bk^j\bg^j)
\right\}_{\alpha\beta}
\frac{1}{\eta} \,,
\end{align}
where the superscript $1/\eta$ denotes that Eq. \eqref{eq: diag 6 ML intermediate result} is the rapidity-divergent part of the diagram~\ref{Fig:diag6}.
To proceed further we use equation of motion (EOM) for the quark background field which, in the dilute target approximation and at lowest order in perturbation theory reduces to the free Dirac equation,  
$\sll \psi (l) =0+ O(\alpha_s)$ 
\footnote{ This follows from the Lorentz transformation  $\alpha_s\int \td ^{4-2\epsilon}y \, e^{iy\cdot l}\langle P| \bar{\psi}(0)\gamma^- \slashed{l}\psi(y)|P \rangle \propto \mathcal{O}(\alpha_s^2)$ for massless quarks. 
%Essentially, applying the Dirac equation shifts some of the rapidity-divergent contributions to a higher order in $\alpha_s$. 
Note that using the Dirac equation in this context requires the phase factor in the Lorentz transformation to be consistent with the quark momentum.}. Then,  one finds
\begin{align}
\label{eq:diracforpsi}
\gamma^- \tx P^- \gamma^+ (- \bk \cdot \bg ) \, \psi (l) =
\bk^j \gamma^- \gamma^j \bl^i\bg^i \, \psi (l)\,
+ O(\alpha_s).
\end{align}
Using Eq. \eqref{eq:diracforpsi} and discarding the terms of order $O(\alpha_s^2)$, the rapidity divergent part of the contribution from diagram~\ref{Fig:diag6} can be further simplified and written as   
\begin{align}
\label{eq:Diag_1a_inter_7}
q^{\textrm{n.r.}}_{\textrm{unsub.}}(\tx, \mathbf{b};\mu^2,\zeta)\Big|^{1/\eta}_{\ref{Fig:diag6}} & =  
- \frac{\alpha_s}{2 \pi} \mu^{2\epsilon} C_F \, [\tx P^-]^\eta \,
 w^2\left[\frac{2 \nu^+}{\nu^-}\right]^{\frac{\eta}{2}} \,
\frac{1}{\eta}\, 
\int \td^{4 - 2 \epsilon} y \,
\big\langle P\big|\T\, {\overline\psi} (0)\, \frac{\gamma^-}{2}\, \psi (y) \big|P\big\rangle_c \, 
e^{i \tx P^- y^+}\, 
 \nonumber \\
&
\times\int \frac{\td^{2-2\epsilon} \bl}{(2\pi)^{2-2\epsilon}} 
\int \frac{\td l^+}{(2 \pi)} \, e^{i (l^+ y^- - \bl\cdot(\by + \bbperp))}  
\int \frac{\td^{2-2\epsilon} \bk}{(2\pi)^{2-2\epsilon}}
\frac{e^{i \bk\cdot\bbperp} [\bk^2 - 2 \bk\cdot\bl]}
{[\bk^2]^{\frac{\eta}{2} + 1} [2 \tx P^- l^+ - (\bl - \bk)^2]} \,.
\end{align}
Let us focus on the integral over the transverse momentum $\bk$ in Eq. \eqref{eq:Diag_1a_inter_7}. By completing the square in the numerator under the the transverse momentum $\bk$ integral in Eq. \eqref{eq:Diag_1a_inter_7}, one gets 
\begin{align}
\int \frac{\td^{2-2\epsilon} \bk}{(2\pi)^{2-2\epsilon}}
\frac{e^{i \bk\cdot\bbperp} [\bk^2 - 2 \bk\cdot\bl]}
{[\bk^2]^{\frac{\eta}{2} + 1} [2 \tx P^- l^+ - (\bl - \bk)^2]} 
=
\int \frac{\td^{2-2\epsilon} \bk}{(2\pi)^{2-2\epsilon}}
\frac{e^{i \bk\cdot\bbperp}}{[\bk^2]^{\frac{\eta}{2} + 1}}
\bigg[
- 1 + \frac{l^2}{[2 \tx P^- l^+ - (\bl - \bk)^2]} 
\bigg]\,,
\label{eq:diag_1a_k_int}
\end{align}
where we have used $l^- = \tx P^-$. Using again the EOM for the quark background field, which reduces at lowest order to the free Dirac equation, one has $\sll [\sll \psi (l) = 0+ O(\alpha_s)] \Rightarrow l^2 \psi (l) = O(\alpha_s)$. Hence, the contribution of each component of the spinor field multiplied by $l^2$ is shifted to higher order in $\alpha_s$. Consequently, the contribution from the second term in the square bracket on the right hand side of Eq. \eqref{eq:diag_1a_k_int} can be discared for our purposes, and one obtains 
\begin{align}
\label{eq:Diag_1a_inter_8}
q^{\textrm{n.r.}}_{\textrm{unsub.}}(\tx, \mathbf{b};\mu^2,\zeta)\Big|^{1/\eta}_{\ref{Fig:diag6}} & =  
- \frac{\alpha_s}{2 \pi} \mu^{2\epsilon} C_F \, [\tx P^-]^\eta \,
 w^2\left[\frac{2 \nu^+}{\nu^-}\right]^{\frac{\eta}{2}} \,
\frac{1}{\eta}\, 
\int \td^{4 - 2 \epsilon} y \,
\big\langle P\big|\T\, {\overline\psi} (0)\, \frac{\gamma^-}{2}\, \psi (y) \big|P\big\rangle_c \, 
e^{i \tx P^- y^+}\, 
 \nonumber \\
&
\times\int \frac{\td^{2-2\epsilon} \bl}{(2\pi)^{2-2\epsilon}} 
\int \frac{\td l^+}{(2 \pi)} \, e^{i (l^+ y^- - \bl\cdot(\by + \bbperp))}  
(-1) \int \frac{\td^{2-2\epsilon} \bk}{(2\pi)^{2-2\epsilon}}
\frac{e^{i \bk\cdot\bbperp}}{[\bk^2]^{\frac{\eta}{2} + 1}}
\end{align}
for the rapidity divergent part of the contribution from diagram~\ref{Fig:diag6}. Note that the transverse momentum integral over $\k$ in Eq. \eqref{eq:Diag_1a_inter_8} is IR divergent. However, as will be shown later in this subsection, when combined with the rapidity divergent contribution from diagram~\ref{Fig:diag4}, the IR divergences will cancel and be replaced by the UV divergence expected for the CSS evolution. 

On the other hand, by collecting the rapidity finite contributions already present in Eq. \eqref{eq:Diag_1a_inter_6}, together with the ones originating from the second term in Eq. \eqref{eq:eta_expand} and setting $\eta=0$, one obtains
\begin{align}
\label{eq:diag_1a_fin_1}
q^{\textrm{n.r.}}_{\textrm{unsub.}} &(\tx, \mathbf{b};\mu^2,0)\Big|_{\ref{Fig:diag6}}^{\rm fin.}  =  
 \frac{\alpha_s \mu^{2\epsilon} C_F}{2\pi} \,  
\int \td^{4 - 2 \epsilon} y \,
\big\langle P\big|\T\, {\overline\psi}_\alpha (0)\,\psi_\beta (y) \big|P\big\rangle_c \, e^{i \tx P^- y^+}\, 
\int \frac{\td^{2-2\epsilon} \bl}{(2\pi)^{2-2\epsilon}} 
\int \frac{\td l^+}{(2 \pi)} \, e^{i (l^+ y^- - \bl\cdot(\by+\bbperp))} 
\nonumber \\
& 
\times\int \frac{\td^{2-2\epsilon} \bk}{(2\pi)^{2-2\epsilon}}
\,\frac{e^{i \bk\cdot\bbperp}e^{-i \tx P^- y^+}\, }{\bk^2\left[2 \tx P^- l^+ - (\bl - \bk)^2\right]} \Bigg\{
 \gamma^- \left[\tx P^- \gamma^+ - (\bl - \bk)^i\bg^i\right]
  \nonumber \\
&
\times 
\int_0^1 \td z\,\left[
\frac{e^{i \tx P^- y^+\frac{1}{z}}}{z^2}\left[ \frac{2 \gamma^+\tx P^- \frac{1-z}{z} - \bk^i\bg^i}{ \frac{1-z}{z}-\frac{ \bk^2}{\left[2 \tx P^- l^+ - (\bl - \bk)^2\right]}}-2\tx P^-\gamma^+\right] -\frac{\bk^i\gamma^i}{1-z}\left[e^{i\tx P^- y^+\frac{1}{z}}-e^{i\tx P^- y^+}\right]\right]
 \Bigg\}_{\alpha\beta}\,
\end{align}
for the contribution of the diagram~\ref{Fig:diag6} that is free of the rapidity divergences. By realizing that 
\begin{align}
    \left[\frac{2 \gamma^+\tx P^- \frac{1-z}{z} - \bk^i\bg^i}{ \frac{1-z}{z}-\frac{ \bk^2}{\left[2 \tx P^- l^+ - (\bl - \bk)^2\right]}}-2\tx P^-\gamma^+\right]
    =
    \frac{\frac{ 2\tx P^- \gamma^+\bk^2}{\left[2 \tx P^- l^+ - (\bl - \bk)^2\right]}- \bk^i\bg^i}{ \frac{1-z}{z}-\frac{ \bk^2}{\left[2 \tx P^- l^+ - (\bl - \bk)^2\right]}}
\end{align}
the two types of terms appearing on the right hand side of Eq. \eqref{eq:diag_1a_fin_1} takes the following form: 
%By writing  
\begin{align}
\label{simpf_1}
     \gamma^- \left[\tx P^- \gamma^+ - (\bl-\bk) ^i\bg^i\right]\bk^i\gamma^i=  \gamma^- \left[-\bk^i\gamma^i\slashed{l}-(\bk-\bl)^2+\bl^2\right]\,, 
\end{align}
\begin{align}
\label{simpf_2}
     \gamma^- \left[\tx P^- \gamma^+ - (\bl-\bk) ^i\bg^i\right]\tx P^- \gamma^+=  \gamma^- \left[-(\bl-\bk)^i\gamma^i\slashed{l}+\bl^2+\bk^i\gamma^i\bl^j\gamma^j\right]\,.  
\end{align}
Using Eqs. \eqref{simpf_1} and \eqref{simpf_2}, and employing one more time the Dirac equation for the background quark field which allows us to discard the terms that are proportional to $\slashed{l}$, the integral over the transverse momentum $\bk$ integral can be further simplified and the  rapidity finite contribution to diagram~\ref{Fig:diag6} reads 
\begin{align}
\label{eq:diag_1a_fin_2}
q^{\textrm{n.r.}}_{\textrm{unsub.}} &(\tx, \mathbf{b};\mu^2,0)\Big|_{\ref{Fig:diag6}}^{\rm fin.}  =  
 \frac{\alpha_s \mu^{2\epsilon} C_F}{2\pi} \,  
\int \td^{4 - 2 \epsilon} y \,
\big\langle P\big|\T\, {\overline\psi}_\alpha (0)\,\psi_\beta (y) \big|P\big\rangle_c \, 
%e^{i \tx P^- y^+}\, 
\int \frac{\td^{2-2\epsilon} \bl}{(2\pi)^{2-2\epsilon}} 
\int \frac{\td l^+}{(2 \pi)} \, e^{i (l^+ y^- - \bl\cdot(\by+\bbperp))} 
\nonumber \\
& \hspace{-1.5cm}
\times 
\gamma^- \,\int \frac{\td^{2-2\epsilon} \bk}{(2\pi)^{2-2\epsilon}}
\,\frac{e^{i \bk\cdot\bbperp} }{\bk^2\left[2 \tx P^- l^+ - (\bl - \bk)^2\right]} 
\int_0^1 \td z\,
\nonumber\\
& \hspace{-1.5cm}\times
\bigg[
\frac{e^{i \tx P^- y^+\frac{1}{z}}}{z^2} \frac{ 2\frac{1-z}{z} (\bl^2+\bk^i\gamma^i\bl^j\gamma^j)\bk^2 - (\bl^2-(\bk-\bl)^2)\left[2 \tx P^- l^+ - (\bl - \bk)^2\right]}{ \left[2 \tx P^- l^+ - (\bl - \bk)^2\right]\frac{1-z}{z}- \bk^2} 
%\nonumber \\
%&
-
(\bl^2 -(\bk-\bl)^2)\frac{e^{i\tx P^- y^+\frac{1}{z}}-e^{i\tx P^- y^+}}{1-z}\bigg]\,,
\end{align}
which is free from divergences in the limit $\bk\to 0$ and in the limit $z\to 1$, and therefore it is completely finite.

%====================================================================================
\subsubsection{Diagram~\ref{Fig:diag4}}
\label{subsubsection: diagram 4}
%====================================================================================

\begin{figure}[h!]
{%
       \includegraphics[width=0.40\textwidth]{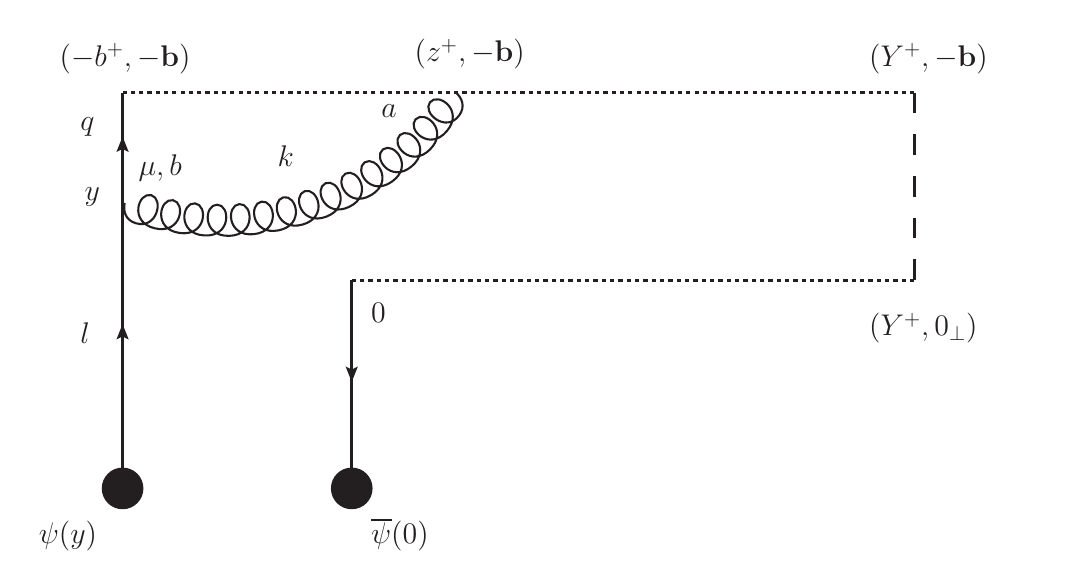}
     }

\caption{Gluon emission from the quark to the upper part of the gauge link - diagram~\ref{Fig:diag4}.}
\end{figure}

Diagram~\ref{Fig:diag4} corresponds to radiation of a gluon from the quark field which is then absorbed by the upper longitudinal Wilson line of the staple. After the spacetime translation \eqref{def:q_op_def_T_ord_transl}, its contribution reads 
\begin{align}
q^{\textrm{n.r.}}_{\textrm{unsub.}}&(\tx, \mathbf{b};\mu^2)\Big|_{\ref{Fig:diag4}}  =
\lim_{Y^+\rightarrow +\infty}
\int \frac{\td b^+}{(2\pi)}\, 
e^{-i\tx P^-b^+} \nonumber\\
& \hspace{3cm}\times 
\big\langle P\big| \T \Big[ 
{\overline\psi}(0) \, \frac{\gamma^-}{2} \, 
\left\{- i \mu^\epsilon g \int_{-b^+}^{Y^+} \td z^+
 t^a \delta A_a^- (z^+, -\bbperp , 0^-)\right\}
\, 
\delta\Psi(-b^+, -\mathbf{b}, 0^-) \Big]
\big|P\big\rangle_c \,.
\end{align}
Expressing the correlators in terms of quark and gluon propagators, the contribution from diagram~\ref{Fig:diag4} can be written as 
\begin{align}
q^{\textrm{n.r.}}_{\textrm{unsub.}}(\tx, \mathbf{b};\mu^2)\Big|_{\ref{Fig:diag4}}  
= 
&
 - g^2 \mu^{2 \epsilon} C_F \, \lim_{Y^+\rightarrow +\infty}
\int \frac{\td b^+}{(2\pi)}\, e^{-i\tx P^-b^+} 
\int \td^{4 - 2 \epsilon} y \, 
\big\langle P\big|{\overline\psi} (0)  \, \frac{\gamma^-}{2} \,
S_{0,F}(- b^+ , - \bbperp , 0^- ; y) \gamma_\mu \, \psi (y) \big|P\big\rangle_c
\nonumber \\
&
\times\int_{-b^+}^{Y^+} \td z^+\, 
G_{0,F}^{- \mu} (z^+, -\bbperp,0^-; y)
\nonumber \\
& \hspace{-2cm}=
- g^2 \mu^{2 \epsilon} C_F \, \lim_{Y^+\rightarrow +\infty}
\int \frac{\td b^+}{(2\pi)}\, e^{-i\tx P^-b^+} 
\int \td^{4 - 2 \epsilon} y \,\big\langle P\big|  {\overline\psi}_\alpha (0)  \, \psi_\beta (y) \big| P\big\rangle \int_{-b^+}^{Y^+} \td z^+\, 
\int \frac{\td^{4-2\epsilon} q}{(2\pi)^{4-2\epsilon}} 
\int \frac{\td^{4-2\epsilon} k}{(2\pi)^{4-2\epsilon}} 
\nonumber \\
&
\times\bigg\{
 \frac{\gamma^-}{2} \frac{i\slashed{q}}{q^2+i0} \gamma_\mu
 \bigg\}_{\alpha\beta} \tilde{G}_{0,F}^{- \mu} (k) \, e^{i q^- (y^+ +b^+)} \, e^{i q^+ y^-} \, e^{- i \bq\cdot (\by +\bbperp)} \, e^{i k^- (y^+ - z^+)} \, e^{i k^+ y^-} 
e^{-i \bk\cdot (\by +\bbperp)}\,.
\end{align}
Integrations over $z^+$ and $b^+$ can be carried out straightforwardly, yielding 
\begin{align}
q^{\textrm{n.r.}}_{\textrm{unsub.}}(\tx, \mathbf{b};\mu^2)\Big|_{\ref{Fig:diag4}} = &  
- i g^2 \mu^{2 \epsilon} C_F \, \lim_{Y^+\rightarrow +\infty} 
\int \td^{4 - 2 \epsilon} y \, 
\big\langle P\big|  {\overline\psi}_\alpha (0)  \, \psi_\beta (y) \big| P\big\rangle 
\int \frac{\td^{4-2\epsilon} q}{(2\pi)^{4-2\epsilon}} 
\int \frac{\td^{4-2\epsilon} k}{(2\pi)^{4-2\epsilon}} 
\frac{1}{k^-} 
\bigg\{
 \frac{\gamma^-}{2} \frac{i\slashed{q}}{q^2+i0} \gamma_\mu
 \bigg\}_{\alpha\beta}
\nonumber \\
\times 
\,
\tilde{G}_{0,F}^{- \mu} (k)\,
& 
e^{i (q^- + k^-) y^+ } \, e^{i (q^+ + k^+ )y^-} \, 
e^{- i (\bq + \bk)\cdot (\by + \bbperp)} 
\bigg[
\delta (q^- - \tx P^-) e^{- i k^- Y^+} - \delta (q^- + k^- - \tx P^-)
\bigg]\,.
\end{align}
Using the gluon propagator contraction in the projectile light-cone gauge 
\begin{align}
\gamma_\mu \, \tilde{G}_{0,F}^{- \mu} (k) =
\frac{i}{k^2 + i0} \frac{1}{\left[n\cdot k\right]} 
\left[2 k^- \gamma^+ - \bk^i\gamma^i\right]
\end{align}
and applying the same change of variables introduced in Eq. \eqref{eq:cov-q-to-l}, the contribution from diagram~\ref{Fig:diag4} can be written as 
\begin{align}
q^{\textrm{n.r.}}_{\textrm{unsub.}}(\tx, \mathbf{b};\mu^2)\Big|_{\ref{Fig:diag4}} = & \, 
i g^2 \mu^{2 \epsilon} C_F \, \lim_{Y^+\rightarrow +\infty} 
\int \td^{4 - 2 \epsilon} y \, 
\big\langle P\big|  {\overline\psi}_\alpha (0)  \, \psi_\beta (y) \big| P\big\rangle 
\int \frac{\td^{4-2\epsilon} k}{(2\pi)^{4-2\epsilon}} \, 
\int \frac{\td^{2-2\epsilon} \bl}{(2\pi)^{2-2\epsilon}} \, 
\int\frac{\td l^+}{(2\pi)} \, 
\int\frac{\td q^-}{(2\pi)} \,
\nonumber \\
&
\times\frac{1}{k^- \left[n\cdot k\right]} \, \frac{1}{k^2 + i0} 
\bigg\{
 \frac{\gamma^-}{2} \left[q^- \gamma^+ - \bq^i\gamma^i\right] \,
 \left[2 k^- \gamma^+ - \bk^i\gamma^i\right]
 \bigg\}_{\alpha\beta} \,  
\nonumber \\
&
\times 
\bigg[
\delta (q^- - \tx P^-) e^{- i k^- Y^+} - \delta (q^- + k^- - \tx P^-)
\bigg]
\frac{e^{i (q^- + k^-) y^+ } \, e^{i l^+ y^-} \, 
e^{- i \bl\cdot (\by + \bbperp)}}{\left[2 q^- (l^+ - k^+) - (\bl - \bk)^2 + i0\right]} \,.
\label{eq:diag1b_inter_1}
\end{align}
Using the delta functions in the last line of Eq. \eqref{eq:diag1b_inter_1}, the expression can be rewritten in the following form, which is more convenient for the remainder of the discussion:
\begin{align}
q^{\textrm{n.r.}}_{\textrm{unsub.}}&(\tx, \mathbf{b};\mu^2)\Big|_{\ref{Fig:diag4}} =  \, 
i g^2 \mu^{2 \epsilon} C_F \, \lim_{Y^+\rightarrow +\infty} 
\int \td^{4 - 2 \epsilon} y \, 
\big\langle P\big|  {\overline\psi}_\alpha (0)  \, \psi_\beta (y) \big| P\big\rangle \, e^{i \tx P^- y^+ } 
\int \frac{\td^{4-2\epsilon} k}{(2\pi)^{4-2\epsilon}} \, 
\int \frac{\td^{2-2\epsilon} \bl}{(2\pi)^{2-2\epsilon}} \, 
\int\frac{\td l^+}{(2\pi)} 
\nonumber \\
&
\times\,\int \frac{\td q^-}{(2\pi)} \,\frac{e^{i l^+ y^-} e^{- i \bl\cdot (\by + \bbperp)}}{k^- \left[n\cdot k\right]} 
\frac{1}{\left[k^2 + i0\right]\left[2 q^- (l^+ - k^+) - (\bl - \bk)^2 + i0\right]} \, 
\left\{
 \frac{\gamma^-}{2} \left[q^- \gamma^+ - \bq^i\gamma^i\right] \,
 \left[2 k^- \gamma^+ - \bk^i\gamma^i\right]
 \right\}_{\alpha\beta} 
 \nonumber \\
 &
\times\left[
\delta (q^- - \tx P^-) e^{- i k^- (Y^+ - y^+)} - \delta (q^- + k^- - \tx P^-)
\right]\,.
\label{eq:diag1b_inter_2}
\end{align}
Let us now focus on the $k^+$ integral. The denominator $[n\cdot k]$ which can equivalently be written as $[k^+]$, represents the extra singularity originating from the gluon propagator in the light-cone gauge and it will be treated within the ML prescription given in Eq. \eqref{eq:ML_prescription}. The contour integration over $k^+$ using the ML prescription follows very closely the computation performed for diagram~\ref{Fig:diag6} and the contributions with $k^-$ and $q^-$ of opposite sign vanish because all $k^+$ poles lie on the same side of the real axis. All in all, one finds  
\begin{align}
\label{eq:ML_kplus_1b}
\int \frac{\td k^+}{(2\pi)} 
\left\{
\frac{\theta (k^-)}{k^+ + i0} +\frac{\theta (- k^-)}{k^+ - i0}
\right\}
\frac{1}{(2 k^-) \left[k^+ - \frac{\bk^2 - i0}{2 k^-}\right]} &
\frac{1}{(- 2 q^-) \left[k^+ - l^+ + \frac{(\bl - \bk)^2 - i0}{2 q^-}\right]} \nonumber\\
& =
i \frac{\theta (k^-) \theta (q^-)}{(2 k^-) (- 2 q^-)}
\frac{1}{\left[l^+ - \frac{(\bl - \bk)^2}{2 q^-}\right]}
\frac{1}{\left[l^+ - \frac{(\bl - \bk)^2}{2 q^-} - \frac{\bk^2}{2 k^-}\right]}\,,
\end{align}
where the case of both $k^-$ and $q^-$ negative is incompatible with the delta function constraints in Eq.~\eqref{eq:diag1b_inter_2} and thus does not contribute. Substituting Eq. \eqref{eq:ML_kplus_1b} back into Eq. \eqref{eq:diag1b_inter_1}, one gets 
\begin{align}
&q^{\textrm{n.r.}}_{\textrm{unsub.}}(\tx, \mathbf{b};\mu^2)\Big|_{\ref{Fig:diag4}} =  \, 
g^2 \mu^{2 \epsilon} C_F \, \lim_{Y^+\rightarrow +\infty} 
\int \td^{4 - 2 \epsilon} y \, 
\big\langle P\big|  {\overline\psi}_\alpha (0)  \, \psi_\beta (y) \big| P\big\rangle \,
e^{i \tx P^- y^+ } 
\int \frac{\td^{2-2\epsilon} \bl}{(2\pi)^{2-2\epsilon}} 
\int\frac{\td l^+}{(2\pi)} 
\int \frac{\td^{2-2\epsilon} \bk}{(2\pi)^{2-2\epsilon}}
\int\frac{\td k^-}{(2\pi)}  
\nonumber \\
&
\times\,
\int\frac{\td q^-}{(2\pi)} \frac{e^{i l^+ y^-} e^{- i \bl\cdot (\by + \bbperp)}\, q^-  \theta (k^-) \theta (q^-)}
{k^- \left[2 q^- l^+ - (\bl - \bk)^2\right] 
\left[2 q^- k^- l^+ - k^- (\bl - \bk)^2 - q^- \bk^2\right]} 
\Big\{
 \frac{\gamma^-}{2} \Big[- (\bl - \bk)^i\gamma^i (2 k^- \gamma^+ - \bk^i\gamma^i)
 - q^- \gamma^+ \bk^i\gamma^i\Big]
 \Big\}_{\alpha\beta}\nonumber \\
&\times \left[
\delta (q^- - \tx P^-) e^{- i k^- (Y^+ - y^+)} - \delta (q^- + k^- - \tx P^-)
\right]\,.
\label{eq:diag_1b_inter_3}
\end{align}
At this stage, we should introduce the pure rapidity regulator given in Eq. \eqref{def:pure_rap_reg_OS} in order to proceed with the $k^-$ integration. The first term in the square bracket in the third line of Eq. \eqref{eq:diag_1b_inter_3}, which is proportional to $\delta(q^--{\rm x}P^-)\, e^{-ik^-(\, Y^+-y^+)}$ can be shown to vanish in the $Y^+\to \infty$ limit after inserting the pure rapidity regulator. To demonstrate that, one may rescale the integration variable $k^-\to k^-(Y^+-y^+)$ and show that the resulting integral scales as $(Y^+-y^+)^{-\eta}\to 0$ for positive $\eta$. The non-vanishing contribution in the $Y^+\to\infty$ limit for the diagram~\ref{Fig:diag4} is the second term inside the square bracket in the third line of Eq. \eqref{eq:diag_1b_inter_3} which is proportional to $\delta(q^-+k^--{\rm x}P^-)$. After taking the $Y^+\to\infty$ limit and trivially integrating over $q^-$, this contribution reads  
\begin{align}
q^{\textrm{n.r.}}_{\textrm{unsub.}}&(\tx, \mathbf{b};\mu^2)\Big|_{\ref{Fig:diag4}} \nonumber \\ 
=& \, 
- \alpha_s \mu^{2 \epsilon} C_F \, \int \td^{4 - 2 \epsilon} y \, 
\big\langle P\big|  {\overline\psi}_\alpha (0)  \, \psi_\beta (y) \big| P\big\rangle  
\, e^{i \tx P^- y^+ } 
\int \frac{\td^{2-2\epsilon} \bl}{(2\pi)^{2-2\epsilon}} \, 
\int\frac{\td l^+}{(2\pi)} \, e^{i l^+ y^-} e^{- i \bl\cdot (\by + \bbperp)}
\int \frac{\td^{2-2\epsilon} \bk}{(2\pi)^{2-2\epsilon}} \,
\nonumber \\
&
\times
\int_0^\infty \frac{\td k^-}{(2\pi)} \,\frac{\tx P^- - k^-}{k^-} 
\frac{\theta (\tx P^- - k^-)}
{\left[2 (\tx P^- - k^-) l^+ - (\bl - \bk)^2\right] 
\left[2 (\tx P^- - k^-)  k^- l^+ - k^- (\bl - \bk)^2 - (\tx P^- - k^-) \bk^2\right]} 
\nonumber \\
&
\times\left\{
\gamma^-\left[- (\bl - \bk)^i\gamma^i (2 k^- \gamma^+ - \bk^j\gamma^j) - (\tx P^- - k^-) \gamma^+ \bk^i\gamma^i\right]
 \right\}_{\alpha\beta} \,
\end{align} 
which, after inserting the pure rapidity regulator and changing the integration variable $k^-\to \xi\equiv k^-/{\rm x}P^-$, can be written as 
\begin{align}
q^{\textrm{n.r.}}_{\textrm{unsub.}}&(\tx, \mathbf{b};\mu^2,\zeta)\Big|_{\ref{Fig:diag4}} \nonumber
\\
& \hspace{-1cm}=  \, 
- \frac{\alpha_s \mu^{2 \epsilon} C_F}{2\pi} \,\int \td^{4 - 2 \epsilon} y \, 
\big\langle P\big|  {\overline\psi}_\alpha (0)  \, \psi_\beta (y) \big| P\big\rangle  
\, e^{i \tx P^- y^+ } 
\int \frac{\td^{2-2\epsilon} \bl}{(2\pi)^{2-2\epsilon}} \, 
\int\frac{\td l^+}{(2\pi)} \, 
e^{i l^+ y^-} e^{- i \bl\cdot (\by + \bbperp)}
\int \frac{\td^{2-2\epsilon} \bk}{(2\pi)^{2-2\epsilon}} \,
\nonumber \\
&\hspace{-1cm}\times w^2
\left[\frac{2 \nu^+}{\bk^2 \nu^-}\right]^{\frac{\eta}{2}} 
\int_0^1 \td \xi  \frac{1 - \xi }{\xi } \, \xi ^\eta [\tx P^-]^\eta  
\frac{\left\{\gamma^-\left[- (\bl - \bk)^i\gamma^i (2 \tx P^-\xi  \gamma^+ - \bk^j\gamma^j) - \tx P^-(1 - \xi ) \gamma^+ \bk^i\gamma^i\right]
 \right\}_{\alpha\beta} }
{\left[2 \tx P^- (1 - \xi ) l^+ - (\bl - \bk)^2\right] 
\left[2 \tx P^- \xi  (1 - \xi ) l^+ - \xi  (\bl - \bk)^2 - (1 - \xi ) \bk^2\right]} \,,
\end{align} 
where the rapidity divergence resides in the $\xi\to0$ region and its extraction can be performed using the expansion in $\eta$ introduced in Eq. \eqref{eq:eta_expand} which in this case is performed around $\xi=0$ and it reads 
\begin{align}
    \int_0^1 \td \xi \,  \xi^{\eta-1}f(\xi,\eta) =& \frac{f(0,\eta)}{\eta}+\int_0^1 \td \xi \, \frac{f(\xi,0)-f(0,0) }{\xi} +\mathcal{O}(\eta)\,.
\label{eq: plus distribution in diag 4}
\end{align}
Using Eq. \eqref{eq: plus distribution in diag 4}, the rapidity divergent contribution of diagram~\ref{Fig:diag4} can be written as 
\begin{align}
q^{\textrm{n.r.}}_{\textrm{unsub.}}&(\tx, \mathbf{b};\mu^2,\zeta)\Big|_{\ref{Fig:diag4}}^{1/\eta} \nonumber \\
& =\, 
- \frac{\alpha_s \mu^{2 \epsilon} C_F}{2\pi}\,  [\tx P^-]^\eta \,
\left[\frac{2 \nu^+}{\nu^-}\right]^{\frac{\eta}{2}} \,\frac{ w^2}{\eta} \, \int \td^{4 - 2 \epsilon} y \, 
\big\langle P\big|  {\overline\psi}_\alpha (0)  \, \psi_\beta (y) \big| P\big\rangle  
\, e^{i \tx P^- y^+ } 
\int \frac{\td^{2-2\epsilon} \bl}{(2\pi)^{2-2\epsilon}}  e^{- i \bl\cdot (\by + \bbperp)}
\nonumber \\
&
\times\int\frac{\td l^+}{(2\pi)} \,  \, 
e^{i l^+ y^-} \int \frac{\td^{2-2\epsilon} \bk}{(2\pi)^{2-2\epsilon}}
\frac{1}{[\bk^2]^{1+\frac{\eta}{2}}}\frac{1}{\left[2 \tx P^- l^+ - (\bl - \bk)^2\right]} \left\{
\gamma^- \left[\tx P^- \gamma^+ - (\bl - \bk)^i\bg^i\right] \left[ \bk^j\bg^j\right]
\right\}_{\alpha\beta}\,.
\label{eq:diag_1b_inter_4}
\end{align}
The numerator in Eq. \eqref{eq:diag_1b_inter_4} can be rewritten, in order to apply the EOM for the background field $\slashed{l}\psi(l)=0+ O(\alpha_s)$, as
\begin{align}
     \left\{
\gamma^- \left[\tx P^- \gamma^+ - (\bl - \bk)^i\bg^i\right] \left[ \bk^j\bg^j\right]
\right\}=  \left\{
\gamma^- \left[2\tx P^-l^+-(\bl-\bk)^2-l^2-\bk^i\gamma^i\slashed{l}  \right]
\right\}\,,
\end{align}
which can be substituted back into Eq. \eqref{eq:diag_1b_inter_4} to get the rapidity divergent contribution of diagram~\ref{Fig:diag4} as 
\begin{align}
q^{\textrm{n.r.}}_{\textrm{unsub.}}(\tx, \mathbf{b};\mu^2,\zeta)\Big|_{\ref{Fig:diag4}}^{1/\eta} &=  -\frac{\alpha_s \mu^{2 \epsilon} C_F}{2\pi}\,  [\tx P^-]^\eta \,
\left[\frac{2 \nu^+}{\nu^-}\right]^{\frac{\eta}{2}} \,\frac{ w^2}{\eta} \, \int \td^{4 - 2 \epsilon} y \, 
\big\langle P\big|  {\overline\psi}_\alpha (0)  \, \psi_\beta (y) \big| P\big\rangle  
\, e^{i \tx P^- y^+ } 
\int \frac{\td^{2-2\epsilon} \bl}{(2\pi)^{2-2\epsilon}} \, 
 e^{- i \bl\cdot (\by + \bbperp)}
\nonumber \\
&
\times \int\frac{\td l^+}{(2\pi)}  
e^{i l^+ y^-} \int \frac{\td^{2-2\epsilon} \bk}{(2\pi)^{2-2\epsilon}}
\frac{1}{[\bk^2]^{1+\frac{\eta}{2}}}\left\{\gamma^-\left[1-\frac{l^2+\bk^i\gamma^i\slashed{l}}{\left[2 \tx P^- l^+ - (\bl - \bk)^2\right]} \right]
\right\}_{\alpha\beta}\, \\
& = 
-\frac{\alpha_s \mu^{2 \epsilon} C_F}{2\pi}\,  [\tx P^-]^\eta \,
 w^2\left[\frac{2 \nu^+}{\nu^-}\right]^{\frac{\eta}{2}} \,\frac{1}{\eta} \, \int \td^{4 - 2 \epsilon} y \, 
\big\langle P\big|  {\overline\psi}(0) \gamma^- \, \psi(y) \big| P\big\rangle  
\,e^{i \tx P^- y^+ } \nonumber \\
&
\times\, 
\int \frac{\td^{2-2\epsilon} \bl}{(2\pi)^{2-2\epsilon}} \, 
\int\frac{\td l^+}{(2\pi)} \,  \, 
e^{i l^+ y^-} e^{- i \bl\cdot (\by + \bbperp)}
 \int \frac{\td^{2-2\epsilon} \bk}{(2\pi)^{2-2\epsilon}}
\frac{1}{[\bk^2]^{1+\frac{\eta}{2}}}\,,
%\left[ 1-e^{i \bk\cdot\bbperp}\right]
\label{eq:diag_1b_inter_5}
\end{align}
where in the second equality we have used the fact that the terms proportional $l^2$ and $\slashed{l}$ are shifted to higher order in $\alpha_s$ due to
the EOM for the quark background field. Eq. \eqref{eq:diag_1b_inter_5} is the rapidity divergent contribution of the diagram~\ref{Fig:diag4} which we will combine together with the rapidity divergent contribution of diagram~\ref{Fig:diag6} given in Eq. \eqref{eq:Diag_1a_inter_8} in the next subsection. Note that in the contribution \eqref{eq:diag_1b_inter_5}, the integral over $\bk$ is scaleless and thus vanishes in dimensional regularization, but is formally divergent both in the IR and in the UV.

For the moment, we focus on the rapidity finite part of the diagram~\ref{Fig:diag4}, which is obtained by collecting the rapidity-finite terms in Eq.~\eqref{eq: plus distribution in diag 4}, setting $\eta =0$, and reads   
\begin{align}
q^{\textrm{n.r.}}_{\textrm{unsub.}}&(\tx, \mathbf{b};\mu^2,0)\Big|_{\ref{Fig:diag4}}^{\rm fin.} \nonumber 
\\
= & \, 
- \frac{\alpha_s \mu^{2 \epsilon} C_F}{2\pi} \, \int \td^{4 - 2 \epsilon} y \, 
\big\langle P\big|  {\overline\psi}_\alpha (0)  \, \psi_\beta (y) \big| P\big\rangle  
\, e^{i \tx P^- y^+ } 
\int \frac{\td^{2-2\epsilon} \bl}{(2\pi)^{2-2\epsilon}} \, 
\int\frac{\td l^+}{(2\pi)} \, 
e^{i l^+ y^-} e^{- i \bl\cdot (\by + \bbperp)}
\nonumber \\
&
\times\int \frac{\td^{2-2\epsilon} \bk}{(2\pi)^{2-2\epsilon}} \, \int_0^1 \frac{\td \xi}{\xi }  \, 
\Bigg\{\gamma^- \Bigg[\frac{(1 - \xi )\left[- (\bl - \bk)^i\gamma^i (2 \tx P^-\xi  \gamma^+ - \bk^j\gamma^j) - \tx P^-(1 - \xi ) \gamma^+ \bk^i\gamma^i\right] }
{\left[2 \tx P^- (1 - \xi ) l^+ - (\bl - \bk)^2\right] 
\left[2 \tx P^- \xi  (1 - \xi ) l^+ - \xi  (\bl - \bk)^2 - (1 - \xi ) \bk^2\right]} 
\nonumber \\
& - \frac{\left[- (\bl - \bk)^i\gamma^i ( \bk^j\gamma^j) + \tx P^-\gamma^+ \bk^i\gamma^i\right]
 }
{\left[2 \tx P^- l^+ - (\bl - \bk)^2\right] 
 \bk^2} 
\Bigg]\Bigg\}_{\alpha\beta} \,. 
\label{eq:diag_1b_rap_fin}
\end{align} 
We note that rapidity finite part of diagram~\ref{Fig:diag4} given in Eq. \eqref{eq:diag_1b_rap_fin} is infrared-safe but still contains UV divergences. The detailed computation of the UV-divergent contribution arising from this rapidity-finite part is presented in Appendix~\ref{appendix: calculation of uv divergent part of diag 4}. All in all, the UV-divergent but rapidity-finite contribution of the diagram~\ref{Fig:diag4} reads 
\begin{align}
q^{\textrm{n.r.}}_{\textrm{unsub.}}(\tx, \mathbf{b};\mu^2)\Big|_{\ref{Fig:diag4}}^{\rm fin.} = & \, 
 \frac{\alpha_s C_F}{2\pi}\, (4\pi\mu^2)^{\epsilon}\frac{\Gamma(2-\epsilon)}{\Gamma(1-\epsilon)}\Gamma(\epsilon)\, \int \frac{\td y^+}{2\pi} \, 
\big\langle P\big|  {\overline\psi}(0) \frac{\gamma^-}{2} \, \psi (y^+,-\bbperp,0^-) \big| P\big\rangle  
\, e^{i \tx P^- y^+ } \nonumber\\
=&
\frac{\alpha_s  C_F}{2\pi}\, (4\pi\mu^2)^{\epsilon}\frac{\Gamma(2-\epsilon)}{\Gamma(1-\epsilon)}\Gamma(\epsilon)\,  q_{\rm Bckgd}(x,\bbperp;\mu^2)\,,
\label{eq: UV divergent part of diag 4}
\end{align} 
where we have used the definition of the quark background TMD given in Eq. \eqref{def:q_Bckgd}.

%==================================================================================
\subsubsection{Combined rapidity-divergent contribution from Diagram~\ref{Fig:diag6} and Diagram~\ref{Fig:diag4}}
%==================================================================================

The rapidity-divergent parts of diagram~\ref{Fig:diag6} and diagram~\ref{Fig:diag4} are given in Eqs. \eqref{eq:Diag_1a_inter_8} and \eqref{eq:diag_1b_inter_5}, respectively. Combining these two expressions, we obtain
\begin{align}
q^{\textrm{n.r.}}_{\textrm{unsub.}}(\tx, \mathbf{b};\mu^2,\zeta)\Big|_{\ref{Fig:diag6}+\ref{Fig:diag4}}^{1/\eta} & =  -\frac{\alpha_s \mu^{2 \epsilon} C_F}{2\pi}\,  [\tx P^-]^\eta \,
 w^2\left[\frac{2 \nu^+}{\nu^-}\right]^{\frac{\eta}{2}} \,\frac{1}{\eta} \, \int \td^{4 - 2 \epsilon} y \, 
\big\langle P\big|  {\overline\psi}(0) \gamma^- \, \psi(y) \big| P\big\rangle  
\,e^{i \tx P^- y^+ } \nonumber \\
&
\times\, 
\int \frac{\td^{2-2\epsilon} \bl}{(2\pi)^{2-2\epsilon}} \, 
\int\frac{\td l^+}{(2\pi)} \,  \, 
e^{i l^+ y^-} e^{- i \bl\cdot (\by + \bbperp)}
 \int \frac{\td^{2-2\epsilon} \bk}{(2\pi)^{2-2\epsilon}}
\frac{1}{[\bk^2]^{1+\frac{\eta}{2}}}\left[ 1-e^{i \bk\cdot\bbperp}\right] \, . 
\label{eq:rap_div_comb_1a+1b}
\end{align}
The transverse momentum integral over $\k$ is now IR finite, since the IR divergences cancel between the two diagrams. However, this integral is UV divergent, due to the contribution of the diagram~\ref{Fig:diag4}. The result reads 
\begin{align}
    \int \frac{\td^{2-2\epsilon} \bk}{(2\pi)^{2-2\epsilon}}
\frac{1}{[\bk^2]^{1+\frac{\eta}{2}}}\left[ 1-e^{i \bk\cdot\bbperp}\right]= -\frac{(\pi \bbperp^2)^\epsilon}{4\pi}\left(\frac{\bbperp^2}{4}\right)^{\frac{\eta}{2}}\frac{\Gamma\left(-\frac{\eta}{2}-\epsilon\right)}{\Gamma\left(1+\frac{\eta}{2} \right)}\,. 
\label{eq:trans_int_1a+1b}
\end{align}
Substituting Eq. \eqref{eq:trans_int_1a+1b} into the combined expression for the rapidity-divergent part given in Eq. \eqref{eq:rap_div_comb_1a+1b}, and trivially performing the integrals over $l^+$ and $\l$, followed by those $y^-$ and $\y$ using the resulting delta functions, one obtains  
\begin{align}
q^{\textrm{n.r.}}_{\textrm{unsub.}}(\tx, \mathbf{b};\mu^2,\zeta)\Big|_{\ref{Fig:diag6}+\ref{Fig:diag4}}^{1/\eta}  
= 
&\frac{\alpha_s  C_F}{2\pi}\,  (\pi\mu^{2} \bbperp^2)^\epsilon\,\left[\frac{(\tx P^-)^2\bbperp^2}{2}\frac{ \nu^+}{\nu^-}\right]^{\frac{\eta}{2}}\,\frac{\Gamma\left(-\frac{\eta}{2}-\epsilon\right)}{\Gamma\left(1+\frac{\eta}{2} \right)} \,\frac{ w^2}{\eta} \, \int \frac{\td y^+}{2\pi} \, 
\big\langle P\big|  {\overline\psi} (0) \frac{\gamma^-}{2} \, \psi(y^+, -\bbperp,0^-) \big| P\big\rangle  \,.
\end{align}
Expanding this result in $\eta$ at finite $\epsilon$, one obtains the final result 
\begin{align}   
q^{\textrm{n.r.}}_{\textrm{unsub.}}(\tx, \mathbf{b};\mu^2,\zeta)\Big|_{\ref{Fig:diag6}+\ref{Fig:diag4}}^{1/\eta} = 
&\frac{\alpha_s  C_F}{2\pi}\,  (\pi\mu^{2} \bbperp^2)^\epsilon\,\left[\frac{\zeta \bbperp^2}{4}\right]^{\frac{\eta}{2}}\,\frac{\Gamma\left(-\frac{\eta}{2}-\epsilon\right)}{\Gamma\left(1+\frac{\eta}{2} \right)} \,\frac{ w^2}{\eta} \, q_{\rm Bckgd}(x,\bbperp;\mu^2)  \nonumber \\
= 
&\frac{\alpha_s  C_F}{4\pi}\,  \Gamma (-\epsilon)(\pi\mu^{2} \bbperp^2)^\epsilon\,\left[ \frac{2 w^2}{\eta} +\log \left(\frac{\zeta \bbperp^2}{c_0^2} \right)-\Psi(-\epsilon)+\Psi(1) \right] \, q_{\rm Bckgd}(x,\bbperp;\mu^2) 
\, ,
\label{eq:doube pole from diags 6 and 4}
\end{align}
where we have written the result in terms of the background TMD given in Eq. \eqref{def:q_Bckgd}, the scale $\zeta$ which is defined as 
\begin{align}
\zeta\equiv \frac{2({\rm x}P^-)^2\nu^+}{\nu^-}    
\end{align}
and the constant $c_0\equiv 2e^{-\gamma_E}$.

This result is the direct analog of the radiation-to-infinity contribution found in the target light-cone gauge calculation of Ref.~\cite{Altinoluk:2025ewj}. The key structural difference is that in the present projectile light-cone gauge calculation the double-logarithmic contribution to the CSS kernel arises from radiation absorbed by the longitudinal Wilson lines, whereas in the target light-cone gauge it originated from radiation to the transverse Wilson line at infinity. Despite the difference in the diagrammatic origin, the final result takes exactly the same form as in the target light-cone gauge calculation, as expected from gauge invariance of the CSS evolution equations. We will return to a detailed comparison in Sec.\ref{sec:comparison}.    

%===============================================================
\subsection{Diagram~\ref{Fig:diag5}: gluon emission from the quark to infinity}
\label{sec: Diagram 5}
%===============================================================

\begin{figure}[h!]
{%
       \includegraphics[width=0.40\textwidth]{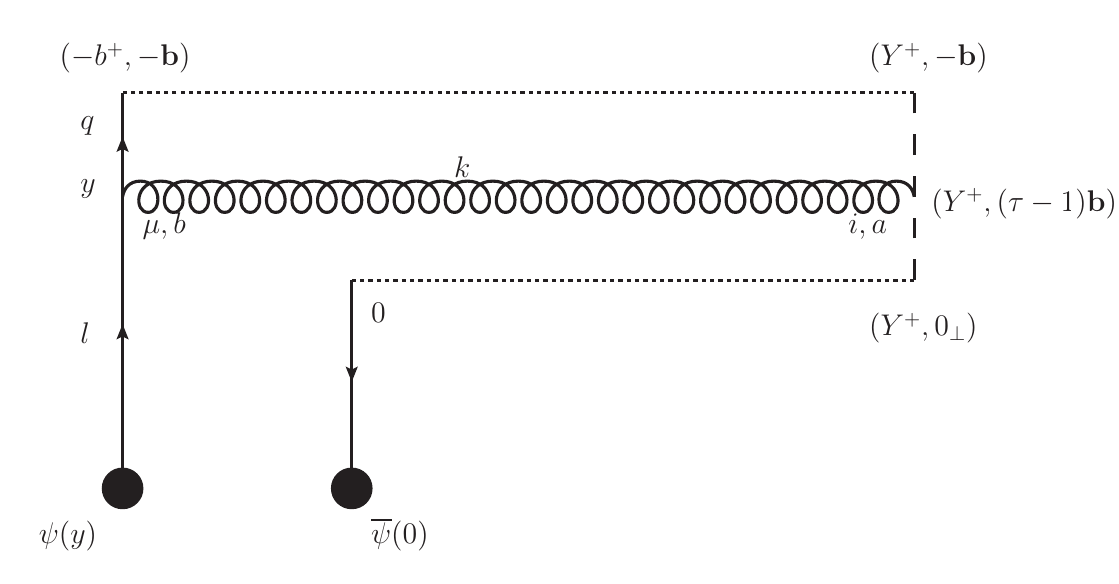}
     }     
\caption{Gluon emission from the quark to infinity - diagram~\ref{Fig:diag5}.}
\end{figure}

Diagram~\ref{Fig:diag5} corresponds to the radiation of a gluon by the background quark field which is subsequently absorbed by the transverse Wilson line at light-cone infinity $Y^+\to\infty$. The longitudinal Wilson lines are set to unity. After performing the space time translation of Eq. \eqref{def:q_op_def_T_ord_transl}, the contribution of this diagram to the quark TMD reads  
\begin{align}
\label{def:q_dist-1-loop-rad-to-infty}
q^{\textrm{n.r.}}_{\textrm{unsub.}}(\tx, \mathbf{b};\mu^2)\Big|_{\ref{Fig:diag5}} & =  
\lim_{Y^+\rightarrow +\infty}
\int \frac{\td b^+}{(2\pi)}\, 
e^{-i\tx P^-b^+} 
\nn
& \times 
\big\langle P\big| \T \Big[ 
{\overline\psi} (0)  \, \frac{\gamma^-}{2} \, 
 (- i g \mu^\epsilon) \int_0^1 d \tau\,  
(-1)\mathbf{b}^i\, t^a\, \delta A^i_a(Y^+, (\tau\!-\!1) \mathbf{b} , 0^-)
%\nn
%& \hspace{8cm}
%\times
\delta \Psi ( - b^+, - \mathbf{b} , 0^-) \Big]
\big|P\big\rangle_c
\, ,
\end{align}
where the transverse path at $Y^+$ has been parametrized as 
\begin{align}
\bx_i \equiv (\tau -1) \bbperp_i\,.
\end{align}
Expressing the two-point correlator of the fluctuation fields in terms of free Feynmann propagators and background field insertion, one finds
\begin{align}
%\label{def:q_dist-1-loop-rad-to-infty}
q^{\textrm{n.r.}}_{\textrm{unsub.}}(\tx, \mathbf{b};\mu^2)\Big|_{\ref{Fig:diag5}} = &
\lim_{Y^+\rightarrow +\infty}
- (- i g \mu^\epsilon)^2 \, C_F \, 
\int \frac{\td b^+}{(2\pi)}\, 
e^{-i\tx P^-b^+}
\int \td^{4 - 2 \epsilon} y \, 
\int_0^1 d \tau\,  
\mathbf{b}^i\,
 G_{0,F}^{i \mu}(Y^+, (\tau \!-\! 1) \mathbf{b} , 0^- ; y) \, 
\nonumber \\
& \hspace{4cm}
\times
\big\langle P\big|  {\overline\psi} (0)  \, \frac{\gamma^-}{2} \, 
S_{0,F}(- b^+ , - \mathbf{b} , 0^- ; y)
 \gamma_\mu \, \psi (y)
\big|P\big\rangle_c
\, .
\end{align}
Using the explicit expressions for the free quark and gluon propagators in momentum space, the contribution to the quark TMD from diagram~\ref{Fig:diag5} can be written as 
\begin{align}
q^{\textrm{n.r.}}_{\textrm{unsub.}}&(\tx, \mathbf{b};\mu^2)\Big|_{\ref{Fig:diag5}}    \nonumber
\\
= &
\lim_{Y^+\rightarrow +\infty}\,
\frac{g^2 \mu^{2\epsilon} C_F}{2}  \, 
\int \frac{\td b^+}{(2\pi)}\, 
e^{-i\tx P^-b^+}  \,
\int_0^1 d \tau\,
\int \td^{4 - 2 \epsilon} y \,
\int \frac{\td^{4-2\epsilon} q}{(2\pi)^{4-2\epsilon}} 
\int \frac{\td^{4-2\epsilon} k}{(2\pi)^{4-2\epsilon}} 
\big\langle P\big|  {\overline\psi}_\alpha (0)  \, 
\big\{\gamma^- i \slq \gamma_\mu\big\}_{\alpha\beta} 
\psi_\beta (y) \big|P\big\rangle_c 
\nonumber \\
&
\times e^{ i q^- (y^+ + b^+)} e^{i q^+ y^-} e^{- i \bq \cdot (\by +\mathbf{b})} \, 
e^{ i k^- (Y^+ - y^+)} e^{i k^+ y^-} e^{- i \bk \cdot [\by  - (\tau -1)\mathbf{b}]}%%
\frac{1}{q^2 + i0} \frac{i}{k^2 + i0} \, 
\mathbf{b}^i\, 
\left[ 
- g^{i \mu} + \frac{n^\mu k^i}{n\cdot k}
\right]\,,
\label{eq:diag_1c_int_1}
\end{align}
where $\alpha,\beta$ denote the Dirac spinor indices of the quark fields. The integral over $b^+$ can be performed straightforwardly, yielding $\delta (q^- \!-\!  \tx P^-)$, which is then used to perform the $q^-$ integration by setting $q^- = \tx P^-$. Furthermore,  we perform the change of variables from $q$ to $l$ which is given in Eq. \eqref{eq:cov-q-to-l}, so that Eq. \eqref{eq:diag_1c_int_1} takes the form 
\begin{align}
q^{\textrm{n.r.}}_{\textrm{unsub.}}&(\tx, \mathbf{b};\mu^2)\Big|_{\ref{Fig:diag5}} \nonumber \\
= &
\lim_{Y^+\rightarrow +\infty}
\alpha_s \mu^{2\epsilon} C_F  \, 
\int \td^{4 - 2 \epsilon} y \,
\big\langle P\big| {\overline\psi}_\alpha (0)  \, \psi_\beta (y) \big|P\big\rangle_c \,  
\int_0^1 \td \tau \, 
\int \frac{\td^{4-2\epsilon} k}{(2\pi)^{4-2\epsilon}} 
\int \frac{\td^{2-2\epsilon} \bl}{(2\pi)^{2-2\epsilon}}
\int \frac{\td l^+}{2 \pi} \,e^{i (\tx P^- + k^-) y^+} \,
\nonumber \\
&
\times 
e^{i l^+ y^-} \, e^{- i \bl\cdot(\by + \bbperp)} 
e^{- i k^- Y^+} e^{i \tau \bk\cdot\bbperp} 
\left( 
\frac{i \gamma^-\left(\tx P^- \gamma^+ \!-\! (\bl - \bk)^i\bg^i \right)\gamma_\mu}
{2\tx P^- (l^+ - k^+) \!-\!(\bl - \bk)^2 \!+\! i0} 
 \right)_{\alpha\beta}
\,
\frac{i \mathbf{b}^i}{k^2 + i0}\, 
\left[ 
- g^{i \mu} + \frac{n^\mu k^i}{n\cdot k}
\right]\,.
\label{eq:diag_1c_full_int_2}
\end{align}
To proceed further, one can simplify the numerator which contracts the Dirac structure with the gluon polarization tensor and it is evaluated as 
\begin{align}
\gamma^- \left[\tx P^- \gamma^+ \!-\! (\bl^j \!-\! \bk^j) \gamma^j \right]  
\gamma_\mu  \,   
\left[- g^{i \mu} + \frac{k^i n^\mu }{[n\!\cdot\! k]}\right] 
%\begin{comment}
=&\,
 - \tx P^- \gamma^- \gamma^+ \gamma^i + (\bl^j \!-\! \bk^j) \gamma^- \gamma^j \gamma^i
 - 
 \frac{\bk^i (\bl^j - \bk^j)}{[k^+]} \gamma^- \gamma^j \gamma^+ \,.
%\end{comment}
\label{eq:dirac-num-rad-infty}
\end{align}
The numerator structure given in Eq. \eqref{eq:dirac-num-rad-infty} contains three distinct terms. The first two, which  are free of extra light-cone gauge  singularity in $k^+$, are referred to as {\it regular} contributions. The third term carries the extra light-cone singularity $1/[k^+]$, which must be treated with the ML prescription \eqref{eq:ML_prescription}. We discuss these two type of contributions separately below.   

%======================================================
\subsubsection{Regular contribution}
%======================================================

For the regular terms, the $k^+$ integral takes the form 
\begin{align}
I_{reg} \equiv 
\int \frac{\td k^+}{2 \pi} \,
\frac{1}{2 k^- \left[k^+ -\frac{\bk^2 - i0}{2 k^-}\right]}
\frac{1}{(- 2 \tx P^-) \left[k^+ - l^+ + \frac{(\bl - \bk)^2 - i0}{2 \tx P^-}\right]}
= \frac{- i \theta (k^-)}
{2\left[2\tx P^- k^- l^+ - \tx P^- \bk^2 - k^- (\bl - \bk)^2\right]}\,,
\end{align}
where the contour has been closed below the real axis using the constraint $Y^+-y^+>0$, and contributions from negative $k^-$ vanish since all poles lie on the same side. The contribution of these {\it regular} terms to diagram~\ref{Fig:diag5} takes the following form  
\begin{align}
q^{\textrm{n.r.}}_{\textrm{unsub.}}(\tx, \mathbf{b};\mu^2)\Big|_{\ref{Fig:diag5}}^{reg} & = 
\lim_{Y^+\rightarrow +\infty}
\frac{\alpha_s \mu^{2\epsilon} C_F}{2}  \, 
\int \td^{4 - 2 \epsilon} y \,
\big\langle P\big| {\overline\psi}_\alpha (0)  \, \psi_\beta (y) \big|P\big\rangle_c \,  
\int_0^1 \td \tau \, 
\int \frac{\td^{2-2\epsilon} \bl}{(2\pi)^{2-2\epsilon}}
\int \frac{\td l^+}{2 \pi} 
\int \frac{\td^{2-2\epsilon} k}{(2\pi)^{2-2\epsilon}} 
\nonumber \\
& \hspace{-1cm}
\times \int_0^\infty \frac{\td k^-}{2 \pi}\, e^{i (\tx P^- + k^-) y^+} \,
e^{i l^+ y^-} \, e^{- i \bl\cdot(\by + \bbperp)} 
e^{- i k^- Y^+} e^{i \tau \bk\cdot\bbperp}  
\frac{i \bbperp^i \left[-\tx P^- \gamma^- \gamma^+ \gamma^i - 
(\bl^j - \bk^j)\gamma^- \gamma^j \gamma^i\right]}
{\left[2\tx P^- k^- l^+ - \tx P^- \bk^2 - k^- (\bl - \bk)^2\right]}  \,.
\end{align}
The subsequent integral over $k^-$ then reads 
\begin{align}
\label{eq:k_minus_1c_reg}
{\cal I}_{1c, \; k^-}^{reg} =
\int_0^\infty \frac{\td k^-}{2 \pi}
\frac{e^{- i k^- (Y^+ - y^+)}}
{\left[\left(2 \tx P^- l^+ - (\bl - \bk)^2\right)k^- - \tx P^- \bk^2\right]}\,.
\end{align}
To analyze the behavior of this integral as $Y^+\to\infty$, we insert the pure rapidity regulator \eqref{def:pure_rap_reg_OS} and for the convenience of the discussion, let us introduce a new variable $t\equiv k^-(Y^+-y^+)$, so that 
\begin{align}
{\cal I}_{1c, \; k^-}^{reg}
= \int_0^\infty \td t 
\frac{t^\eta \, e^{- i t}}{\left[t - \frac{\bk^2}{l^2}\, \tx P^-\, (Y^+ - y^+)\right]}
\frac{1}{\left[Y^+ - y^+\right]^\eta}\Bigg|_{l^-={\rm x}P^-}
\label{1c_reg_k_minus}
\end{align}
for positive $\eta$. Since the integral in Eq. \eqref{1c_reg_k_minus} clearly vanishes in the $Y^+\to\infty$ limit, one may safely conclude that the contributions to diagram~\ref{Fig:diag5} originating from the regular terms vanish. 

%=====================================================================
\subsubsection{Contribution with the extra light-cone gauge singularity}
%=====================================================================
For the term containing the extra light-cone gauge singularity $1/[k^+]$, the $k^+$ integral with the ML prescription \eqref{eq:ml-prescription_lc_var} reads 
\begin{align}
I_{LC} \equiv 
\int \frac{\td k^+}{2 \pi} \,
\frac{1}{2 k^- \left[k^+ -\frac{\bk^2 - i0}{2 k^-}\right]}
\frac{1}{(- 2 \tx P^-) \left[k^+ - l^+ + \frac{(\bl - \bk)^2 - i0}{2 \tx P^-}\right]}
\Bigg\{
\frac{\theta (k^-)}{k^+ + i0} + \frac{\theta (-k^-)}{k^+ -i0}
\Bigg\}\,.
\end{align}
Applying contour integration and performing the shift $\l\to \l+\k$, which is allowed since this integral is nested inside a separate $\l$ integration (see eq. \eqref{eq:diag_1c_full_int_2}), one finds  
\begin{align}
I_{LC} \sim - i \tx P^- \frac{\theta (k^-)}{l^4 \left[k^- - \tx P^- \frac{\bk^2}{l^2}\right]}\,,
\end{align} 
where $l^2=2{\rm x}P^-l^+-\l^2$. The subsequent $k^-$ integration reads 
\begin{align}
{\cal I}^{LC}_{1c, \; k^-}=
\int_0^\infty \frac{\td k^-}{2 \pi}
\frac{e^{- i k^- (Y^+ - y^+)}}
{\left[k^- -  \frac{\bk^2}{l^2} \, \tx P^-\right]}\,,
\end{align}
which has exactly the same structure as the regular $k^-$ integral given in Eq. \eqref{eq:k_minus_1c_reg}, up to a redefinition of the coefficient of $k^-$ in the denominator. Therefore, the same rescaling argument applies. More precisely, after inserting the pure rapidity regulator \eqref{def:pure_rap_reg_OS} and defining $t\equiv k^-(Y^+-y^+)$, one can show that the integral scales as $(Y^+-y^+)^{-\eta}\to0$ for positive $\eta$. Therefore this contribution also vanishes in the $Y^+\to\infty$ limit. 

%========================================================================
\subsubsection{Conclusion for diagram\ref{Fig:diag5}}
%========================================================================
Since both the regular and the light-cone singular contributions to diagram~\ref{Fig:diag5} vanish after taking the $Y^+\to\infty$ limit with the pure rapidity regulator in place, we conclude that 
\begin{align}
q^{\textrm{n.r.}}_{\textrm{unsub.}}(\tx, \mathbf{b};\mu^2)\Big|_{\ref{Fig:diag5}}^{reg} =0\,.
\label{diag_5_final_zero}
\end{align}

We note that this result stands in sharp contrast to the corresponding calculation performed in the target light-cone gauge in Ref. \cite{Altinoluk:2025ewj}. In the target light-cone gauge calculation, it was shown that the analogous diagram (gluon emission from the quark line to the transverse Wilson line at infinity) produced the dominant rapidity-divergent contribution to the CSS kernel, driven by the ghost-like zero mode of the ML prescription. In the present projectile light-cone gauge calculation, this diagram vanishes due to the fact that the phase factor $e^{-ik^-Y^+}$ accompanying the transverse Wilson line at $Y^+$ suppresses both the regular and the light-cone singular contributions in the $Y^+\to\infty$ limit. The CSS kernel is instead generated by the diagrams where gluon is emitted from the quark field and subsequently absorbed by either lower  (diagram~\ref{Fig:diag6}) or upper (diagram~\ref{Fig:diag4}) longitudinal Wilson lines, as computed in the previous subsections.

%=============================================================================================================================================
\subsection{Diagram~\ref{Fig:diag1}: quark-to-quark ladder}
\label{sec: ladder diagram}
%=============================================================================================================================================

\begin{figure}[h!]
{%
       \includegraphics[width=0.40\textwidth]{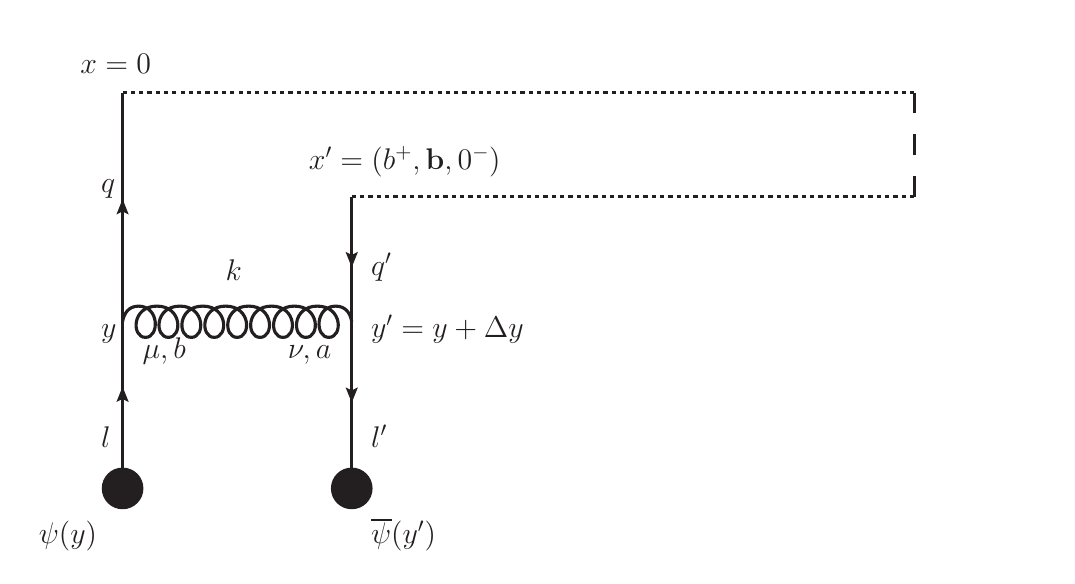}
     }
\caption{Quark-to-quark ladder - diagram~\ref{Fig:diag1}}
\end{figure}

Let us continue our analysis with the ladder diagram presented in diagram~\ref{Fig:diag1} which corresponds to the first term in the expanded expression given in Eq. \eqref{eq: operator expansion in background field}. Written in terms of the quark and gluon propagators, it takes the following form:
\begin{align}
q^{\textrm{n.r.}}_{\textrm{unsub.}}(\tx, \mathbf{b};\mu^2)\Big|_{\ref{Fig:diag1}} &=  
 -g^2 \mu^{2 \epsilon} \, \lim_{Y^+\rightarrow +\infty}
\int \frac{\td b^+}{(2\pi)}\, 
e^{-i\tx P^-b^+}
\int \td^{4-2\epsilon} y \int \td^{4-2\epsilon}y' \,  \delta_{ab} \; G_{0,F}^{\nu\mu}(y',y)\nonumber \\
&
\times
\big\langle P\big|{\overline\psi} (y') \, \gamma_\nu \,  t^a S\, _{0,F}(y' ; b^+, \bbperp, 0^-) \, \frac{\gamma^-}{2} S_{0,F\, }(0 ; y) \gamma_\mu \,  t^b \,
 \psi (y) \big|P\big\rangle_c \, . 
\end{align}
Performing a translation $y^{\prime\mu}\to y^\mu+\Delta y^{\mu}$ with $\Delta y^\mu=y^{\prime\mu}-y^\mu$, and using $\delta^{ab}t^at^b=C_F$, this becomes 
\begin{align}
q^{\textrm{n.r.}}_{\textrm{unsub.}}(\tx, \mathbf{b};\mu^2)\Big|_{\ref{Fig:diag1}} &= 
   -g^2 \mu^{2 \epsilon} C_F \, 
\int \frac{\td b^+}{(2\pi)}\, 
e^{-i\tx P^-b^+} \int \td^{4-2\epsilon} y \int \td^{4-2\epsilon}\Delta y \,G_{0,F}^{\nu\mu}(\Delta y,0)
\big\langle P\big|{\overline\psi}_\alpha  (\Delta y) \,
 \psi_\beta (y) \big|P\big\rangle_c \nonumber \\
&
\times \int \frac{\td^{4-2\epsilon} q}{(2\pi)^{4-2\epsilon}}\int \frac{\td^{4-2\epsilon} q'}{(2\pi)^{4-2\epsilon}} e^{-iq'\cdot(\Delta y+y)} e^{i{q'}^-b^+}e^{-i\mathbf{q}' \cdot \bbperp} e^{iq\cdot y} \left[\gamma_\nu \frac{i\slashed{q'}}{{q'}^2+i0}\frac{\gamma^-}{2} \frac{i\slashed{q}}{{q}^2+i0}\gamma_\mu \right]_{\alpha\beta}\,,
\end{align}
where the quark propagators have been written in momentum space. 
The integrations over $y$ and $b^+$ generate the delta functions $\delta^{(4-2\epsilon)}(q-q')\delta({q'}^- -\tx P^-)$. Performing the remaining integrations over $q'$ and $q^-$ then gives 
\begin{align}
\label{eq:1d_int_1}
q^{\textrm{n.r.}}_{\textrm{unsub.}}(\tx, \mathbf{b};\mu^2)\Big|_{\ref{Fig:diag1}}= &  -\alpha_s \mu^{2 \epsilon} C_F \, 
  \int \td^{4-2\epsilon}\Delta y \,
\big\langle P\big|{\overline\psi}_\alpha  (\Delta y) \,
 \psi_\beta (y) \big|P\big\rangle_c e^{-i\tx P^-\Delta y^+}\int\frac{\td^{4-2\epsilon} k}{(2\pi)^{4-2\epsilon}}e^{-ik\cdot \Delta y}\tilde{G}_{0,F}^{\nu\mu}(k)\nonumber \\
&
\times \int \frac{\td^{2-2\epsilon} \mathbf{q}}{(2\pi)^{2-2\epsilon}}e^{i\mathbf{q}\cdot (\Delta \mathbf{y}-\bbperp)}\int\frac{\td q^+}{(2\pi)} e^{-iq^+\Delta y^-} \left[\gamma_\nu \slashed{q}\gamma^-\slashed{q}\gamma_\mu \right]_{\alpha\beta} \left( \frac{i}{{q}^2+i0}\right)^2  \Bigg|_{q^-=\tx P^-}\,.
\end{align}
The Lorentz structure contracted with the gluon propagator in the projectile light-cone gauge is 
\begin{align}
\label{eq:Lorentz_struc_contracted_1d}
    \left[\gamma_\nu \slashed{q}\gamma^-\slashed{q}\gamma_\mu \right]_{\beta\alpha}\left[-g^{\nu\mu}+\frac{k^\nu n^\mu+n^\nu k^\mu}{[k^+]} \right] \Bigg|_{q^-=\tx P^-} =& \Big\{
    2(1-\delta_s\epsilon)\left[2 (\tx P^-)^2\gamma^+ +\mathbf{q}^2\gamma^- \right]+4\delta_s\epsilon\tx P^- \mathbf{q}^i\gamma^i  \nonumber \\ &+\frac{1}{[k^+]}\left[2\mathbf{q}^2(2k^-\gamma^+-\mathbf{k}^i\gamma^i)-4\tx P^-\mathbf{k}\cdot\mathbf{q}\gamma^+ \right] \Big\}_{\alpha\beta}\,,
\end{align}
where $\delta_s\epsilon$ is defined via $g_{\mu\nu}g^{\mu\nu} \equiv D_s = 4 -2\delta_s\epsilon$, with $\delta_s\equiv 1$ in conventional dimensional regularization. The first line in Eq. \eqref{eq:Lorentz_struc_contracted_1d} contains no light-cone gauge singularity, while the second line carries the extra $1/[k^+]$ pole which must be treated with the ML prescription given in Eq. \eqref{eq:ml-prescription_lc_var}. Upon substituting Eq. \eqref{eq:Lorentz_struc_contracted_1d} into Eq. \eqref{eq:1d_int_1} and performing the changes of variables from Eq.~\eqref{eq:cov-q-to-l}, one arrives at 
\begin{align}
\label{eq:1d_int_1_2}
q^{\textrm{n.r.}}_{\textrm{unsub.}}(\tx, \mathbf{b};\mu^2)\Big|_{\ref{Fig:diag1}}= & \alpha_s \mu^{2 \epsilon} C_F \, 
  \int \td^{4-2\epsilon}\Delta y \,
\big\langle P\big|{\overline\psi}_\alpha  (\Delta y) \,
 \psi_\beta (y) \big|P\big\rangle_c e^{-i\tx P^-\Delta y^+}
 \int \frac{\td^{2-2\epsilon} \mathbf{l}}{(2\pi)^{2-2\epsilon}}e^{i\mathbf{l}\cdot (\Delta \mathbf{y}-\bbperp)}\int\frac{\td l^+}{(2\pi)} e^{-il^+\Delta y^-} 
 \nonumber \\
&
\times 
 \int\frac{\td^{4-2\epsilon} k}{(2\pi)^{4-2\epsilon}}
  \frac{i\, e^{-ik^- \Delta y^+} e^{i\mathbf{k}\cdot \bbperp}}{\left({k}^2+i0\right)\left[2\tx P^-(l^+\!-\!\k^+)-(\bl\!-\!\bk)^2+i0\right]^2}
  \Big\{
    2(1-\delta_s\epsilon)\left[2 (\tx P^-)^2\gamma^+ +(\bl\!-\!\bk)^2\gamma^- \right]
    \nonumber \\ &
  +4\delta_s\epsilon\tx P^- (\bl^i\!-\!\bk^i)\gamma^i
  +\frac{1}{[k^+]}\left[2(\bl\!-\!\bk)^2(2k^-\gamma^+-\mathbf{k}^i\gamma^i)-4\tx P^-\mathbf{k}\!\cdot\!(\bl\!-\!\bk)\gamma^+ \right] \Big\}_{\alpha\beta}  
     \, .
\end{align}
The $k^+$ integration can be performed in the same manner as in the previous cases, for $\tx>0$, as
\begin{align}
\int\frac{\td k^+}{2\pi}
\frac{i}{\left({k}^2+i0\right)\left[2\tx P^-(l^+\!-\!\k^+)-(\bl\!-\!\bk)^2+i0\right]^2}
=&\, \frac{\theta(k^-)}{2k^-}\,
\frac{1}{\left[2\tx P^-l^+\!-\!\frac{\tx P^-}{k^-}\bk^2 -(\bl\!-\!\bk)^2+i0\right]^2}
\label{int_plus_lad_scal}
\end{align}
and with the ML prescription
\begin{align}
\int\frac{\td k^+}{2\pi}
\frac{1}{[k^+]}
\frac{i}{\left({k}^2+i0\right)\left[2\tx P^-(l^+\!-\!\k^+)-(\bl\!-\!\bk)^2+i0\right]^2}
=
\frac{\theta(k^-)}{\bk^2}\,
\Bigg\{
&\,
\frac{1}{\left[2\tx P^-l^+\!-\!\frac{\tx P^-}{k^-}\bk^2 -(\bl\!-\!\bk)^2+i0\right]^2}
\nonumber\\
&\,
-
\frac{1}{\left[2\tx P^-l^+ -(\bl\!-\!\bk)^2+i0\right]^2}
\Bigg\}
\label{int_plus_lad_ML}\, .
\end{align}
Indeed, for $\tx>0$, all poles are on the same side of the real axis for $k^-<0$, in both cases, so that a non-vanishing contribution is obtained only for $k^->0$.
Moreover, the result \eqref{int_plus_lad_ML} can be rewritten as
\begin{align}
&\,
\int\frac{\td k^+}{2\pi}
\frac{1}{[k^+]}
\frac{i}{\left({k}^2+i0\right)\left[2\tx P^-(l^+\!-\!\k^+)-(\bl\!-\!\bk)^2+i0\right]^2}
=
\frac{\theta(k^-) }{2k^-}\,
\frac{2\tx P^-}{\left[2\tx P^-l^+\!-\!\frac{\tx P^-}{k^-}\bk^2 -(\bl\!-\!\bk)^2+i0\right]}
\nonumber\\
&\, \times\,
\frac{1}{\left[2\tx P^-l^+ -(\bl\!-\!\bk)^2+i0\right]}
\Bigg\{
\frac{1}{\left[2\tx P^-l^+\!-\!\frac{\tx P^-}{k^-}\bk^2 -(\bl\!-\!\bk)^2+i0\right]}
+
\frac{1}{\left[2\tx P^-l^+ -(\bl\!-\!\bk)^2+i0\right]}
\Bigg\}
\label{int_plus_lad_ML_2}\, .
\end{align}

Inserting Eqs.~\eqref{int_plus_lad_scal} and \eqref{int_plus_lad_ML_2} into Eq.~\eqref{eq:1d_int_1_2}, one gets
\begin{align}
\label{eq:1d_int_1_3}
& q^{\textrm{n.r.}}_{\textrm{unsub.}}(\tx, \mathbf{b};\mu^2)\Big|_{\ref{Fig:diag1}}
=  
\frac{\alpha_s \mu^{2 \epsilon} C_F}{4\pi} \, 
  \int \td^{4-2\epsilon}\Delta y \,
\big\langle P\big|{\overline\psi}_\alpha  (\Delta y) \,
 \psi_\beta (y) \big|P\big\rangle_c e^{-i\tx P^-\Delta y^+}
 \int \frac{\td^{2-2\epsilon} \mathbf{l}}{(2\pi)^{2-2\epsilon}}e^{i\mathbf{l}\cdot (\Delta \mathbf{y}-\bbperp)}\int\frac{\td l^+}{(2\pi)} e^{-il^+\Delta y^-} 
 \nonumber \\
&
\times 
\int\frac{\td^{2-2\epsilon} \bk}{(2\pi)^{2-2\epsilon}}\, e^{i\mathbf{k}\cdot \bbperp} 
\int_0^{+\infty} \frac{\td k^-}{k^-}\,
e^{-ik^- \Delta y^+}
  \Bigg\{
  \frac{\left[
    2(1-\delta_s\epsilon)\left(2 (\tx P^-)^2\gamma^+ +(\bl\!-\!\bk)^2\gamma^- \right)
    +4\delta_s\epsilon\tx P^- (\bl^i\!-\!\bk^i)\gamma^i\right]}{\left[2\tx P^-l^+\!-\!\frac{\tx P^-}{k^-}\bk^2 -(\bl\!-\!\bk)^2+i0\right]^2}
    \nonumber \\ 
    & \hspace{1cm}
  +\frac{2\tx P^-
  \left[2(\bl\!-\!\bk)^2(2k^-\gamma^+-\mathbf{k}^i\gamma^i)-4\tx P^-\mathbf{k}\!\cdot\!(\bl\!-\!\bk)\gamma^+ \right]}{\left[2\tx P^-l^+\!-\!\frac{\tx P^-}{k^-}\bk^2 -(\bl\!-\!\bk)^2+i0\right]\left[2\tx P^-l^+ -(\bl\!-\!\bk)^2+i0\right]} 
  \nonumber\\
  &\hspace{2cm}\times\, 
  \left[
  \frac{1}{\left[2\tx P^-l^+\!-\!\frac{\tx P^-}{k^-}\bk^2 -(\bl\!-\!\bk)^2+i0\right]}
+
\frac{1}{\left[2\tx P^-l^+ -(\bl\!-\!\bk)^2+i0\right]}
\right]
  \Bigg\}_{\alpha\beta}  
     \, .
\end{align}
Finally, using the change of variables from Eq.~\eqref{cov:k_minus_2_z} leads to
\begin{align}
\label{eq:1d_int_1_4}
& q^{\textrm{n.r.}}_{\textrm{unsub.}}(\tx, \mathbf{b};\mu^2)\Big|_{\ref{Fig:diag1}}
=  
\frac{\alpha_s \mu^{2 \epsilon} C_F}{4\pi} \, 
  \int \td^{4-2\epsilon}\Delta y \,
\big\langle P\big|{\overline\psi}_\alpha  (\Delta y) \,
 \psi_\beta (y) \big|P\big\rangle_c 
 \int \frac{\td^{2-2\epsilon} \mathbf{l}}{(2\pi)^{2-2\epsilon}}e^{i\mathbf{l}\cdot (\Delta \mathbf{y}-\bbperp)}\int\frac{\td l^+}{(2\pi)} e^{-il^+\Delta y^-} 
 \nonumber \\
&
\times 
\int\frac{\td^{2-2\epsilon} \bk}{(2\pi)^{2-2\epsilon}}\, e^{i\mathbf{k}\cdot \bbperp} 
\int_0^{1} \frac{\td z}{z}\,
e^{-i\frac{\tx P^-}{z}\Delta y^+}
  \Bigg\{
  (1\!-\!z)\frac{\left[
    2(1-\delta_s\epsilon)\left(2 (\tx P^-)^2\gamma^+ +(\bl\!-\!\bk)^2\gamma^- \right)
    +4\delta_s\epsilon\tx P^- (\bl^i\!-\!\bk^i)\gamma^i\right]}{\left[2(1\!-\!z)\tx P^-l^+\!-\!z\bk^2 -(1\!-\!z)(\bl\!-\!\bk)^2+i0\right]^2}
    \nonumber \\ 
    & \hspace{1cm}
  +\frac{2\tx P^-
  \left[2(\bl\!-\!\bk)^2(2\tx P^-
  \frac{(1\!-\!z)}{z}  
  \gamma^+-\mathbf{k}^i\gamma^i)-4\tx P^-\mathbf{k}\!\cdot\!(\bl\!-\!\bk)\gamma^+ \right]}{\left[2(1\!-\!z)\tx P^-l^+\!-\!z\bk^2 -(1\!-\!z)(\bl\!-\!\bk)^2+i0\right]\left[2\tx P^-l^+ -(\bl\!-\!\bk)^2+i0\right]} 
  \nonumber\\
  &\hspace{2cm}\times\, 
  \left[
  \frac{(1\!-\!z)}{\left[2(1\!-\!z)\tx P^-l^+\!-\!z\bk^2 -(1\!-\!z)(\bl\!-\!\bk)^2+i0\right]}
+
\frac{1}{\left[2\tx P^-l^+ -(\bl\!-\!\bk)^2+i0\right]}
\right]
  \Bigg\}_{\alpha\beta}  
     \, .
\end{align}
Remarkably, the integrand is regular for $z\rightarrow 1$, so that no rapidity regulator is needed. Moreover, the phase in $\tx P^-\Delta y^+/z$ suppresses the regime $z\rightarrow 0$, so that the integration over $z$ is convergent, for $\tx>0$. The only possible divergence could be from the UV regime for the integral over $\bk$. Such UV divergence would come from the contribution in the second line of Eq.~\eqref{eq:1d_int_1_4}, but only for $\bbperp=0$. Indeed for $\bbperp\neq 0$, the $\bbperp$ dependent phase factor is enough to regularize that potential UV divergence. 

We therefore conclude that, for nonzero $\b$, the ladder diagram~\ref{Fig:diag1} is completely finite and does not require any  UV or rapidity regularization. It thus contributes only to finite NLO corrections to the quark TMD and not to its CSS evolution. This result is similar to the corresponding result in the target light-cone gauge computation of Ref.~\cite{Altinoluk:2025ewj}, where the same ladder diagram was shown to be finite for nonzero $\b$ within the ML prescription. We further note that at $\b=0$, the quark TMD reduces to the collinear quark PDF, and the resulting UV divergence is associated with the DGLAP evolution of the parton distribution, in agreement with Ref.~\cite{Altinoluk:2023dww}. In the present work, we focus on the CSS evolution of the TMD and therefore restrict ourselves to $\b\neq 0$ throughout.

%=============================================================================================================================================
\subsection{Diagram~\ref{Fig:diag11}: transverse Wilson line self-energy at infinity}
\label{sec: WLSE at infinity diagram}
%=============================================================================================================================================

Concerning the diagram \ref{Fig:diag11}, since the light-cone vector $n^{\mu}$ has no transverse components by definition, the gluon propagator reduces in that case to the Feynman gauge propagator, which was already the case in the target light-cone gauge in Ref.~\cite{Altinoluk:2025ewj}. Hence, we can simply reuse the result for that diagram obtained in appendix B from Ref.~\cite{Altinoluk:2025ewj}, which is
 \begin{align}
      q^{\textrm{n.r.}}_{\textrm{unsub.}}(\tx, \mathbf{b};\mu^2)\big|_{\ref{Fig:diag11}}
      &= \frac{\alpha_s C_F}{2 \pi} 
      \frac{\Gamma(1\!-\!\epsilon)}{\epsilon(1\!-\!2\epsilon)}\,
      \big(\pi\mu^2\mathbf{b}^2\big)^\epsilon\;
      q_{\textrm{Bckgd}}(\tx, \mathbf{b};\mu^2) 
      \, .
\label{SE_WL_at_inf}
 \end{align}
Here, the $\epsilon$ pole corresponds to a UV divergence. That contribution actually has a UV power divergence, which appears in dimensional regularization as the $(1\!-\!2\epsilon)$ denominator.

%==========================================================
\subsection{Diagrams~\ref{Fig:diag12}, \ref{Fig:diag13}, \ref{Fig:diag14}}
\label{sec: Diags 12 13 14}
%==========================================================
We now turn to the three remaining one-loop diagrams that are specific to the projectile light-cone gauge and do not appear in the target light-cone gauge computation. These are the Wilson line self-energy on the lower segment of the gauge link (diagram~\ref{Fig:diag12}), the gluon emission from the upper segment of the gauge link extending to infinity (diagram~\ref{Fig:diag13}), and the gluon exchange between the upper and lower longitudinal Wilson lines (diagram~\ref{Fig:diag14}). As we show below, all three diagrams vanish by the same mechanism: the pole structure in $k^+$ arising from the light-cone gauge gluon propagator in the projectile gauge places all poles on the same side of the real axis, causing the corresponding contour integrals to vanish identically.

\subsubsection{Diagram~\ref{Fig:diag12}: Wilson line self-energy on the lower part of the gauge link}

\begin{figure}[h!]
{%
       \includegraphics[width=0.40\textwidth]{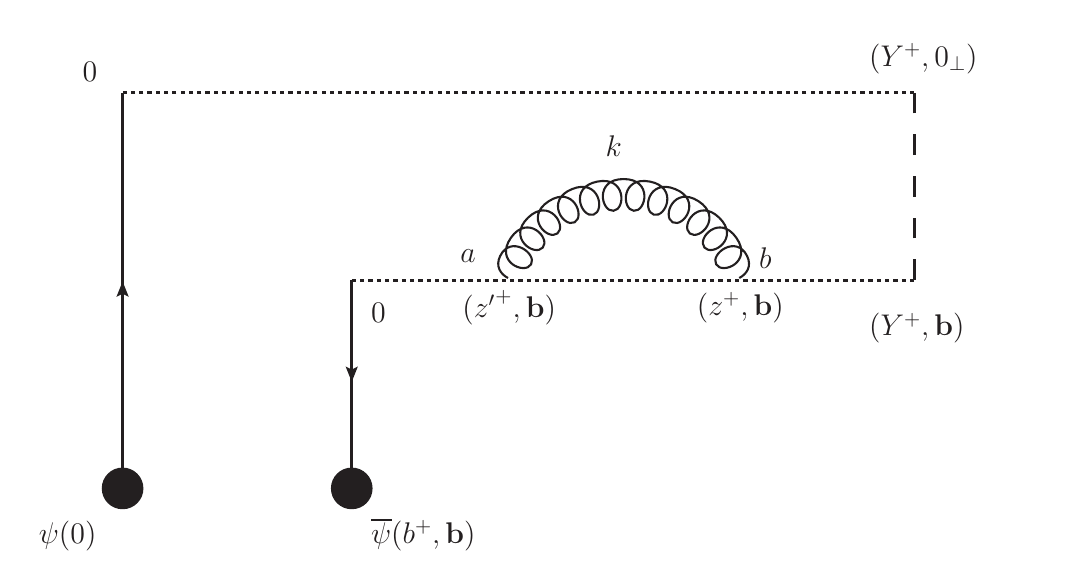}
     }  

\caption{Wilson line self-energy diagram in the lower part of the gauge link - 
diagram~\ref{Fig:diag12}}
\end{figure}

Diagram~\ref{Fig:diag12} corresponds to the one-loop self-energy of the lower longitudinal Wilson line, running from $0$ to $Y^+$ at $\mathbf{b}$. It arises from the fifth term in the background field expansion \eqref{eq: operator expansion in background field} and can be written as
\begin{align}
q^{\textrm{n.r.}}_{\textrm{unsub.}}(x, \mathbf{b}; \mu^2)\Big|_{\ref{Fig:diag12}}
& = \lim_{Y^+\to+\infty}
\int\frac{db^+}{(2\pi)}\,e^{-ixP^-b^+}
\Big\langle P\Big|\mathcal{T}\Big[
\psi(b^+,\mathbf{b},0^-)
\frac{\gamma^-}{2}
\nonumber \\
& \times
\left(+i\mu^\epsilon g\int_{b^+}^{Y^+}dz^+\,\delta A^-_a(z^+,\mathbf{b},0^-)t^a\right)
\left(+i\mu^\epsilon g\int_{b^+}^{z^+}dz'^+\,\delta A^-_b(z'^+,\mathbf{b},0^-)t^b\right)
\psi(0)
\Big]\Big|P\Big\rangle_c.
\label{eq:diag1f_start}
\end{align}
Contracting the two fluctuation fields into a free gluon propagator and performing the color algebra, this reduces to
\begin{align}
q^{\textrm{n.r.}}_{\textrm{unsub.}}(x, \mathbf{b}; \mu^2)\Big|_{\ref{Fig:diag12}}
& = -g^2\mu^{2\epsilon}C_F
\lim_{Y^+\to+\infty}
\int\frac{db^+}{(2\pi)}\,e^{-ixP^-b^+}
\Big\langle P\Big|\psi(b^+,\mathbf{b},0^-)\frac{\gamma^-}{2}\psi(0)\Big|P\Big\rangle_c
\nonumber \\
&\times 
\int_{b^+}^{Y^+}dz^+
\int_{b^+}^{z^+}dz'^+\,
G^{--}_{0,F}(z^+,\mathbf{b},0^-;z'^+,\mathbf{b},0^-)\,,
\label{eq:diag1f_propagator}
\end{align}
where the gluon propagator component $G^{--}_{0,F}$ in the projectile light-cone gauge, evaluated in momentum space, reads
\begin{align}
\tilde{G}^{--}_{0,F}(k) = \frac{i}{k^2+i0}\frac{2k^-}{[k^+]}\,.
\label{eq:Gmm_prop}
\end{align}
Substituting Eq. \eqref{eq:Gmm_prop} and writing out the $z'^+$ and $z^+$ integrations explicitly, the expression for diagram~\ref{Fig:diag12} becomes
\begin{align}
q^{\textrm{n.r.}}_{\textrm{unsub.}}(x, \mathbf{b}; \mu^2)\Big|_{~\ref{Fig:diag12}}
& = -g^2\mu^{2\epsilon}C_F
\lim_{Y^+\to+\infty}
\int\frac{db^+}{(2\pi)}\,e^{-ixP^-b^+}
\Big\langle P\Big|\psi(b^+,\mathbf{b},0^-)\frac{\gamma^-}{2}\psi(0)\Big|P\Big\rangle_c
\nonumber\\
& \times 
\int\frac{d^{4-2\epsilon}k}{(2\pi)^{4-2\epsilon}}
\frac{i}{k^2+i0}\frac{2k^-}{[k^+]}
\int_{b^+}^{Y^+}dz^+
\int_{b^+}^{z^+}dz'^+\,
e^{ik^-(z'^+-z^+)}\,.
\label{eq:diag1f_momentum}
\end{align}
The key observation is that the $k^+$ integral, which must be performed first as required by the ML prescription, takes the form
\begin{align}
\int\frac{dk^+}{(2\pi)}
\frac{i}{k^+-\frac{k^2-i0}{2k^-}}
\left(\frac{\theta(k^-)}{k^++i0}+\frac{\theta(-k^-)}{k^+-i0}\right)\,.
\label{eq:kplus_1f}
\end{align}
For positive $k^-$, the poles at $k^+=0-i0$ and $k^+=k^2/(2k^-)-i0$ both lie below the real axis, so closing 
the contour above yields zero. For negative $k^-$, the poles at $k^+=0+i0$ and $k^+=k^2/(2k^-)+i0$ both lie 
above the real axis, so closing the contour below yields zero. In both cases all poles lie on the same side of 
the real axis, and the $k^+$ integral vanishes identically. We therefore conclude that
\begin{align}
q^{\textrm{n.r.}}_{\textrm{unsub.}}(x, \mathbf{b}; \mu^2)\Big|_{\ref{Fig:diag12}} = 0\,.
\label{eq:diag1f_zero}
\end{align}
%
%============================================================
\subsubsection{Diagram~\ref{Fig:diag13}: Gluon emission from the upper part of the gauge link to infinity}
%=====================================================

\begin{figure}[h!]
{%
       \includegraphics[width=0.40\textwidth]{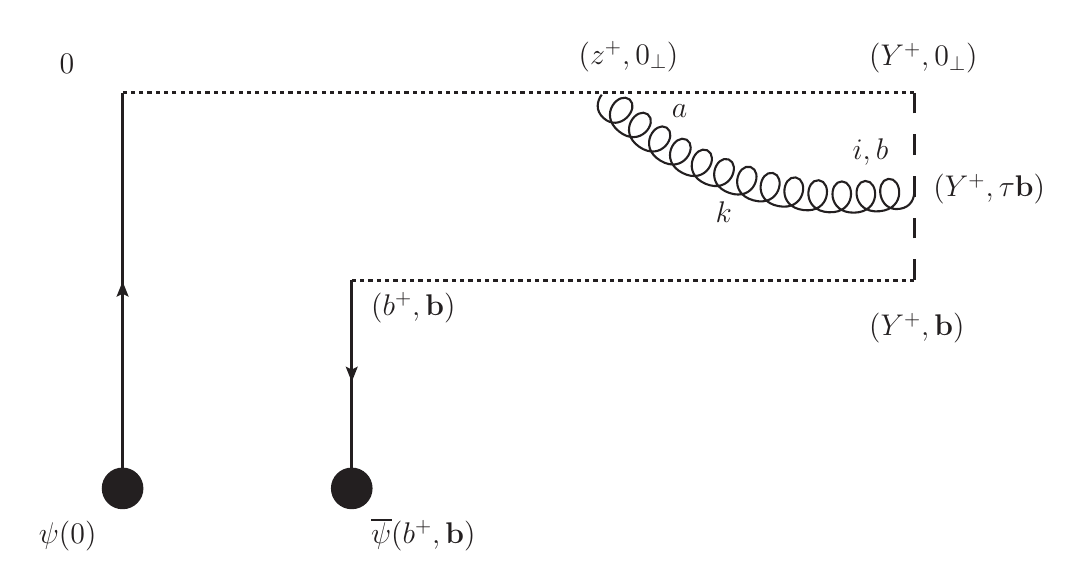}
     } 
   
\caption{Emission of gluon from the upper part of the gauge link to infinity - diagram~\ref{Fig:diag13}.}
\end{figure}

Diagram~\ref{Fig:diag13} corresponds to gluon emission from the lower longitudinal Wilson line at $0_\perp$ to the  transverse Wilson line at light-cone infinity $Y^+$. It arises from the seventh term in the background field expansion \eqref{eq: operator expansion in background field} and can be written as
\begin{align}
q^{\textrm{n.r.}}_{\textrm{unsub.}}(x, \mathbf{b}; \mu^2)\Big|_{\ref{Fig:diag13}}
& = \lim_{Y^+\to+\infty}
\int\frac{db^+}{(2\pi)}\,e^{-ixP^-b^+}
\Big\langle P\Big|\mathcal{T}\Big[
\psi(b^+,\mathbf{b},0^-)
\frac{\gamma^-}{2}
\left(-i\mu^\epsilon g\int_0^1 d\tau\,\mathbf{b}^i\,\delta A^b_i(Y^+,\tau\mathbf{b},0^-)t^b\right)
\nonumber \\
& \hspace{5cm}
\times 
\left(-i\mu^\epsilon g\int_0^{Y^+}dz^+\,\delta A^-_a(z^+,0_\perp,0^-)t^a\right)
\psi(0)
\Big]\Big|P\Big\rangle_c.
\label{eq:diag1g_start}
\end{align}
Contracting the fluctuation fields and performing the color algebra, Eq. \eqref{eq:diag1g_start}  becomes
\begin{align}
q^{\textrm{n.r.}}_{\textrm{unsub.}}(x, \mathbf{b}; \mu^2)\Big|_{\ref{Fig:diag13}}
& = -g^2\mu^{2\epsilon}C_F
\lim_{Y^+\to+\infty}
\int\frac{db^+}{(2\pi)}\,e^{-ixP^-b^+}
\Big\langle P\Big|\psi(b^+,\mathbf{b},0^-)
\frac{\gamma^-}{2}\psi(0)\Big|P\Big\rangle_c
\nonumber \\ 
& \hspace{5cm}
\times 
\int_0^{Y^+}dz^+
\int_0^1 d\tau\,
G^{i-}_{0,F}(Y^+,\tau\mathbf{b},0^-;z^+,0_\perp,0^-)\,,
\label{eq:diag1g_propagator}
\end{align}
where the gluon propagator in Eq. \eqref{eq:diag1g_propagator} in the projectile light-cone gauge reads in momentum space
\begin{align}
\tilde{G}^{i-}_{0,F}(k) = \frac{i}{k^2+i0}\frac{k^i}{[k^+]}\,.
\label{eq:Gim_prop}
\end{align}
Writing out the position-space integrations explicitly in momentum space, one arrives at
\begin{align}
q^{\textrm{n.r.}}_{\textrm{unsub.}}(x, \mathbf{b}; \mu^2)\Big|_{\ref{Fig:diag13}}
& = -g^2\mu^{2\epsilon}C_F
\lim_{Y^+\to+\infty}
\int\frac{db^+}{(2\pi)}\,e^{-ixP^-b^+}
\Big\langle P\Big|\psi(b^+,\mathbf{b},0^-)
\frac{\gamma^-}{2}\psi(0)\Big|P\Big\rangle_c
\nonumber \\
& \hspace{3cm}
\times 
\int_0^{Y^+}dz^+
\int_0^1 d\tau
\int\frac{d^{4-2\epsilon}k}{(2\pi)^{4-2\epsilon}}\,
e^{ik^-(z^+-Y^+)}e^{i\tau\mathbf{k}\cdot\mathbf{b}}
\frac{i\mathbf{b}^i}{k^2+i0}\frac{k^i}{[k^+]}\,.
\label{eq:diag1g_momentum}
\end{align}
The $k^+$ integral that needs to be performed is identical to the one given in Eq. \eqref{eq:kplus_1f}.  
By the same pole argument, all poles lie on the same 
side of the real axis for any sign of $k^-$, and the integral vanishes. Therefore,
\begin{align}
q^{\textrm{n.r.}}_{\textrm{unsub.}}(x, \mathbf{b}; \mu^2)\Big|_{\ref{Fig:diag13}} = 0\,.
\label{eq:diag1g_zero}
\end{align}
%

%=====================================
\subsubsection{Diagram~\ref{Fig:diag14}: Gluon exchange between the two longitudinal Wilson lines}
%=====================================

\begin{figure}[h!]
{%
       \includegraphics[width=0.40\textwidth]{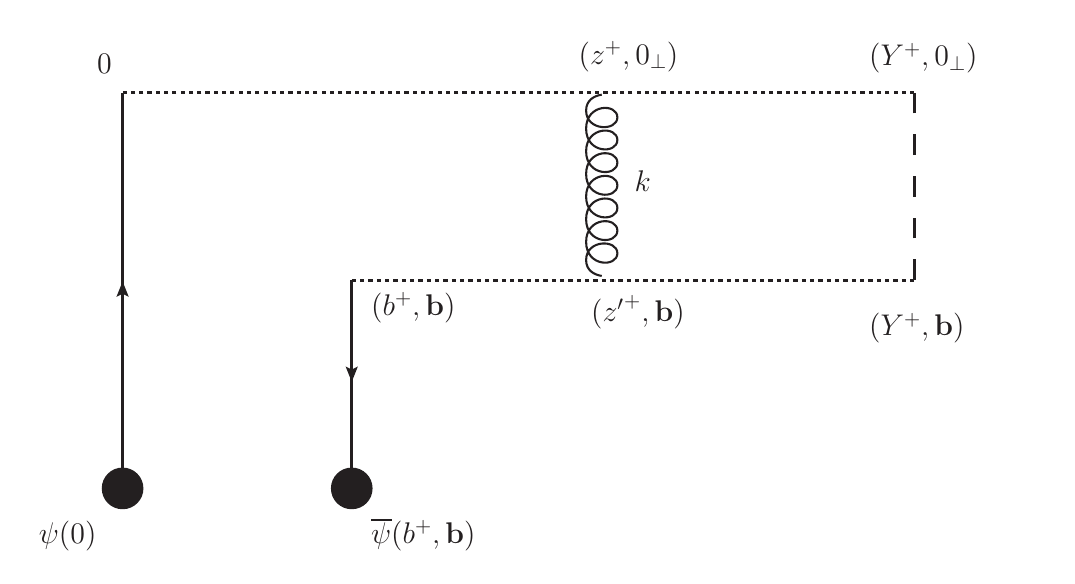}
     }

\caption{Gluon exchange between two light-like Wilson lines in the gauge link - diagram\ref{Fig:diag14}.}
\end{figure}

Diagram~\ref{Fig:diag14} corresponds to the exchange of a gluon between the lower longitudinal Wilson line (running from  $b^+$ to $Y^+$ at $\mathbf{b}$) and the upper longitudinal Wilson line (running from $0$ to $Y^+$ at $0_\perp$). It arises from the eighth term in the background field expansion \eqref{eq: operator expansion in background field} and reads
\begin{align}
q^{\textrm{n.r.}}_{\textrm{unsub.}}(x, \mathbf{b}; \mu^2)\Big|_{\ref{Fig:diag14}}
& = \lim_{Y^+\to+\infty}
\int\frac{db^+}{(2\pi)}\,e^{-ixP^-b^+}
\Big\langle P\Big|\mathcal{T}\Big[
\psi(b^+,\mathbf{b},0^-)
\frac{\gamma^-}{2}
\left(+i\mu^\epsilon g\int_{b^+}^{Y^+}dz'^+\,\delta A^-_a(z'^+,\mathbf{b},0^-)t^a\right)
\nonumber \\ 
& \hspace{5cm}
\times 
\left(-i\mu^\epsilon g\int_0^{Y^+}dz^+\,\delta A^-_b(z^+,0_\perp,0^-)t^b\right)
\psi(0)
\Big]\Big|P\Big\rangle_c.
\label{eq:diag1h_start}
\end{align}
Contracting the two fluctuation fields and performing the color algebra, one obtains
\begin{align}
q^{\textrm{n.r.}}_{\textrm{unsub.}}(x, \mathbf{b}; \mu^2)\Big|_{\ref{Fig:diag14}}
& = g^2\mu^{2\epsilon}C_F
\lim_{Y^+\to+\infty}
\int\frac{db^+}{(2\pi)}\,e^{-ixP^-b^+}
\Big\langle P\Big|\psi(b^+,\mathbf{b},0^-)
\frac{\gamma^-}{2}\psi(0)\Big|P\Big\rangle_c
\nonumber \\
& \hspace{5cm}
\times 
\int_{b^+}^{Y^+}dz'^+
\int_0^{Y^+}dz^+\,
G^{--}_{0,F}(z'^+,\mathbf{b},0^-;z^+,0_\perp,0^-)\,.
\label{eq:diag1h_propagator}
\end{align}
The $G^{--}_{0,F}$ propagator component is again given by \eqref{eq:Gmm_prop}. Writing the integrations 
explicitly in momentum space, the $k^+$ integral that must be performed is once again of the form 
\eqref{eq:kplus_1f}, and vanishes by the same pole argument. We therefore have,
\begin{align}
q^{\textrm{n.r.}}_{\textrm{unsub.}}(x, \mathbf{b}; \mu^2)\Big|_{\ref{Fig:diag14}} = 0\,.
\label{eq:diag1h_zero}
\end{align}

%=====================================
\subsubsection{Summary for diagrams~\ref{Fig:diag12}, \ref{Fig:diag13} and \ref{Fig:diag14}}
%=====================================

All three diagrams~\ref{Fig:diag12}, \ref{Fig:diag13} and \ref{Fig:diag14} vanish identically in the projectile light-cone gauge. The vanishing 
mechanism is the same in each case: the gluon propagator components $\tilde{G}^{--}_{0,F}$ and $\tilde{G}^{i-}_{0,F}$, which are the only ones that can appear in these diagrams, carry the extra light-cone gauge singularity $1/[k^+]$ from the ML prescription. When the $k^+$ integral is performed first, as required by the ML prescription, the resulting pole structure places all poles on the same side of the real axis regardless of the sign of $k^-$, causing the contour integral to vanish. This is a direct consequence of working in the projectile light-cone gauge $A^+=0$ with the ML prescription, and has no analog in the target light-cone gauge calculation of Ref.~\cite{Altinoluk:2025ewj}, where these types of longitudinal Wilson line diagrams do not arise at all. We also note that the same vanishing mechanism applies to several of the soft factor diagrams evaluated in the following subsection, as well as to the symmetric counterparts of diagrams~\ref{Fig:diag12} and \ref{Fig:diag13} listed in Appendix~\ref{appendix_symmetric_diagrams}.

%===========================================================
\subsection{Soft factor}
\label{subsec:Soft_Factor}
%===========================================================

%%%%%%%%%%%%%%%%%%%%%%%%%%%%%%%%%%%%%%%%%%%%%%%%%%%%%%%%%%%%%%%%%%%%%%%%%%%%%%%%%%%%%%
\begin{figure}[t]
\subfloat[ \label{Fig:Soft factor a}]{%
       \includegraphics[width=0.35\textwidth]{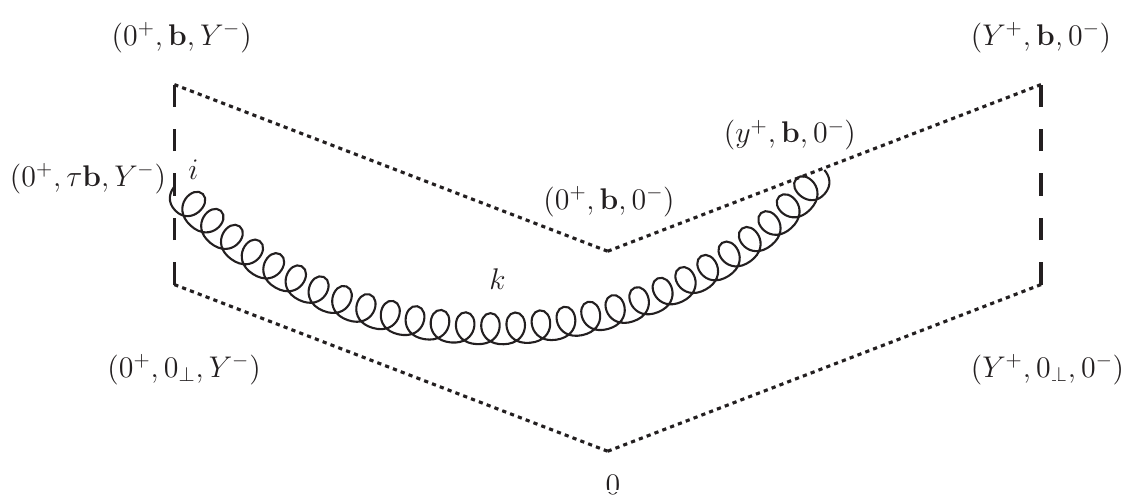}
     }
\hfill
\subfloat[ \label{Fig:Soft factor b}]{%
       \includegraphics[width=0.35\textwidth]{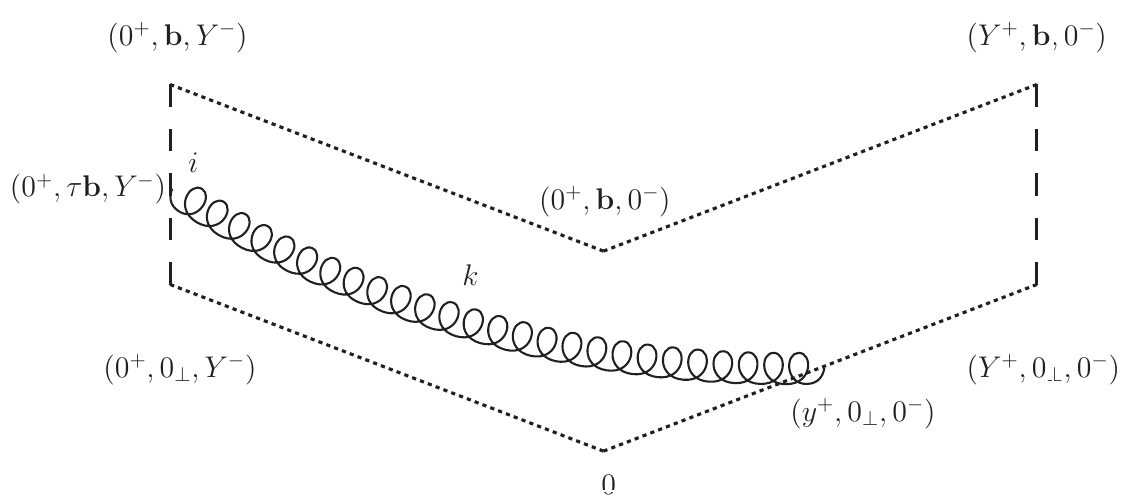}
     }

\subfloat[ \label{Fig:Soft factor c}]{%
       \includegraphics[width=0.35\textwidth]{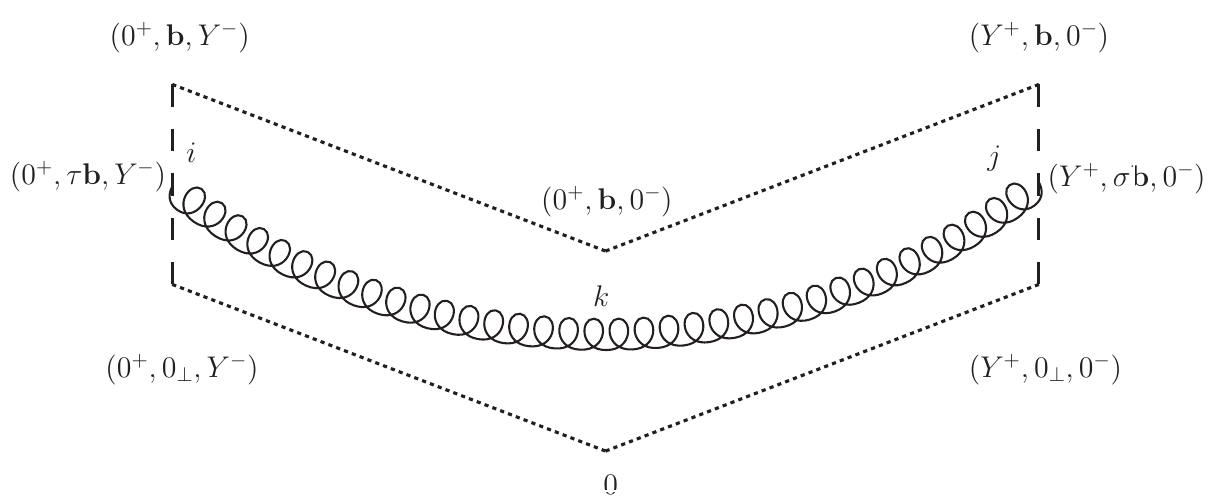}
     }
\hfill
\subfloat[ \label{Fig:Soft factor d}]{%
       \includegraphics[width=0.35\textwidth]{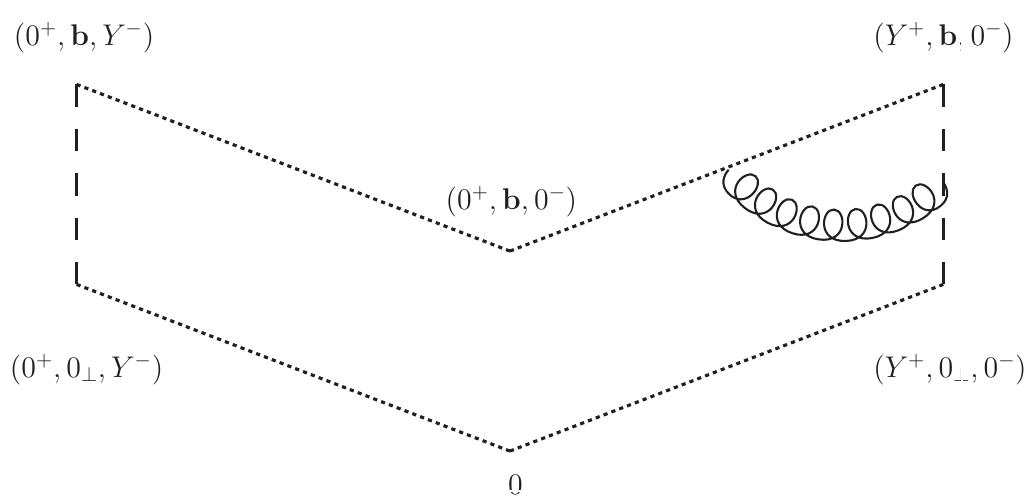}
     }

\subfloat[ \label{Fig:Soft factor e}]{%
       \includegraphics[width=0.35\textwidth]{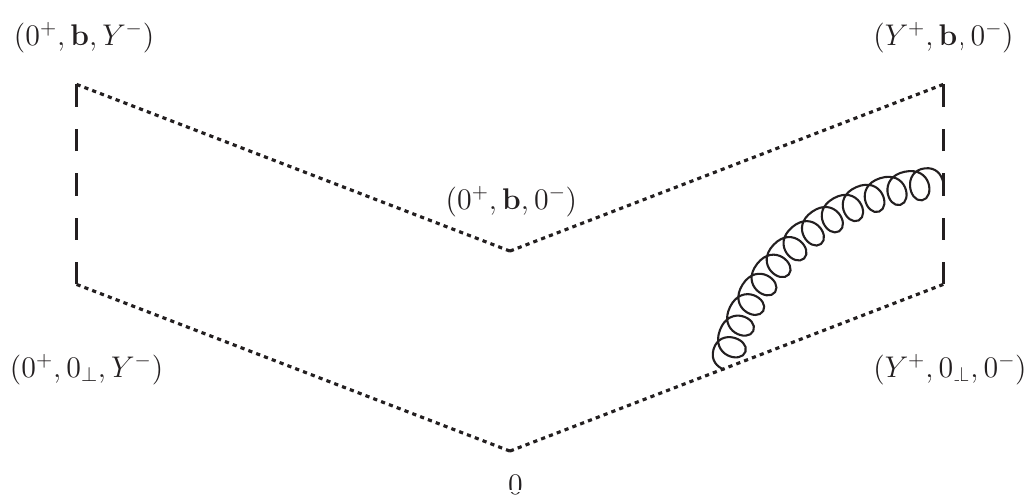}
     }
\hfill
\subfloat[ \label{Fig:Soft factor f}]{%
       \includegraphics[width=0.35\textwidth]{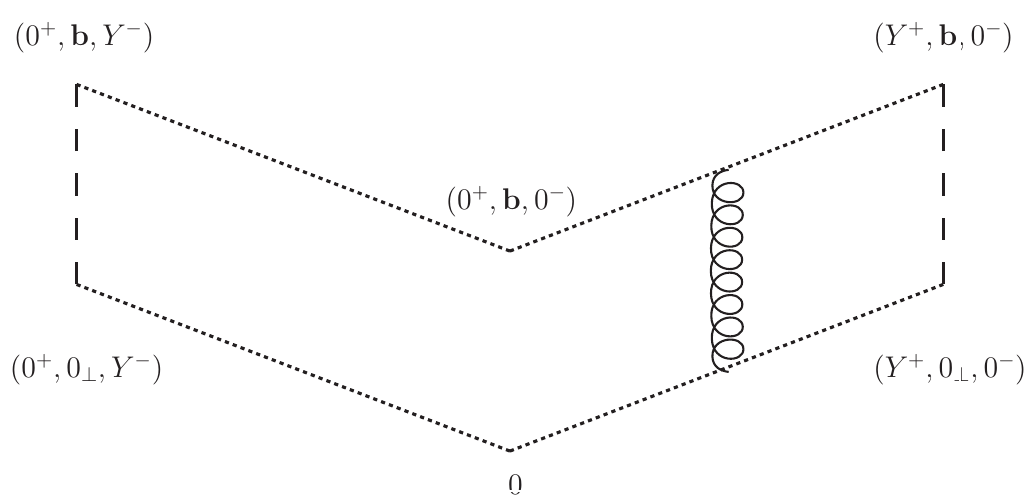}
     }

\subfloat[ \label{Fig:Soft factor g}]{%
       \includegraphics[width=0.35\textwidth]{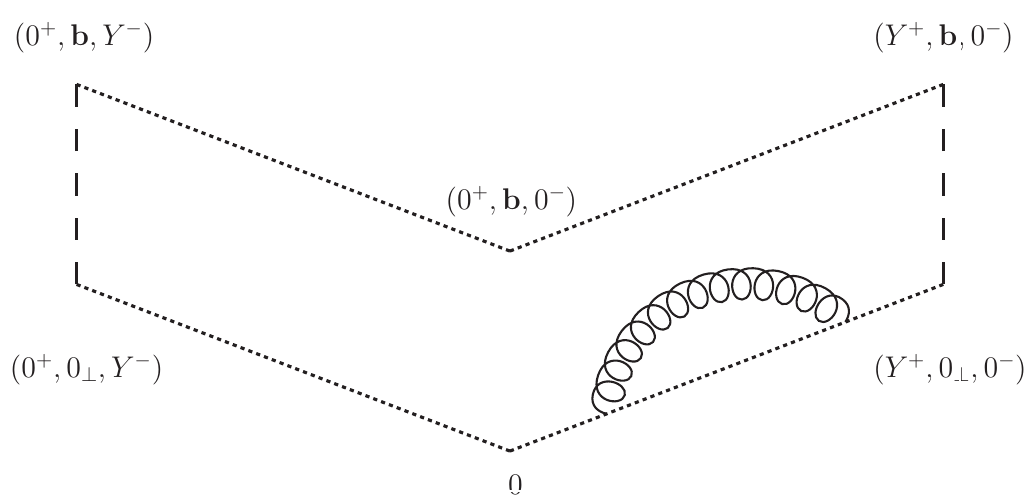}
     }
\hfill
\subfloat[ \label{Fig:Soft factor h}]{%
       \includegraphics[width=0.35\textwidth]{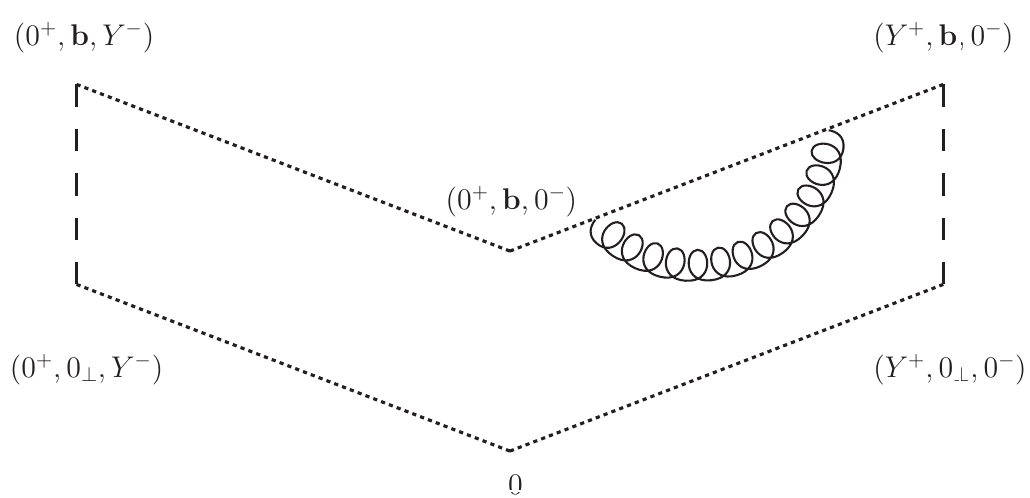}
     }

\subfloat[ \label{Fig:Soft factor transverse wilson line Y+}]{%
       \includegraphics[width=0.35\textwidth]{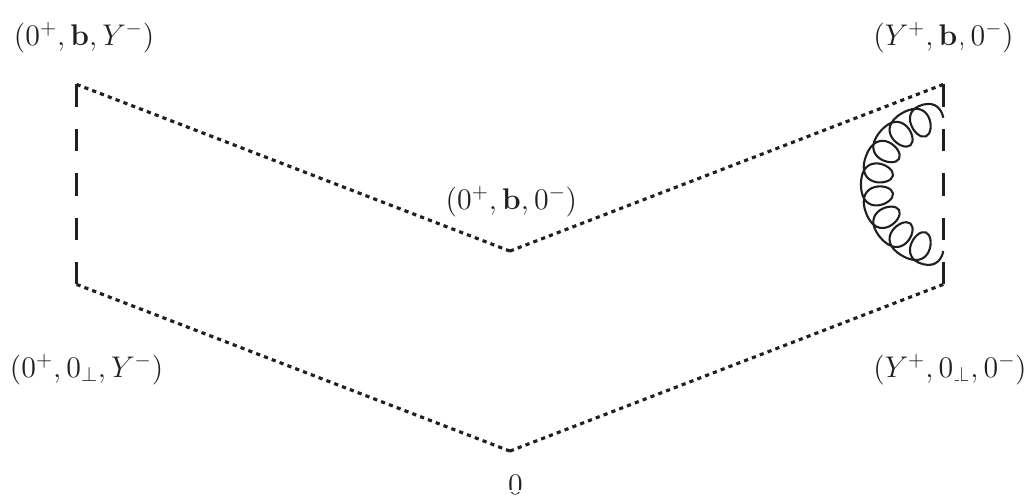}
     }
\hfill
\subfloat[ \label{Fig:Soft factor transverse wilson line Y-}]{%
       \includegraphics[width=0.35\textwidth]{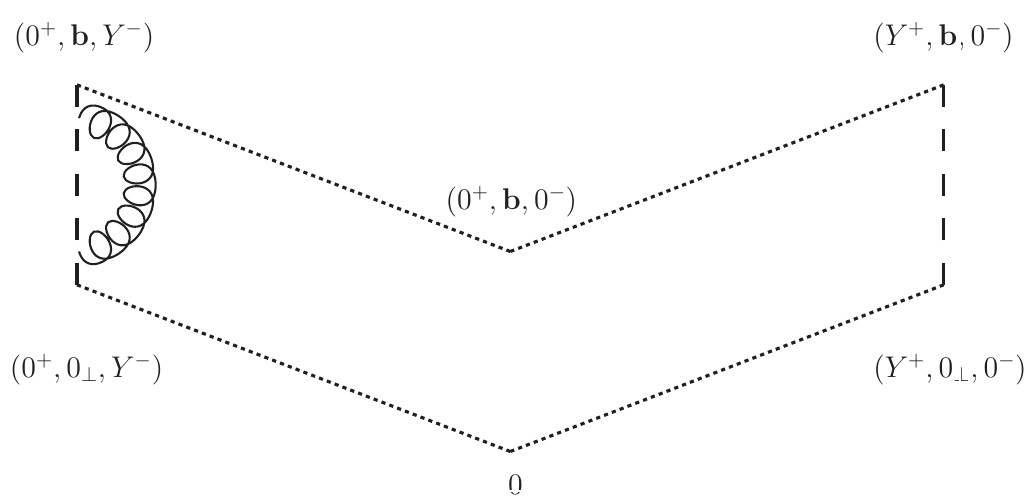}
     }
\caption{ Diagrams contributing to the soft factor.\label{Fig: Soft factor diagrams} }
\end{figure}
%%%%%%%%%%%%%%%%%%%%%%%%%%%%%%%%%%%%%%%%%%%%%%%%%%%%%%%%%%%%%%%%%%%%%%%%%%%%%%%%%%%%%%
%%%%%%%%%%%%%%%%%%%%%%%%%%%%%%%%%%%%%%%%%%%%%%%%%%%%%%%%%%%%%%%%%%%%%%%%%%%%%%%%%%%%%%

As discussed in Sec.~\ref{sec:setup}, the operator definition of the quark TMD receives contributions from Wilson-line self-interactions that do not belong to the physical TMD and must therefore be subtracted. Following the standard treatment~\cite{Collins:2011zzd,Boussarie:2023izj}, this subtraction is implemented through the introduction of a soft factor, whose square root is used to normalize the TMD operator. At one-loop order, the soft factor is given in Eq. \eqref{def:soft_factor}
% %
% %
where the integration is performed along a closed contour $\ell$ in the $(x^+,x^-)$ plane at fixed 
transverse separation $\mathbf{b}$, with light-like segments of the gauge staple extending to $Y^\pm\to+\infty$ and trace is taken over the fundamental indices. The one-loop expansion of the corresponding Wilson loop reads 
\begin{align}
S(\mathbf{b}) &= 1
+ \frac{1}{N_c}\tr_F\bigg[
\Big\langle 0 \Big|\mathcal{T}\,\frac{1}{2}
\left(-ig\mu^\epsilon\oint_\ell dx^\mu\,t^a A^a_\mu(x)\right)^2
\Big|0\Big\rangle\bigg] + \mathcal{O}(g^3)
\nonumber\\
&= 1 - \frac{g^2\mu^{2\epsilon}}{2N_c}\tr_F(t^at^b)
\oint_\ell dx^\mu\oint_\ell dy^\nu
\langle 0|\mathcal{T}\,A^a_\mu(x)A^b_\nu(y)|0\rangle + \mathcal{O}(g^3)
\nonumber\\
&= 1 - \frac{g^2\mu^{2\epsilon}}{2}C_F
\oint_\ell dx^\mu\oint_\ell dy^\nu\,
G_{0,F,\mu\nu}(x,y) + \mathcal{O}(g^3)\,,
\label{eq:soft_factor_expanded}
\end{align}
where in the last step we used $\tr_F(t^at^b)/N_c\to C_F$ in the large $N_c$ limit and as usual the two-point correlator of gluon fields is expressed in terms of free Feynman gluon propagator.  In the projectile light-cone gauge $A^+=0$, the non-vanishing components of the free gluon 
propagator in momentum space are
\begin{align}
\tilde{G}^{--}_{0,F}(k) &= \frac{i}{k^2+i0}\frac{2k^-}{[k^+]}\,,
\label{eq:prop_mm}\\
\tilde{G}^{i-}_{0,F}(k) = \tilde{G}^{-i}_{0,F}(k) &= \frac{i}{k^2+i0}\frac{\k^i}{[k^+]}\,,
\label{eq:prop_im}\\
\tilde{G}^{ij}_{0,F}(k) &= \frac{-ig^{ij}}{k^2+i0}\,,
\label{eq:prop_ij}
\end{align}
where $[k^+]$ singularity is regulated via the ML prescription \eqref{eq:ML_prescription}. We note that the propagators \eqref{eq:prop_mm} and \eqref{eq:prop_im} involve the extra light-cone gauge singularity $1/[k^+]$, whose $k^+$ pole structure is the same as in diagrams~\ref{Fig:diag12}, \ref{Fig:diag13} and \ref{Fig:diag14} analyzed in Sec.~\ref{sec: Diags 12 13 14}. As shown there, the $k^+$ integral with this pole structure vanishes identically for any $k^-$. Therefore, the diagrams in Fig.~\ref{Fig: Soft factor diagrams} that involve the propagators \eqref{eq:prop_mm} or \eqref{eq:prop_im}, namely diagrams~\ref{Fig:Soft factor d}, \ref{Fig:Soft factor e}, \ref{Fig:Soft factor f}, \ref{Fig:Soft factor g} and \ref{Fig:Soft factor h},  give a vanishing contribution to the soft factor by exactly the same argument. The remaining non-trivial contributions come from diagrams~\ref{Fig:Soft factor a}, \ref{Fig:Soft factor b}, \ref{Fig:Soft factor c}, \ref{Fig:Soft factor transverse wilson line Y+} and \ref{Fig:Soft factor transverse wilson line Y-}, which we evaluate explicitly in the rest of this subsection.

%======================================
\subsubsection{Diagram~\ref{Fig:Soft factor a}}

The contribution to the soft factor given by the diagram~\ref{Fig:Soft factor a} can be written as
\begin{align}
S(\mathbf{b})\Big|_{\ref{Fig:Soft factor a}}
= -\frac{1}{2}g^2\mu^{2\epsilon}C_F
\lim_{Y^\pm\to+\infty}
\int\frac{d^{4-2\epsilon}k}{(2\pi)^{4-2\epsilon}}
\frac{i\, e^{-ik^+Y^-}}{k^2+i0}
\frac{(-\k^i)}{[k^+]}
\int_0^{Y^+}dy^+\,e^{ik^-y^+}
\int_0^1 d\tau\,\mathbf{b}^i\,e^{-i\tau\mathbf{k}\cdot\mathbf{b}}\,.
\label{eq:soft_2a_start}
\end{align}
After performing the integrations over $y^+$ and $\tau$, one can reorganize the expression as
\begin{align}
S(\mathbf{b})\Big|_{\ref{Fig:Soft factor a}}
= \frac{1}{2}g^2\mu^{2\epsilon}C_F
\lim_{Y^\pm\to+\infty}
\int\frac{d^{2-2\epsilon}\mathbf{k}}{(2\pi)^{2-2\epsilon}}
i\left(e^{-i\mathbf{k}\cdot\mathbf{b}}-1\right)
\int\frac{dk^-}{2\pi}\frac{-i}{k^-}
\left[1-e^{-ik^-Y^+}\right]
\int\frac{dk^+}{2\pi}
\frac{ie^{ik^+Y^-}}{k^2+i0}\frac{1}{[k^+]}\,.
\label{eq:soft_2a_reorganized}
\end{align}
The contour integration over $k^+$ must be performed first, as required by the ML prescription. Closing the 
contour above the real axis so that the phase factor $e^{ik^+Y^-}$ with $Y^->0$ remains convergent, one finds
\begin{align}
\int\frac{dk^+}{2\pi}
\frac{ie^{ik^+Y^-}}{k^2+i0}
\left(\frac{\theta(k^-)}{k^++i0}+\frac{\theta(-k^-)}{k^+-i0}\right)
= \frac{\theta(-k^-)}{\k^2}
\left(1-e^{i\frac{k^2}{2k^-}Y^-}\right)\,,
\label{eq:soft_2a_kplus}
\end{align}
where contributions from positive $k^-$ vanish since all poles lie below the real axis when the contour is 
closed above. Substituting \eqref{eq:soft_2a_kplus} into \eqref{eq:soft_2a_reorganized} and inserting the 
pure rapidity regulator \eqref{def:pure_rap_reg_OS} to handle the $1/k^-$ singularity, one obtains
\begin{align}
S(\mathbf{b})\Big|_{\ref{Fig:Soft factor a}}
= \frac{1}{2}g^2\mu^{2\epsilon}C_F
\lim_{Y^\pm\to+\infty}
 w^2\left(\frac{2\nu^+}{\nu^-}\right)^{\eta/2}
\int\frac{d^{2-2\epsilon}\mathbf{k}}{(2\pi)^{2-2\epsilon}}
\frac{\big[e^{-i\mathbf{k}\cdot\mathbf{b}}-1\big]}{[\k^2]^{1+\eta/2}}
\int_{-\infty}^0\frac{dk^-}{2\pi}
\frac{|k^-|^\eta}{k^-}
\left[1-e^{-ik^-Y^+}\right]
\left[1-e^{i\frac{\k^2}{2k^-}Y^-}\right]\,.
\label{eq:soft_2a_regulated}
\end{align}
The integrand in \eqref{eq:soft_2a_regulated} contains three types of terms in the $k^-$ integration: a scaleless 
term (proportional to $|k^-|^{\eta-1}$ alone), and terms proportional to the phase factors 
$e^{-ik^-Y^+}$, $e^{i\frac{k^2}{2k^-}Y^-}$, and their product. The scaleless term vanishes in pure rapidity regularization, by analogy to the vanishing of scaleless integrals in dimensional
regularization. The remaining terms, which involve one or both of the phase factors, are shown to vanish in 
the $Y^\pm\to+\infty$ limit in Appendix~\ref{appendix: Vanishing integrals in the soft factor}, where the three contributing 
integrals $I_1$, $I_2$, and $I_3$ are evaluated explicitly. We therefore conclude that
\begin{align}
S(\mathbf{b})\Big|_{\ref{Fig:Soft factor a}} = 0\,.
\label{eq:soft_2a_zero}
\end{align}

%======================================
\subsubsection{Diagram~\ref{Fig:Soft factor b}}
%======================================

The contribution from diagram~\ref{Fig:Soft factor b}  differs from that of diagram~\ref{Fig:Soft factor a} by an overall sign, which arises from the 
reversed direction of integration along the upper longitudinal Wilson line, and by an additional phase factor 
$e^{i\mathbf{k}\cdot\mathbf{b}}$ reflecting the transverse displacement. Specifically,
\begin{align}
S(\mathbf{b})\Big|_{\ref{Fig:Soft factor b}}
= -\frac{1}{2}g^2\mu^{2\epsilon}C_F
\lim_{Y^\pm\to+\infty}
\int\frac{d^{4-2\epsilon}k}{(2\pi)^{4-2\epsilon}}
\frac{ie^{-ik^+Y^-}}{k^2+i0}
\frac{(-\k^i)}{[k^+]}
\int_{Y^+}^0 dy^+\,e^{ik^-y^++i\mathbf{k}\cdot\mathbf{b}}
\int_0^1 d\tau\,\mathbf{b}^i\,e^{-i\tau\mathbf{k}\cdot\mathbf{b}}\,.
\label{eq:soft_2b_start}
\end{align}
The reversal of integration limits over $y^+$ introduces a relative minus sign with respect to diagram~\ref{Fig:Soft factor a}, 
while the extra factor $e^{i\mathbf{k}\cdot\mathbf{b}}$ does not affect the $k^-$ integrations. By the same 
argument used for diagram~\ref{Fig:Soft factor a}, including the contour integration over $k^+$ and insertion of the rapidity 
regulator, this contribution also vanishes in the $Y^\pm\to+\infty$ limit:
\begin{align}
S(\mathbf{b})\Big|_{\ref{Fig:Soft factor b}} = 0\,.
\label{eq:soft_2b_zero}
\end{align}
%

%======================================
\subsubsection{Diagram~\ref{Fig:Soft factor c}}
%======================================

Diagram~\ref{Fig:Soft factor c} involves a gluon connecting the two transverse Wilson lines at $Y^+$ and $Y^-$, located at 
transverse positions $\tau\mathbf{b}$ and $\sigma\mathbf{b}$ respectively. Its contribution reads
\begin{align}
S(\mathbf{b})\Big|_{\ref{Fig:Soft factor c}}
= -\frac{1}{2}g^2\mu^{2\epsilon}C_F
\lim_{Y^\pm\to+\infty}
\int\frac{d^{4-2\epsilon}k}{(2\pi)^{4-2\epsilon}}
\frac{i}{k^2+i0}
\int_0^1 d\sigma\,\mathbf{b}^j\,e^{-ik^-Y^++i\sigma\mathbf{k}\cdot\mathbf{b}}
\int_0^1 d\tau\,\mathbf{b}^i\,e^{ik^+Y^--i\tau\mathbf{k}\cdot\mathbf{b}}\,.
\label{eq:soft_2c_start}
\end{align}
Performing the contour integration over $k^+$ in the same way as for diagram~\ref{Fig:Soft factor a}, and inserting the rapidity 
regulator \eqref{def:pure_rap_reg_OS}, one finds
\begin{align}
S(\mathbf{b})\Big|_{\ref{Fig:Soft factor c}}
& = \frac{1}{4}g^2\mu^{2\epsilon}C_F
\lim_{Y^\pm\to+\infty}
 w^2\left(\frac{2\nu^+}{\nu^-}\right)^{\eta/2}
\int_0^1 d\sigma\,\mathbf{b}^j\,e^{+i\sigma\mathbf{k}\cdot\mathbf{b}}
\int_0^1 d\tau\,\mathbf{b}^i\,e^{-i\tau\mathbf{k}\cdot\mathbf{b}}
\nonumber \\
& 
\times 
\int\frac{d^{2-2\epsilon}\mathbf{k}}{(2\pi)^{2-2\epsilon}}
\frac{1}{[\k^2]^{1+\eta/2}}
\int\frac{dk^-}{2\pi}
|k^-|^\eta \, \frac{\theta(-k^-)}{k^-}
e^{i\frac{\k^2}{2k^-}Y^-}e^{-ik^-Y^+}\,.
\label{eq:soft_2c_regulated}
\end{align}
The $k^-$ integral in \eqref{eq:soft_2c_regulated} is precisely the integral $I_3$ analyzed in 
Appendix~\ref{appendix: Vanishing integrals in the soft factor}, which is shown to vanish in the $Y^\pm\to+\infty$ limit. 
Therefore,
\begin{align}
S(\mathbf{b})\Big|_{\ref{Fig:Soft factor c}} = 0\,.
\label{eq:soft_2c_zero}
\end{align}
% 

%===================================
\subsubsection{Diagrams~\ref{Fig:Soft factor transverse wilson line Y+} and \ref{Fig:Soft factor transverse wilson line Y-}}
%===================================

The remaining non-vanishing contributions to the soft factor come from diagrams~\ref{Fig:Soft factor transverse wilson line Y+} and \ref{Fig:Soft factor transverse wilson line Y-}  which corresponds to the gluon loop on the two space-like Wilson lines at light-cone infinity $Y^+\to+\infty$ and $Y^-\to+\infty$, respectively. These diagrams involve 
only the transverse gluon propagator given in Eq. \eqref{eq:prop_ij}, and does not carry an extra 
light-cone gauge singularity. The calculation is therefore identical in structure to the Wilson line 
self-energy at infinity computed in the target light-cone gauge of Ref.~\cite{Altinoluk:2025ewj}. Adding the 
contributions from diagrams~\ref{Fig:Soft factor transverse wilson line Y+} and \ref{Fig:Soft factor transverse wilson line Y-} together gives
\begin{align}
S(\mathbf{b})\Big|_{\ref{Fig:Soft factor transverse wilson line Y+} + \ref{Fig:Soft factor transverse wilson line Y-}}
= -2g^2\mu^{2\epsilon}C_F\,\mathbf{b}^2
\int\frac{d^{4-2\epsilon}k}{(2\pi)^{4-2\epsilon}}
\frac{i}{k^2+i0}
\int_0^1 d\sigma
\int_0^1 d\tau\,
e^{i(\tau-\sigma)\mathbf{k}\cdot\mathbf{b}}\,.
\label{eq:soft_2ij_start}
\end{align}
This integral is evaluated in the same way as in the target light-cone gauge computation (see  Appendix B of Ref.~\cite{Altinoluk:2025ewj} for a detailed calculation) using Schwinger parametrization and dimensional regularization, yielding
\begin{align}
S(\mathbf{b})\Big|_{\ref{Fig:Soft factor transverse wilson line Y+} + \ref{Fig:Soft factor transverse wilson line Y-}}
= \frac{\alpha_s C_F}{\pi}(\pi\mu^2 \bbperp^2)^\epsilon
\frac{\Gamma(1-\epsilon)}{\epsilon(1-2\epsilon)}
+ \mathcal{O}(\alpha_s^2)\,.
\label{eq:soft_2ij_result}
\end{align}

%=============================
\subsubsection{Total soft factor}
%=============================

Combining all contributions, the one-loop soft factor in the projectile light-cone gauge in pure rapidity regularization is
\begin{align}
S(\mathbf{b})
= 1 + \frac{\alpha_s C_F}{\pi}(\pi\mu^2 \bbperp^2)^\epsilon
\frac{\Gamma(1-\epsilon)}{\epsilon(1-2\epsilon)}
+ \mathcal{O}(\alpha_s^2)\,.
\label{eq:soft_factor_total}
\end{align}
This result is identical to the soft factor used in the target light-cone gauge calculation of Ref.~\cite{Altinoluk:2025ewj}. Due to the symmetry between the $+$ and $-$ directions in the definition of the contour for the soft factor, not only the result but also the whole calculation of the soft factor is identical between the target light-cone gauge and the projectile light-cone gauge, up to the interchange of the role of the $+$ and $-$ directions. The soft factor was not explictly calculated in Ref.~\cite{Altinoluk:2025ewj}, but our calculation in this section provides the justification for its expression \eqref{eq:soft_factor_total} used there.
The soft factor \eqref{eq:soft_factor_total} will be used in the next subsection to assemble the total NLO correction to the quark TMD.

%============================================================
\subsection{Total NLO corrections to the quark TMD from the operator definition}
\label{subsec:total_NLO}
%============================================================

Having computed all the individual diagram contributions in the preceding subsections, we are now ready to assemble the total one-loop correction to the quark TMD. Let us first summarize the outcome of each contribution:

\begin{itemize}
\item Diagrams~\ref{Fig:diag6} and \ref{Fig:diag4} (Sec.~\ref{sec: diag 6 and 4}): These produce the dominant rapidity-divergent  contribution, given in Eq.~\eqref{eq:doube pole from diags 6 and 4}, together with a UV-divergent but  rapidity-finite contribution from diagram~\ref{Fig:diag4}, given in Eq.~\eqref{eq: UV divergent part of diag 4}. The rapidity-finite  part of diagram~\ref{Fig:diag6} is also UV-finite and contributes only to the finite NLO remainder.

\item Diagram~\ref{Fig:diag5} (Sec.~\ref{sec: Diagram 5}): Vanishes in the $Y^+\to+\infty$ limit after insertion of  the pure rapidity regulator, as shown in Eq.~\eqref{diag_5_final_zero}.

\item Ladder diagram \ref{Fig:diag1} (Sec.~\ref{sec: ladder diagram}): Completely finite at non-zero $\mathbf{b}$, with no  rapidity or UV divergences. It contributes only to the finite NLO remainder.

\item Diagram~\ref{Fig:diag11} (Wilson line self-energy at infinity): This diagram  produces a UV-divergent contribution~\eqref{SE_WL_at_inf} that is canceled by the soft factor, as discussed below.

\item Diagrams~\ref{Fig:diag12}, \ref{Fig:diag13} and \ref{Fig:diag14} (Sec.~\ref{sec: Diags 12 13 14}): All vanish identically due to the $k^+$ pole  structure of the light-cone gauge propagator combined with the ML prescription, as shown in  Eqs.~\eqref{eq:diag1f_zero}, \eqref{eq:diag1g_zero}, and \eqref{eq:diag1h_zero}.

\item Symmetric counterparts (Appendix~\ref{appendix_symmetric_diagrams}): The diagrams listed in  Appendix~\ref{appendix_symmetric_diagrams} are the symmetric counterparts of diagrams~\ref{Fig:diag6}, \ref{Fig:diag4},  \ref{Fig:diag5}, \ref{Fig:diag12} and \ref{Fig:diag13}. By the symmetry of the TMD operator under exchange of the quark and antiquark fields, each symmetric diagram gives exactly the same contribution as its counterpart. The symmetric counterparts of diagrams~ \ref{Fig:diag5}, \ref{Fig:diag12} and \ref{Fig:diag13} therefore vanish by the same arguments as the originals, while those of diagrams~\ref{Fig:diag6} and \ref{Fig:diag4} double the contributions already computed.
\end{itemize}

The Wilson line self-energy at infinity, diagram~\ref{Fig:diag11}, requires special treatment. As is standard in the definition of TMDs, the self-energy contributions to the Wilson lines are removed by dividing the TMD operator expectation value by the square root of the soft factor $S(\mathbf{b})$ defined in Sec.~\ref{subsec:Soft_Factor}. At one loop, the soft factor computed in Eq.~\eqref{eq:soft_factor_total} reads
\begin{align}
\sqrt{S(\mathbf{b})} = 1 + \frac{\alpha_s C_F}{2\pi}(\pi\mu^2 \bbperp^2)^\epsilon
\frac{\Gamma(1-\epsilon)}{\epsilon(1-2\epsilon)} + \mathcal{O}(\alpha_s^2)\,,
\label{eq:sqrt_soft}
\end{align}
and its contribution precisely cancels the UV-divergent contribution of diagram~\ref{Fig:diag11}, as expected. In practice, therefore we simply discard the contribution from diagram~\ref{Fig:diag11} in what follows below, following the same procedure adopted in Ref.~\cite{Altinoluk:2025ewj}.

Collecting the non-vanishing contributions and accounting for the factor of two from the symmetric  diagrams, the total unsubtracted quark TMD normalized by the square root of the soft factor at one loop reads
\begin{align}
\frac{1}{\sqrt{S(\mathbf{b})}}\,
q^{\textrm{n.r.}}_{\textrm{unsub.}}(x, \mathbf{b}; \mu^2, \zeta)
=&\left\{
1
+ \frac{\alpha_s C_F}{2\pi}\Gamma(-\epsilon)(\pi\mu^2 \bbperp^2)^\epsilon
\left(\frac{2 w^2}{\eta}
+ \log\frac{\zeta \bbperp^2}{c_0^2}
- \Psi(-\epsilon) + \Psi(1)\right)
\right.
\nonumber\\
&\left.
\quad
+\frac{\alpha_s C_F}{\pi}(4\pi\mu^2)^\epsilon
\frac{\Gamma(2-\epsilon)}{\Gamma(1-\epsilon)}\Gamma(\epsilon)
\right\}
q_{\textrm{Bckgd}}(x,\mathbf{b};\mu^2)
+ \text{finite NLO} + \mathcal{O}(\alpha_s^2)\,,
\label{eq:total_NLO}
\end{align}
where $c_0 \equiv 2e^{-\gamma_E}$ and the finite NLO terms are independent of $\zeta$ and therefore do not contribute to CSS evolution. The first line of \eqref{eq:total_NLO} originates from the combined rapidity-divergent contribution of diagrams~\ref{Fig:diag6} and \ref{Fig:diag4} given in Eq.~\eqref{eq:doube pole from diags 6 and 4} and their symmetric counterparts. The second line originates from the UV-divergent but rapidity-finite 
contribution of diagram~\ref{Fig:diag4} given in Eq.~\eqref{eq: UV divergent part of diag 4} and its symmetric counterpart. 

It is instructive to compare the structure of \eqref{eq:total_NLO} with the analogous result obtained in the target light-cone gauge in Ref.~\cite{Altinoluk:2025ewj}.
In both cases the total NLO correction has the same coefficient of the rapidity pole $1/\eta$ and the same logarithm $\log(\zeta \b^2/c_0^2)$. 
The diagrammatic origin of these divergent contributions differs significantly between the two gauges. In the target light-cone  gauge, the rapidity-divergent term arises from a single diagram (gluon emission from the quark to the transverse Wilson line at infinity and its antiquark counterpart), driven by the ghost-like zero-mode of the ML prescription. In the projectile light-cone  gauge, that diagram vanishes (diagram~\ref{Fig:diag5}), and the rapidity-divergent contribution is instead generated by two diagrams (\ref{Fig:diag6} and \ref{Fig:diag4}) describing gluon emission to the longitudinal Wilson lines, which are non-trivial only in the projectile gauge. 
In contrast, the UV-divergent but rapidity-finite contribution proportional to $\Gamma(\epsilon)$ in Eq.~\eqref{eq:total_NLO} is absent from the total NLO correction in the target light-cone gauge, whereas this contribution arises from diagram~\ref{Fig:diag4} in the projectile light-cone gauge. 
Despite the differences in the divergent contributions in these two gauges, the gauge invariance in the CSS evolution is preserved, as we will show in the next Section. 
We will return to a more detailed comparison in Sec.~\ref{sec:comparison}.

Finally, we note that the result \eqref{eq:total_NLO} is expressed entirely in terms of the background TMD $q_{\textrm{Bckgd}}(x,\mathbf{b};\mu^2)$ defined in Eq.~\eqref{def:q_Bckgd}, up to finite NLO corrections. This is a direct consequence of working in the dilute target limit and restricting to the leading-order background field insertions. The renormalization of this result, leading to the CSS evolution equations, is carried out in the next section.

%=============================================================================================
\section{Renormalization of the fields and extraction of the CSS evolution}
\label{sec:CSS}
%=============================================================================================
In this section we carry out the UV and rapidity renormalization of the quark TMD and extract the CSS evolution equations. The starting point is the total one-loop result assembled in Sec.~\ref{subsec:total_NLO} given in Eq.~\eqref{eq:total_NLO}. We proceed in two steps; first we subtract the rapidity divergences, then we remove the remaining UV divergences. 

%==========================================
\subsection{Rapidity Subtraction}
\label{subsec:rap_sub}
%===========================================
The rapidity divergences in \eqref{eq:total_NLO} appear as  $1/\eta$ poles. They are subtracted by introducing a rapidity renormalization factor $Z_{\textrm{rap.}}$, defined 
in the minimal subtraction scheme as
\begin{align}
Z_{\textrm{rap.}}
=&\, 
1- \frac{\alpha_s C_F}{\pi}\,  
\Gamma(-\epsilon)
\big[\pi \mu^2 \mathbf{b}^2\big]^{\epsilon}
\frac{w^2}{\eta}
+O(\alpha_s^2)
\label{Eq:Z_rap_one_loop_ren}
\, ,
\end{align}
at finite $\epsilon$. The rapidity subtracted quark TMD (still not UV-renormalized) is then defined as  
\begin{align}
\label{def:sub_unsub_rel_nr}
q^{\textrm{n.r.}}_{\textrm{sub.}}(\tx, \mathbf{b};\mu^2,\zeta)
= &\, \lim_{\eta\rightarrow 0}\,
Z_{\textrm{rap.}}\: 
\frac{1}{\sqrt{{\cal{S}}(\mathbf{b})}}\:
q^{\textrm{n.r.}}_{\textrm{unsub.}}(\tx, \mathbf{b};\mu^2)
\, .
\end{align}
It is useful to express $Z_{\rm rap.}$ in terms of the bare coupling $\alpha_s^{(0)}$, using the relation
\begin{align}
\alpha_s^{(0)}=&\,
\mu^{2\epsilon}\,  \alpha_s(\mu^2)\,
{Z_3}^{-1}\,
=\mu^{2\epsilon}\,  \alpha_s(\mu^2)\, \Big[1+O(\alpha_s)\Big]
\label{Eq:bare_renorm_transfo_alpha_s}
\, ,
\end{align}
which gives
\begin{align}
Z_{\textrm{rap.}}
=&\, 
1- \frac{\alpha_s^{(0)} C_F}{\pi}\,  
\Gamma(-\epsilon)
\big[\pi \mathbf{b}^2\big]^{\epsilon}
\frac{w^2}{\eta}
+O(\alpha_s^2)
\label{Eq:Z_rap_one_loop_ren_Bare_Coupling}
\, .
\end{align}
Written in terms of the bare coupling in this way, $Z_{\rm rap.}$ is independent of the renormalization scale $\mu$. 

After performing the rapidity subtraction \eqref{def:sub_unsub_rel_nr} and taking $\eta \to 0$ limit at finite $\epsilon$, the $1/\eta$ poles cancel and one obtains 
\begin{align}
q^{\textrm{n.r.}}_{\textrm{sub.}}(x, \mathbf{b}; \mu^2, \zeta)
=&\left\{
1
+ \frac{\alpha_s C_F}{2\pi}\Gamma(-\epsilon)(\pi\mu^2 \b^2)^\epsilon
\left[
\log\left(\frac{\zeta \b^2}{c_0^2}\right)
- \Psi(-\epsilon) + \Psi(1)\right]
\right.
\nonumber\\
&\left.
\quad
+\frac{\alpha_s C_F}{\pi}(4\pi\mu^2)^\epsilon
\frac{\Gamma(2-\epsilon)}{\Gamma(1-\epsilon)}\Gamma(\epsilon)
\right\}
q_{\textrm{Bckgd}}(x,\mathbf{b};\mu^2)
+ \text{finite NLO} + \mathcal{O}(\alpha_s^2)\,,
\label{eq:total_NLO_rap_sub}
\end{align}
where the finite NLO terms are independent of $\zeta$ and do not contribute to to the CSS evolution. Expanding Eq. \eqref{eq:total_NLO_rap_sub} around $\epsilon=0$, one finds that the UV divergences remain as poles in $\epsilon$. This can be realized by noting that 
\begin{align}
\Gamma(-\epsilon)(\pi\mu^2 \b^2)^\epsilon
\left[\log\left(\frac{\zeta \b^2}{c_0^2}\right) - \Psi(-\epsilon) + \Psi(1)\right]
= \frac{S_\epsilon}{\epsilon^2} + \frac{S_\epsilon}{\epsilon}\log\frac{\mu^2}{\zeta}
+ \mathcal{O}(\epsilon^0)\,,
\label{eq:UV_poles}
\end{align}
and
\begin{align}
(4\pi\mu^2)^\epsilon\frac{\Gamma(2-\epsilon)}{\Gamma(1-\epsilon)}\Gamma(\epsilon)
= \frac{S_\epsilon}{\epsilon} + \mathcal{O}(\epsilon^0)\,,
\label{eq:UV_poles_2}
\end{align}
where $S_\epsilon \equiv (4\pi e^{-\gamma_E})^\epsilon$ absorbs the 
universal constants of the $\overline{\textrm{MS}}$ scheme. All in all, after this expansion the rapidity subtracted but still not UV-renormalized quark TMD reads 
\begin{align}
    q_{\rm sub.}^{\rm n.r}(x,\bbperp;\mu^2,\zeta)=  q_{\rm Bckgd}(x,\bbperp;\mu^2)  \left\{1+
\frac{\alpha_s  C_F}{2\pi}\, \left[ \frac{S_\epsilon}{\epsilon^2} +\frac{S_\epsilon}{\epsilon}\log \left(\frac{\mu^2}{\zeta} \right)+2\frac{S_\epsilon}{\epsilon} \right] \right\}+\text{finite NLO}+\mathcal{O}(\alpha_s^2)\,.
\label{eq:after_rap_sub}
\end{align}
%

%================================
\subsection{UV renormalization and renormalization constants}
%================================

The UV divergences found in Eq.~\eqref{eq:after_rap_sub} are the extra divergences appearing in the quark TMD as defined from a composite operator, beyond the standard UV divergences associated with the renormalization of the fields and coupling. However, the $\mu$ dependence of the TMDs is determined by all UV divergences, including the ones associated with fields and coupling renormalization. The effect of running coupling will arise only at order $\alpha_s^2$ in the evolution equation for the TMDs, so that we only need to discuss the UV renormalization effects for the quark field.    

 As discussed in detail in Ref.~\cite{Altinoluk:2025ewj}, following the renormalization procedure established in Refs.~\cite{Bassetto:1985dr,Bassetto:1987sw}, the renormalization of the quark and gluon fields in the light-cone gauge with the ML prescription does not proceed by simple multiplicative constants due to the breaking of Lorentz invariance by the gauge condition. Instead, the good and bad components of the quark spinor are renormalized by different factors. In the projectile light-cone gauge $A^+=0$, with $n^\mu = g^{\mu+}$ and $\bar{n}^\mu = g^{\mu-}$, the good components of the quark field (projected by $\bar{n}\!\!\!/\,n\!\!\!/ / (2\bar{n}\cdot n)$) are renormalized by $\sqrt{Z_2\tilde{Z}_2}$, while the bad components (projected by $n\!\!\!/\,\bar{n}\!\!\!/ / (2\bar{n}\cdot n)$) are renormalized by $\sqrt{Z_2/\tilde{Z}_2}$. The quark TMD operator \eqref{def:q_op_def_T_ord} involves the matrix $\gamma^-$, which in the projectile gauge projects onto the good components of the quark field. The relation between the bare quark TMD and the one defined in terms of renormalized fields and coupling (but not further renormalized)  is therefore
\begin{align}
q^{(0)}_{\textrm{sub.}}(x, \mathbf{b}; \zeta)
= Z_2\, \tilde{Z}_2 \, 
q^{\textrm{n.r.}}_{\textrm{sub.}}(x, \mathbf{b}; \mu^2, \zeta)\,,
\label{eq:bare_TMD}
\end{align}
at the level of rapidity-subtracted distributions.
Note that the bare but rapidity-subtracted quark TMD $q^{(0)}_{\textrm{sub.}}$ is 
independent of $\mu$ by construction, since it is defined as a series in bare perturbation theory. The one-loop renormalization constants for the quark field in the 
$\overline{\textrm{MS}}$ scheme are~\cite{Bassetto:1985dr}
\begin{align}
Z_2 &= 1 + \frac{\alpha_s C_F}{4\pi}\frac{S_\epsilon}{\epsilon} + \mathcal{O}(\alpha_s^2),
\label{eq:Z2}
\\
\tilde{Z}_2 &= 1 - \frac{\alpha_s C_F}{2\pi}\frac{S_\epsilon}{\epsilon} + \mathcal{O}(\alpha_s^2)\,,
\label{eq:Z2tilde}
\end{align}
from which one finds
\begin{align}
\frac{Z_2}{\tilde{Z}_2} &= 1 + \frac{3\alpha_s C_F}{4\pi}\frac{S_\epsilon}{\epsilon}
+ \mathcal{O}(\alpha_s^2)\,,
\label{eq:Z2_ratio}
\\
Z_2\tilde{Z}_2 &= 1 - \frac{\alpha_s C_F}{4\pi}\frac{S_\epsilon}{\epsilon}
+ \mathcal{O}(\alpha_s^2)\,.
\label{eq:Z2_product}
\end{align}
Substituting Eq.\eqref{eq:after_rap_sub} and \eqref{eq:Z2_product} into Eq. \eqref{eq:bare_TMD}, one obtains the bare (but rapidity subtracted) quark TMD as
\begin{align}
   q^{(0)}_{\textrm{sub.}}(x, \mathbf{b}; \zeta)=q_{\rm Bckgd}(x,\bbperp;\mu^2)  \left\{1+
\frac{\alpha_s  C_F}{2\pi}\, \left[ \frac{S_\epsilon}{\epsilon^2} +\left(\log \left(\frac{\mu^2}{\zeta} \right)+\frac{3}{2}\right)\frac{S_\epsilon}{\epsilon} \right]\right\}+\text{finite NLO}+\mathcal{O}(\alpha_s^2)\,.
\label{eq:bare_TMD_explicit}
\end{align}
The fully renormalized quark TMD is then defined as 
\begin{align}
q(x, \mathbf{b}; \mu^2, \zeta) = Z_{\textrm{UV}}\,q^{(0)}_{\textrm{sub.}}(x, \mathbf{b}; \zeta)
= Z_{\textrm{UV}}\, Z_2\, \tilde{Z}_2\, 
q^{\textrm{n.r.}}_{\textrm{sub.}}(x, \mathbf{b}; \mu^2, \zeta)\,,
\label{eq:full_ren_TMD}
\end{align}
where $Z_{\textrm{UV}}$ is the UV renormalization factor that removes all of the UV divergences, in the 
$\overline{\textrm{MS}}$ scheme. 
Substituting Eqs. \eqref{eq:bare_TMD_explicit} into Eq. \eqref{eq:full_ren_TMD}, and requiring that $q(x,\mathbf{b};\mu^2,\zeta)$ 
be finite at $\epsilon=0$, one determines $Z_{\textrm{UV}}$ to be
\begin{align}
    Z_{\rm UV}=1-
\frac{\alpha_s  C_F}{2\pi}\, \left[ \frac{S_\epsilon}{\epsilon^2} +\left(\log \left(\frac{\mu^2}{\zeta} \right)+\frac{3}{2}\right)\frac{S_\epsilon}{\epsilon} \right]+\mathcal{O}(\alpha_s^2)\, . 
\label{eq:ZUV}
\end{align}
%
%================================
\subsection{Extraction of the CSS evolution equations}
%================================
Since the bare, but rapidity subtracted quark TMD $q_{\rm sub.}^{(0)}$ is by construction independent of the renormalization scale $\mu$, it follows straightforwardly from Eq. \eqref{eq:full_ren_TMD} that the fully renormalized quark TMD satisfies
\begin{align}
\label{eq:dmudlogZUV}
\mu^2\frac{d}{d\mu^2}\log \Big(
q(\tx, \mathbf{b};\mu^2,\zeta) \Big)
=&\, 
\mu^2\frac{d}{d\mu^2} 
\log Z_{UV} \, . 
\end{align}
The renormalization group equation for the coupling at finite $\epsilon$ reads 
\begin{align}
\mu^2\frac{d}{d\mu^2}\alpha_s(\mu^2) = -\epsilon\,\alpha_s(\mu^2) + \mathcal{O}(\alpha_s^2)\,,
\label{eq:RGE_coupling}
\end{align}
which follows from the $\mu$-independence of the bare coupling as expressed in Eq.~\eqref{Eq:bare_renorm_transfo_alpha_s}. Upon substituting Eqs.~\eqref{eq:RGE_coupling} and \eqref{eq:ZUV} into Eq.~\eqref{eq:dmudlogZUV}, we find
\begin{align}
\mu^2\frac{d}{d\mu^2}\log \Big(
q(\tx, \mathbf{b};\mu^2,\zeta) \Big)
=& 
- \frac{C_F}{2\pi}\,
\left[\frac{S_{\epsilon}}{\epsilon^2}
+\left(
\log\left(\frac{\mu^2}{\zeta}\right) 
+\frac{3}{2}\right)
\frac{S_{\epsilon}}{\epsilon}
\right]\,
\mu^2\frac{d}{d\mu^2}\alpha_s(\mu^2)
- \frac{\alpha_s(\mu^2) C_F}{2\pi}\,
\frac{S_{\epsilon}}{\epsilon}
+O(\alpha_s^2)
\nonumber\\
=&\,
\frac{\alpha_s C_F}{2\pi}\,
\left[
\log\left(\frac{\mu^2}{\zeta}\right) 
+\frac{3}{2}
\right]S_{\epsilon}
+O(\alpha_s^2)\,.
\label{Eq:CSS_RG_eps}
\end{align}
Taking the $\epsilon \to 0$ we get $S_\epsilon\to 1$, and one then recovers the first CSS equation: 
\begin{align}
\mu^2\frac{d}{d\mu^2}
q(\tx, \mathbf{b};\mu^2,\zeta) 
=&\, 
\left\{
\frac{\alpha_s C_F}{2\pi}\,
\left[
\log\left(\frac{\mu^2}{\zeta}\right) 
+\frac{3}{2}
\right]
+O(\alpha_s^2)\right\} q(\tx, \mathbf{b};\mu^2,\zeta) \, .
\label{Eq:CSS_RG}
\end{align}
To extract the second CSS equation governing the $\zeta$ dependence, we differentiate the fully renormalized TMD given in Eq. \eqref{eq:full_ren_TMD} with respect to $\zeta$. Since both $Z_2\, \tilde{Z}_2$ and the finite NLO terms in Eq. \eqref{eq:after_rap_sub} are independent of $\zeta$, one obtains 
\begin{align}
\zeta\frac{d}{d\zeta}\log q(x, \mathbf{b}; \mu^2, \zeta)
&= \zeta\frac{d}{d\zeta}\log Z_{\textrm{UV}}
+ \zeta\frac{d}{d\zeta}\log q^{\textrm{n.r.}}_{\textrm{sub.}}(x, \mathbf{b}; \mu^2, \zeta)
\nonumber\\
&= \frac{\alpha_s C_F}{2\pi}\frac{S_\epsilon}{\epsilon}
+ \frac{\alpha_s C_F}{2\pi}\Gamma(-\epsilon)(\pi\mu^2 \bbperp^2)^\epsilon
+ \mathcal{O}(\alpha_s^2)
\nonumber\\
&= \frac{\alpha_s C_F}{2\pi}\frac{S_\epsilon}{\epsilon}
\left[1 - \Gamma(1-\epsilon)e^{\epsilon\Psi(1)}
\left(\frac{\mu^2 \bbperp^2}{c_0^2}\right)^\epsilon\right]
+ \mathcal{O}(\alpha_s^2)\,,
\label{eq:zeta_deriv_log}
\end{align}
at finite $\epsilon$, where in the last step we have combined the two terms using the identity $\Gamma(-\epsilon) = -\Gamma(1-\epsilon)/\epsilon$. Taking the $\epsilon\to 0$ limit, one recovers the second CSS equation
\begin{align}
\zeta\frac{d}{d\zeta}
q(\tx, \mathbf{b};\mu^2,\zeta) 
=&\, 
\left\{
-\frac{\alpha_s C_F}{2\pi}\,
\log\left(\frac{\mu^2 \mathbf{b}^2}{c_0^2}\right)
+O(\alpha_s^2)\right\} 
q(\tx, \mathbf{b};\mu^2,\zeta) 
\label{Eq:CSS_RAD}
\, .
\end{align}

Equations \eqref{Eq:CSS_RG} and \eqref{Eq:CSS_RAD} are the one-loop CSS evolution equations for the quark TMD, obtained here in the projectile light-cone gauge $A^+=0$ using the background field formalism with the ML prescription and the pure rapidity regulator. These are identical in form to the CSS equations obtained in the target light-cone gauge $A^-=0$ in Ref.~\cite{Altinoluk:2025ewj}, as well as to the standard results in the literature~\cite{Collins:1981uk,Collins:1984kg,Collins:2011zzd}. This gauge-invariance of the CSS equations is therefore a non-trivial consistency check of the our approach. Despite the fact that the individual diagrams contributing to the quark TMD differ substantially between the two gauges, with the rapidity-divergent contribution arising from radiation to the longitudinal Wilson lines in the projectile light-cone gauge rather than from radiation to the transverse Wilson line at infinity in the target light-cone gauge, the resulting evolution equations are identical. This confirms that the CSS evolution is a genuine gauge-invariant feature  of the quark TMD, and that the background field formalism with the ML prescription provides a consistent framework for deriving it in either gauge. A detailed comparison of the two computations is presented in Sec.~\ref{sec:comparison}.

%=============================================================================================
\section{Comparison with the target light-cone gauge}
\label{sec:comparison}
%=============================================================================================
The present paper and our earlier work Ref.~\cite{Altinoluk:2025ewj} together constitute a systematic derivation of the CSS evolution equations for the quark TMD within the background field formalism in both light-cone gauges. While the final results, namely the CSS evolution equations \eqref{Eq:CSS_RG} and \eqref{Eq:CSS_RAD}, are identical in both gauges as required by gauge invariance, the two calculations differ substantially in their diagrammatic structure, in the dynamics generating the divergences, and in the role played by the various components of the gauge link and the gluon propagator. In this section we provide a detailed side-by-side comparison of the two computations, and argue that the projectile gauge calculation is not just a consistency check but a necessary and qualitatively distinct step toward establishing a unified framework connecting the CGC effective theory and TMD factorization.

%=======================
\subsection{Diagrammatic structure and origin of divergences}
\label{subsec:comp_tech}
%=======================

The most immediate structural difference between the two computations concerns the gauge link in the TMD operator definition \eqref{def:q_op_def_T_ord}. In the target light-cone gauge $A^-=0$, the gauge link vector $n^\mu = g^{\mu-}$ is aligned with the dominant momentum component of the target, and the two longitudinal Wilson lines in the staple reduce to identity color matrices. The entire non-trivial gauge link structure is concentrated in the single transverse Wilson line at light-cone infinity $Y^+$, given in Eq.~\eqref{def:Wilson_para}. As a consequence, at one loop only three types of diagrams may contribute non-trivially to the quark TMD in the target light-cone gauge: gluon emission from the quark (and antiquark) to the transverse Wilson line at infinity, the quark-to-quark ladder diagram, and the Wilson line self-energy at infinity. In the projectile light-cone gauge $A^+=0$, by contrast, the longitudinal Wilson lines along $x^-$ are non-trivial and generate a richer set of one loop diagrams. In addition to the counterparts of the target light-cone gauge diagrams, the projectile light-cone gauge calculation involves diagrams describing gluon emission to the lower longitudinal Wilson line (diagram~\ref{Fig:diag6}), gluon emission to the upper longitudinal Wilson line (diagram~\ref {Fig:diag4}), Wilson line self-energies on the longitudinal parts of the staple (diagram~\ref{Fig:diag12}), gluon emission from the longitudinal Wilson line to the transverse Wilson line at infinity (diagram~\ref{Fig:diag13}), and gluon exchange between the two longitudinal Wilson lines (diagram~\ref{Fig:diag14}). This richer diagrammatic structure is a direct reflection of the more complex gauge link geometry in the projectile light-cone gauge and has no analog in the target light-cone gauge computation.

 Both calculations employ the ML prescription to regulate the extra light-cone gauge singularity $1/[n\cdot k]$ in the gluon propagator. However, the direction of the light-cone vector $n^\mu$ is opposite in the two cases, and this has important consequences for the order of integration in the loop integrals and for the mechanism generating the rapidity divergences.

In the target light-cone gauge, $n^\mu = g^{\mu-}$ and the ML prescription regulates the singularity at $k^-=0$, requiring the $k^-$ pole integration to be performed first. Rapidity divergences subsequently arise in the $k^+\to+\infty$ regime, and are regulated by the pure rapidity regulator factor \eqref{def:pure_rap_reg}. In the projectile light-cone gauge, $n^\mu = g^{\mu+}$ and the ML prescription regulates the singularity at $k^+=0$, requiring the $k^+$ pole integration to be performed first. Rapidity divergences then arise in the $k^-\to 0$ regime, and are regulated by the on-shell analog \eqref{def:pure_rap_reg_OS} of the pure rapidity regulator. This interchange of the roles of $k^+$ and $k^-$ propagates through the entire calculation and changes the analytic structure of virtually every loop integral.

A particularly remarkable consequence of this interchange concerns the ghost-like zero-mode of the ML prescription. In the target light-cone gauge, this zero-mode, corresponding to the pole at $k^-=0$ in the decomposition \eqref{eq:Feyn-prop_2},  was identified in Ref.~\cite{Altinoluk:2025ewj} as the sole source of the double-logarithmic contribution to the CSS kernel. It appeared in the diagrams describing gluon emission from the background quark to the transverse Wilson line at infinity, and its contribution survived the $Y^+\to+\infty$ limit precisely because because it was not sensitive to the phase suppression responsible for eliminating all other contributions. In the projectile light-cone gauge, the analogous diagram (diagram~\ref{Fig:diag5}) involves a phase factor $e^{-ik^-Y^+}$ that suppresses both the regular and the zero-mode contributions in the $Y^+\to+\infty$ limit. The zero-mode of the projectile light-cone gauge ML prescription (now at $k^+=0$) instead plays a different role. It causes the $k^+$ integrals in diagrams~\ref{Fig:diag12}, \ref{Fig:diag13} and \ref{Fig:diag14} to vanish identically, by placing all poles on the same side of the real axis. The double-logarithmic contribution to the CSS kernel is therefore generated in the projectile light-cone gauge not by the zero-mode, but by the diagrams involving the longitudinal Wilson lines (\ref{Fig:diag6} and\ref{Fig:diag4}), which are entirely absent in the target gauge. This represents a genuine and non-trivial redistribution of the dynamics between diagrams, driven by the gauge choice.

One of the more subtle differences between the two calculations concerns the UV-divergent but rapidity-finite contribution (written explicitly in Eq. \eqref{eq:UV_poles_2}) in the NLO result given in Eq. \eqref{eq:total_NLO}. In the target light-cone gauge computation of Ref.~\cite{Altinoluk:2025ewj}, this contribution does not exist explicitly. In the projectile light-cone gauge, it comes from the diagram~\ref{Fig:diag4} which corresponds to gluon emission from the quark to the upper longitudinal Wilson line and contributes to UV renormalization factor $Z_{UV}$. We emphasize that this diagram has no counter part in the target light-cone gauge. 

A concise diagram-by-diagram comparison is provided in Table~\ref{tab:comparison}, using the notation ``\textrm{Rap. div.}'' for a contribution containing a $1/\eta$ rapidity pole (and accompanying $1/\epsilon$ UV poles via the $\Gamma(-\epsilon)$ prefactor); ``\textrm{UV div.}'' for a rapidity-finite contribution with a $1/\epsilon$ UV pole; and ``\textrm{Finite}'' for a contribution that is both rapidity-finite and UV-finite at~$\mathbf{b}\neq 0$.
\begin{table}%[b]
\centering
\renewcommand{\arraystretch}{1.2}
\begin{tabular}{lll}
\hline\hline
\textbf{Contribution} & \textbf{Target gauge ($A^-=0$)} & \textbf{Projectile gauge ($A^+=0$)} \\
\hline
Gluon emission to trans. & Rapidity divergent   & Vanishes
 \\ 
Wilson line at $\infty$ (diagram~\ref{Fig:diag5}) 
& (derives CSS) 
& ($Y^+\to\infty$ phase suppression) \\
\hline
Gluon emission to lower & Absent & Rapidity divergent \\
 long. Wilson line (diagram\ref{Fig:diag6}) 
& (long. Wilson line $= \mathbf{1}$) 
& (contributes to CSS) \\
\hline
Gluon emission to upper &  Absent & Rap. div. (contributes to CSS)\\
 long. Wilson line (diagram~\ref{Fig:diag4})
& (long. Wilson line $= \mathbf{1}$) 
& Rap. fin. but UV div. part (contributes to $Z_{UV}$) \\
\hline
Quark-to-quark ladder& Finite  & Finite \\
(diagram~\ref{Fig:diag1})& (contributes to finite NLO) &  (contributes to finite NLO) \\
\hline
Long. Wilson line & Absent & Vanishes \\
self-energy (diagram~\ref{Fig:diag12})&  (long. Wilson line $= \mathbf{1}$) 
& (due to ML prescription pole structure) \\
\hline
Emission from long. Wilson& Absent & Vanishes  \\
line to $\infty$ (diagram~\ref{Fig:diag13}) &  (long. Wilson line $= \mathbf{1}$) & (due to ML prescription pole structure) \\
\hline
Gluon exchange between long.  & Absent & Vanishes \\
Wilson lines (diagram~\ref{Fig:diag14}) &  (long. Wilson line $= \mathbf{1}$) & (due to ML prescription pole structure) \\
\hline
Trans. Wilson line self-energy & UV div. & UV div. \\ 
at infinity (diagram~\ref{Fig:diag11}) 
& Canceled by soft factor 
& Canceled by soft factor \\
\hline
Soft factor $S(\b)$ & UV div. (determined by& UV div. (same calculation up to \\ 
{} &  trans. propagator only)  
& the interchange of $+$ and $-$ components) \\
\hline
\hline
\end{tabular}
\caption{Diagram-by-diagram comparison of the one-loop quark TMD calculation in the target 
light-cone gauge $A^-=0$ within the ML prescription (Ref.~ \cite{Altinoluk:2025ewj}) and in the projectile light-cone gauge $A^+=0$ 
(this work). Despite the very different diagrammatic structures, the final CSS evolution equations 
are identical, providing a non-trivial gauge-invariance check.}
\label{tab:comparison}
\end{table}

%===================================
\subsection{Significance for the CGC-TMD connection}
%==================================
At first sight, the calculation in projectile light-cone gauge may appear to serve only as a consistency check of the corresponding analysis in target light-cone gauge. Because gauge invariance requires the CSS equations to be independent of the choice of gauge, one expects their final form to be identical in both cases, suggesting that the overall outcome may appear largely predetermined. This view, however, misses an important aspect of the present work. Gauge invariance is a property of physical observables rather than of individual perturbative contributions, and its correct realization must be demonstrated through an explicit calculation. The nontrivial redistribution of contributions among diagrams, discussed in Sec.~\ref{subsec:comp_tech}, cannot be inferred from the target light-cone gauge analysis alone. In particular, this includes the vanishing of diagrams~\ref{Fig:diag5}, \ref{Fig:diag12}, \ref{Fig:diag13}, and \ref{Fig:diag14}; the shift of the rapidity-finite but UV-divergent contribution to diagram~\ref{Fig:diag4}; and the realization of gauge independence of the soft factor through the ML prescription. Each of these features provides a nontrivial check of the consistency of the background field formalism combined with the ML prescription in projectile light-cone gauge. Taken together, they indicate that the framework is internally consistent and correctly reproduces the expected gauge-invariant structure.

The deeper significance of the present calculation, however, lies not in the verification of gauge invariance itself, but in the gauge choice in which it is carried out. The projectile light-cone gauge $A^+=0$ is a natural gauge choice in the CGC framework~\cite{Gelis:2010nm,Albacete:2014fwa,Blaizot:2016qgz}, where small-$x$ evolution equations and particle production processes are commonly formulated~\cite{Dumitru:2005gt,Chirilli:2011km,Altinoluk:2011qy}. By contrast, the TMD factorization framework has predominantly been developed in Feynman gauge or in the target light-cone gauge, where the gauge-link structure simplifies and longitudinal Wilson lines become trivial. This difference in customary gauge choices has complicated the construction of a fully transparent connection between the CGC and TMD descriptions of high-energy QCD. The present calculation addresses this issue at the level of TMD evolution by deriving the CSS equations within the background field formalism in projectile light-cone gauge. This shows that the TMD evolution equations can be consistently formulated in the same gauge used in CGC applications, at least within the scope of the present analysis.

The background field formalism employed here is closely aligned with the CGC framework, where the classical background field represents the  low rapidity modes, corresponding to a dense target, while perturbative quantum fluctuations describe high rapidity modes. In the present work, we restrict ourselves to the dilute limit of the target, in which the background field is treated perturbatively and each background field insertion is suppressed by a power of $g$. The CGC framework is applicable in the dense regime where gluon saturation effects become important and the background field is strong. Extending the present analysis to that regime, and understanding how the evolution equations are modified in the presence of saturation effects, lies beyond the scope of this work. Nevertheless, the dilute limit provides a natural starting point, and we show that the TMD evolution equations can be consistently derived within the background field formalism in projectile light-cone gauge, which is a standard gauge choice in many CGC applications, already at this level. This suggests a possible path toward a unified computational perspective on TMD factorization and CGC dynamics within their respective domains of validity.

%=============================================================================================
\section{Conclusions and outlook}
\label{sec:conc}
%=============================================================================================
We have computed the one-loop corrections to the quark TMD in the projectile light-cone gauge $A^+=0$ within the background field formalism, using the ML prescription for the light-cone propagator singularity and the rapidity regulator of Ref.~\cite{Ebert:2018gsn}. The resulting CSS evolution equations agree with the standard results~\cite{Collins:1981uk,Collins:1984kg,Collins:2011zzd} and with the corresponding calculation in the target light-cone gauge~\cite{Altinoluk:2025ewj}, in accordance with gauge invariance.

Although the final results are identical, the two formulations exhibit a significantly different intermediate structure. In the projectile light-cone gauge, longitudinal Wilson lines generate additional diagrammatic contributions, many of which vanish due to the $k^+$ pole structure of the ML prescription. The organization of rapidity divergences differs from that in the target light-cone gauge.  
On the other hand, the soft factor is derived in the same way in both light-cone gauges, up to a trivial exchange of the $+$ and $-$ light-cone directions. This reveals how CSS evolution emerges from distinct but equivalent diagrammatic realizations in different gauge choices, providing a useful consistency check of the background field formalism in light-cone gauges employing ML prescription.

A key motivation of this work is the formulation of TMD evolution in a framework similar to the one used in the CGC formalism. We have demonstrated that the CSS equations can be derived in the background field formalism in projectile light-cone gauge, the natural gauge choice in CGC studies. This establishes a direct one-loop realization of TMD evolution in a CGC-compatible setting.

This opens a pathway toward a more unified treatment of TMD factorization and CGC dynamics. Natural extensions include gluon TMDs, Sudakov effects in the presence of saturation, and connections to beyond-eikonal formulations. It may also help clarify the relation between CGC-based approaches to evolution and standard TMD factorization in their overlapping regimes of validity.

\acknowledgements
TA thanks Baruch College, CUNY and JJM thanks NCBJ for hospitality during the mutual visits when this work was performed. TA is supported in part by the National Science Centre (Poland) under the research Grant No. 2023/50/E/ST2/00133
(SONATA BIS 13). GB is supported in part by the National Science Centre (Poland) under the research
Grant No. 2020/38/E/ST2/00122 (SONATA BIS 10). This material is based upon
work supported by the U.S. Department of Energy, Office of Science, Office of Nuclear Physics, within the framework
of the Saturated Glue (SURGE) Topical Theory Collaboration. JJM is supported by the US DOE Office of Nuclear Physics through Grant No. DE-SC0002307. MT is supported in part by the National Science Centre (Poland) under the research Grant No. 2024/53/B/ST2/00968
(OPUS 27).

\appendix

\section{Symmetric diagrams}
\label{appendix_symmetric_diagrams}

%%%%%%%%%%%%%%%%%%%%%%%%%%%%%%%%%%%%%%%%%%%%%%%%%%%%%%%%%%%%%%%%%%%%%%%%%%%%%%%%%%%%%%
\begin{figure}[t]
\subfloat[ Diagram of gluon emission from the antiquark to the upper part of the gauge link \label{Fig:diag7}]{%
       \includegraphics[width=0.40\textwidth]{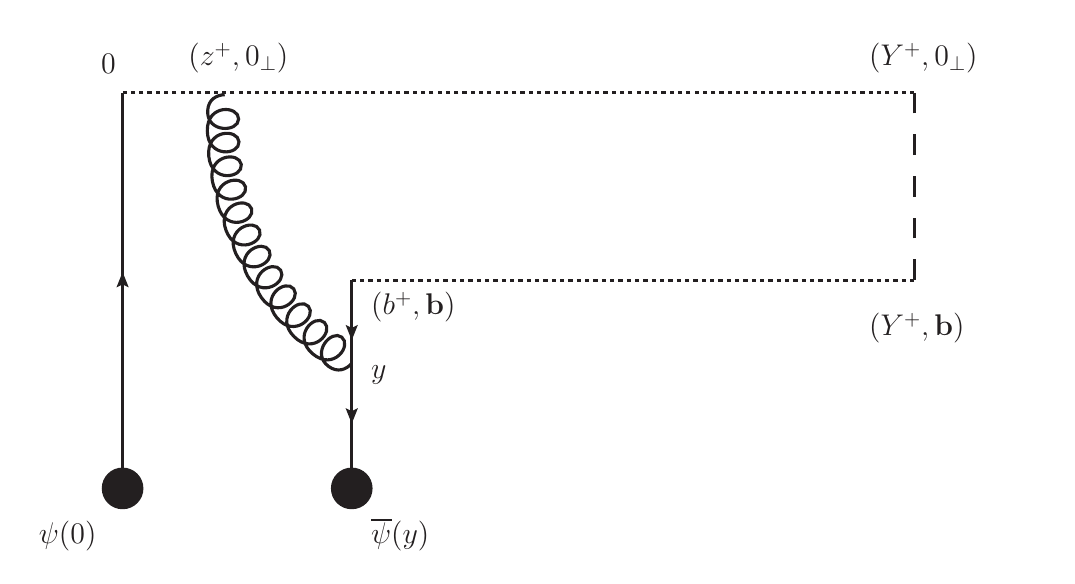}
     }
\hfill
\subfloat[ Diagram of gluon emission from the antiquark to the lower part of the gauge link\label{Fig:diag9}]{%
       \includegraphics[width=0.40\textwidth]{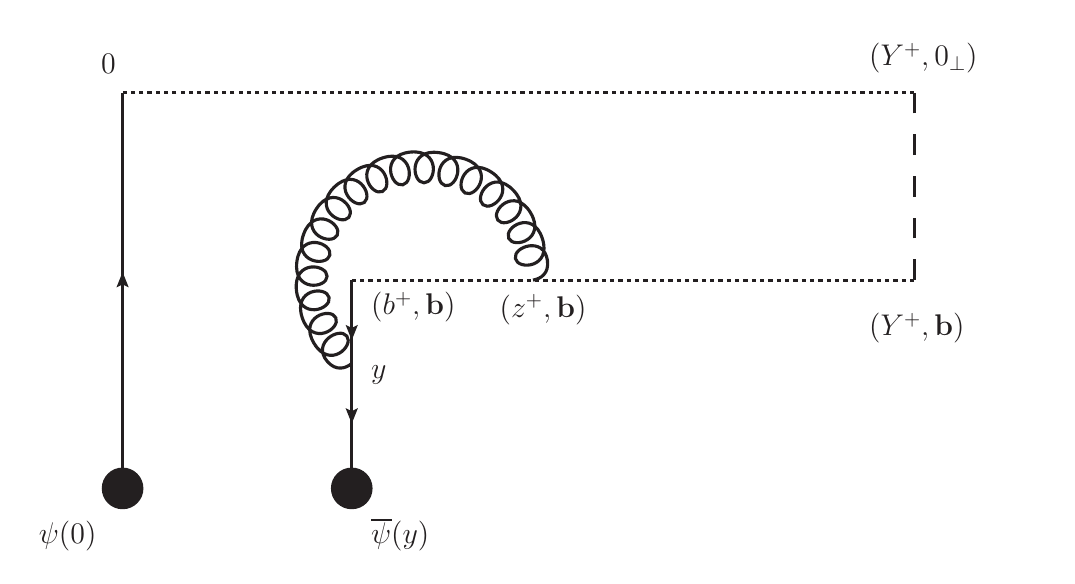}
     }

\subfloat[Diagram of gluon emission from the antiquark to infinity \label{Fig:diag8}]{%
       \includegraphics[width=0.40\textwidth]{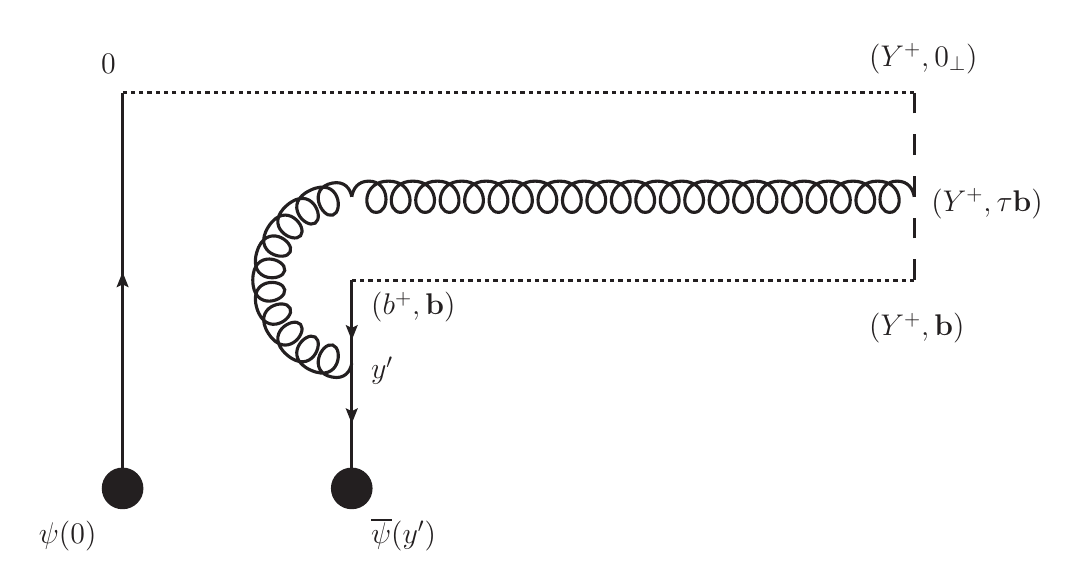}
     }
\hfill
\subfloat[Light-like Wilson line self-energy diagram
\label{Fig:diag10}]{%
       \includegraphics[width=0.40\textwidth]{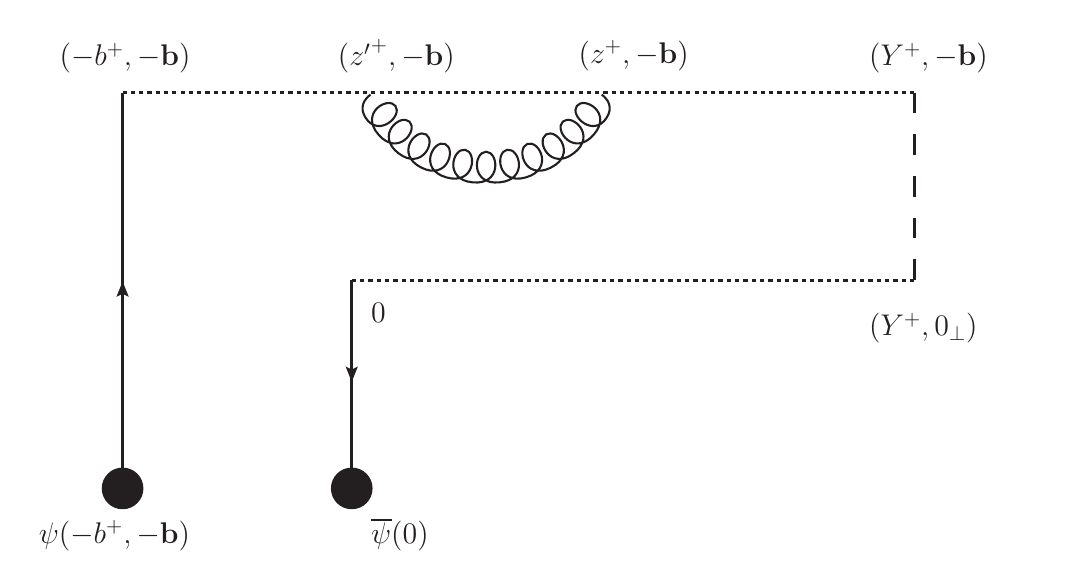}
     }  

\subfloat[Emission of gluon from the lower part of the gauge link to infinity \label{Fig:diag15}]{%
       \includegraphics[width=0.40\textwidth]{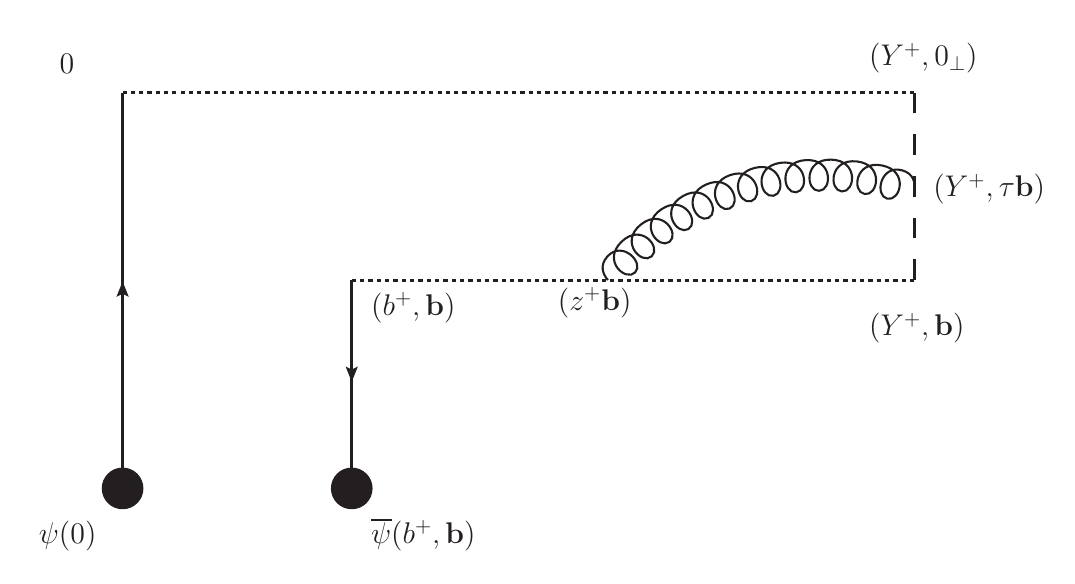}
     } 
\caption{\label{Fig:real_diags 2} The remaining NLO diagrams in the expansion of the quark TMD around its background contribution in the projectile light-cone gauge. These diagrams are symmetric to the diagrams~\ref{Fig:diag6}, \ref{Fig:diag4}, \ref{Fig:diag5}, \ref{Fig:diag12}, and \ref{Fig:diag13}.}
\end{figure}
%%%%%%%%%%%%%%%%%%%%%%%%%%%%%%%%%%%%%%%%%%%%%%%%%%%%%%%%%%%%%%%%%%%%%%%%%%%%%%%%%%%%%%
%%%%%%%%%%%%%%%%%%%%%%%%%%%%%%%%%%%%%%%%%%%%%%%%%%%%%%%%%%%%%%%%%%%%%%%%%%%%%%%%%%%%%%

The remaining NLO corrections which were omitted from Eq.~\eqref{eq: operator expansion in background field} and Fig.~\ref{Fig:real_diags 1}:
\begin{align} 
& 
\big\langle P\big| \T \Big[ 
{\overline{\delta\Psi}}(b^+, \mathbf{b}, 0^-) \, \frac{\gamma^-}{2} \, 
\left\{- i \mu^\epsilon g \int_0^1 d \tau\,  
\mathbf{b}^i\, t^a\, \delta A_i^a (Y^+, \tau \mathbf{b} , 0^-)\right\} 
\, 
\psi(0) \Big]
\big|P\big\rangle_c
\nonumber\\
&\,
+
\big\langle P\big| \T \Big[ 
{\overline{\delta\Psi}}(b^+, \mathbf{b}, 0^-) \, \frac{\gamma^-}{2} \, 
\left\{+ i \mu^\epsilon g \int_{b^+}^{Y^+} \td z^+
 t^a \delta A_a^- (z^+, \mathbf{b} , 0^-)\right\}
\, 
\psi(0) \Big]
\big|P\big\rangle_c
\nonumber\\
&\,
+
\big\langle P\big| \T \Big[ 
{\overline{\delta\Psi}}(b^+, \mathbf{b}, 0^-) \, \frac{\gamma^-}{2} \, 
\left\{- i \mu^\epsilon g \int_{0}^{Y^+} \td z^+
 t^a \delta A_a^- (z^+, 0_\perp , 0^-)\right\}
\, 
\psi(0) \Big]
\big|P\big\rangle_c
\nonumber\\
&\,
+
\big\langle P\big| \T \Big[ 
{\overline{\psi}}(b^+, \mathbf{b}, 0^-) \, \frac{\gamma^-}{2} \, 
\frac{1}{2} {\cal{P}}\left\{- i \mu^\epsilon g \int_{0}^{Y^+} \td z^+
 t^a \delta A_a^- (z^+,0_\perp , 0^-)\right\}^2 
\, 
\psi(0) \Big]
\big|P\big\rangle_c
\nonumber\\
&\,
+
\big\langle P\big| \T \Big[ 
{\overline\psi}(b^+, \mathbf{b}, 0^-) \, \frac{\gamma^-}{2} \, 
\left\{+ i \mu^\epsilon g \int_{b^+}^{Y^+} \td z^+
 t^a \delta A_a^- (z^+, \bbperp , 0^-)\right\}
\left\{- i \mu^\epsilon g \int_0^1 d \tau\,  
\mathbf{b}^i\, t^b\, \delta A_i^b (Y^+, \tau \mathbf{b} , 0^-)\right\}
\, 
\psi(0) \Big]
\big|P\big\rangle_c\,.
\end{align}
These can be associated with corresponding Feynman diagrams presented in Fig.~\ref{Fig:real_diags 2}.

\section{Calculation of the UV-divergent part of the diagram~\ref{Fig:diag4}}
\label{appendix: calculation of uv divergent part of diag 4}

The rapidity-finite, but UV-divergent, part of the diagram~\ref{Fig:diag4} reads
\begin{align}
q^{\textrm{n.r.}}_{\textrm{unsub.}}&(\tx, \mathbf{b};\mu^2,0)\Big|_{\ref{Fig:diag4}}^{\rm fin.} \nonumber 
\\
= & \, 
- \frac{\alpha_s \mu^{2 \epsilon} C_F}{2\pi} \, \int \td^{4 - 2 \epsilon} y \, 
\big\langle P\big|  {\overline\psi}_\alpha (0)  \, \psi_\beta (y) \big| P\big\rangle  
\, e^{i \tx P^- y^+ } 
\int \frac{\td^{2-2\epsilon} \bl}{(2\pi)^{2-2\epsilon}} \, 
\int\frac{\td l^+}{(2\pi)} \, 
e^{i l^+ y^-} e^{- i \bl\cdot (\by + \bbperp)}
\nonumber \\
&
\times\int \frac{\td^{2-2\epsilon} \bk}{(2\pi)^{2-2\epsilon}} \, \int_0^1 \frac{\td \xi}{\xi }  \, 
\Bigg\{\gamma^- \Bigg[\frac{(1 - \xi )\left[- (\bl - \bk)^i\gamma^i (2 \tx P^-\xi  \gamma^+ - \bk^j\gamma^j) - \tx P^-(1 - \xi ) \gamma^+ \bk^i\gamma^i\right] }
{\left[2 \tx P^- (1 - \xi ) l^+ - (\bl - \bk)^2\right] 
\left[2 \tx P^- \xi  (1 - \xi ) l^+ - \xi  (\bl - \bk)^2 - (1 - \xi ) \bk^2\right]} 
\nonumber \\
& - \frac{\left[- (\bl - \bk)^i\gamma^i ( \bk^j\gamma^j) + \tx P^-\gamma^+ \bk^i\gamma^i\right]
 }
{\left[2 \tx P^- l^+ - (\bl - \bk)^2\right] 
 \bk^2} 
\Bigg]\Bigg\}_{\alpha\beta} \,,
\label{eq: appendix eta-finite part in diag 4}
\end{align} 
where we can simplify the first term in the $\bk$ integral by writing:
\begin{align}
   & \frac{1}
{\left[2 \tx P^- (1 - \xi ) l^+ - (\bl - \bk)^2\right] 
\left[2 \tx P^- \xi  (1 - \xi ) l^+ - \xi  (\bl - \bk)^2 - (1 - \xi ) \bk^2\right]} \nonumber\\
&= -\frac{1}{1-\xi}\frac{1}{\bk^2}\left[\frac{1}{2 \tx P^- (1 - \xi ) l^+ - (\bl - \bk)^2}-\frac{\xi}{\xi\left[2 \tx P^- (1 - \xi ) l^+ - (\bl - \bk)^2\right]-(1-\xi)\bk^2}\right]\,.
\end{align}
With this, the $\bk$ integral reads
\begin{align}
& \int \frac{\td^{2-2\epsilon} \bk}{(2\pi)^{2-2\epsilon}} \, \frac{1}{\bk^2} \,
\Bigg[
-\frac{\left[- (\bl - \bk)^i\gamma^i (2 \tx P^-\xi  \gamma^+ - \bk^j\gamma^j) - \tx P^-(1 - \xi ) \gamma^+ \bk^i\gamma^i\right]}{2 \tx P^- (1 - \xi ) l^+ - (\bl - \bk)^2 } \nonumber \\
&
+\frac{\xi\left[- (\bl - \bk)^i\gamma^i (2 \tx P^-\xi  \gamma^+ - \bk^j\gamma^j) - \tx P^-(1 - \xi ) \gamma^+ \bk^i\gamma^i\right]}{\xi\left[2 \tx P^- (1 - \xi ) l^+ - (\bl - \bk)^2\right]-(1-\xi)\bk^2}
 - \frac{\left[- (\bl - \bk)^i\gamma^i ( \bk^j\gamma^j) + \tx P^-\gamma^+ \bk^i\gamma^i\right]
 }{2 \tx P^- l^+ - (\bl - \bk)^2 } 
\Bigg]\,,
\end{align} 
where the first and third terms are individually IR divergent, but their divergences cancel each other. Thus, the rapidity-finite contribution of the diagram~\ref{Fig:diag4} is IR-safe. Here the UV divergence arises from the terms in the numerator which are proportional to $\bk^2$.
Standard evaluation of this integral with the Feynman parametrization gives
\begin{align}
    & (4\pi)^{\epsilon-1}\frac{\Gamma(2-\epsilon)}{\Gamma(1-\epsilon)}\Gamma(\epsilon)\,\int_0^1 \td t \, \left[\Delta^{-\epsilon}_1(\xi,t) -\xi\Delta^{-\epsilon}_2(\xi,t) -\Delta^{-\epsilon}_1(0,t) \right]+\text{ finite}\,,
\end{align}
where $t$ arises from the Feynman parametrization. Here we have defined 
\begin{align}
    \Delta_1(\xi,t)\equiv& -t^2\bl^2 -t\left[ 2\tx P^- l^+(1-\xi)-\bl^2\right]\,, \\
    \Delta_2(\xi,t)\equiv&  -t^2\xi^2\bl^2 -t\xi\left[ 2\tx P^- l^+(1-\xi)-\bl^2\right]\,.
\end{align}
Since we are only interested in the UV-divergent result, we can use $\Delta^{-\epsilon}_{1,2}=1+\mathcal{O}(\epsilon)$. Plugging everything in Eq.~\eqref{eq: appendix eta-finite part in diag 4}, we get
\begin{align}
q^{\textrm{n.r.}}_{\textrm{unsub.}}&(\tx, \mathbf{b};\mu^2,0)\Big|_{\ref{Fig:diag4}}^{\rm fin.} \nonumber 
\\
= & \, 
- \frac{\alpha_s \mu^{2 \epsilon} C_F}{2\pi} \, \int \td^{4 - 2 \epsilon} y \, 
\big\langle P\big|  {\overline\psi}_\alpha (0)  \, \psi_\beta (y) \big| P\big\rangle  
\, e^{i \tx P^- y^+ } 
\int \frac{\td^{2-2\epsilon} \bl}{(2\pi)^{2-2\epsilon}} \, 
\int\frac{\td l^+}{(2\pi)} \, 
e^{i l^+ y^-} e^{- i \bl\cdot (\by + \bbperp)}
\nonumber \\
&
\times\ (4\pi)^{\epsilon-1}\frac{\Gamma(2-\epsilon)}{\Gamma(1-\epsilon)}\Gamma(\epsilon)\, \int_0^1 \frac{\td \xi}{\xi }  \, 
\left\{\gamma^- \left[1-\xi-1+\mathcal{O}(\epsilon)
\right]\right\}_{\alpha\beta}  \nonumber \\
= 
& \, 
 \frac{\alpha_s \mu^{2 \epsilon} C_F}{2\pi} \, (4\pi)^{\epsilon-1}\frac{\Gamma(2-\epsilon)}{\Gamma(1-\epsilon)}\Gamma(\epsilon)\,  \int \td^{4 - 2 \epsilon} y \, 
\big\langle P\big|  {\overline\psi} (0) \gamma^- \, \psi (y) \big| P\big\rangle  
\, e^{i \tx P^- y^+ } 
\int \frac{\td^{2-2\epsilon} \bl}{(2\pi)^{2-2\epsilon}} \, 
\int\frac{\td l^+}{(2\pi)} \, 
e^{i l^+ y^-} e^{- i \bl\cdot (\by + \bbperp)} \nonumber \\
= 
& \, 
 \frac{\alpha_s  C_F}{2\pi} \, (4\pi\mu^2)^{\epsilon}\frac{\Gamma(2-\epsilon)}{\Gamma(1-\epsilon)}\Gamma(\epsilon)\,  \int \frac{\td y^+}{2\pi}\, 
\big\langle P\big|  {\overline\psi} (0) \frac{\gamma^-}{2} \, \psi(y^+,-\bbperp,0^-) \big| P\big\rangle  
\, e^{i \tx P^- y^+ } 
\,,
\end{align} 
where we have neglected the finite contributions.

%%%%%%%%%%%%%%%%%%%%%%%%%%%%%%%%%%%%%%%%%%%%%%%%%%%%%%%%%%%%%%%%%%%%%%%%%%%%%%%%%%%%%%
%%%%%%%%%%%%%%%%%%%%%%%%%%%%%%%%%%%%%%%%%%%%%%%%%%%%%%%%%%%%%%%%%%%%%%%%%%%%%%%%%%%%%%
\section{Vanishing integrals in the soft factor}
\label{appendix: Vanishing integrals in the soft factor}

In evaluating the digram a of the soft factor, we encounter the following integral over $k^-$:
\begin{align}
      &\int_{-\infty}^0\frac{\td k^-}{2\pi}\frac{|k^-|^\eta}{k^-}\left[1-e^{-ik^-Y^+} \right]\,
    \left[ 1-e^{i\frac{\bk^2}{2k^-}Y^-}\right] \nonumber\\
    &
    =-\int_{0}^\infty\frac{\td k^-}{2\pi}(k^-)^{\eta-1}\left[1-e^{+ik^-Y^+} \right]\,
    \left[ 1-e^{-i\frac{\bk^2}{2k^-}Y^-}\right] \,,
\end{align}
which can be expressed in therms of three different integrals: 
\begin{align}
    I_1\equiv & \int_{0}^\infty\frac{\td k^-}{2\pi}(k^-)^{\eta-1}e^{+ik^-Y^+} \,, \\
    I_2 \equiv & \int_{0}^\infty\frac{\td k^-}{2\pi}(k^-)^{\eta-1}e^{-i\frac{\bk^2}{2k^-}Y^-} \,, \\
    I_3\equiv & \int_{0}^\infty\frac{\td k^-}{2\pi}(k^-)^{\eta-1} e^{+ik^-Y^+}\,e^{-i\frac{\bk^2}{2k^-}Y^-}\,, 
\end{align}
together with the scaleless integral which is zero in pure rapidity regularization.
The integral $I_1$ can be shown to vanish in the limit $Y^+\to \infty$, for a positive $\eta$, following the same argument as for the similar integral in Section~\ref{subsubsection: diagram 4}.

By using Cauchy's integral theorem, we can write the integral $I_2$ as 
\begin{align}
   I_2&= \int_0^{i\infty} \frac{\td k^-}{2\pi}(k^-)^{\eta-1}e^{i\frac{\bk^2}{2k^-}Y^-}\,,
    \label{eq: k^- int in S^a}
\end{align}
where we closed the contour in the first quadrant of the complex plane along the positive real and imaginary axes. Let us then define $t\equiv ik^-$ so that Eq.~\eqref{eq: k^- int in S^a} can be written as
\begin{align}
    \frac{i^\eta }{2\pi}\int_0^{\infty} \td tt^{\eta-1}e^{-\frac{\bk^2}{2t}Y^-}
    =\frac{i^\eta}{2\pi}\left(\frac{\bk^2 Y^-}{2} \right)^\eta 
    \Gamma(-\eta)
    \,,
\end{align}
which approaches zero in the limit $Y^-\to \infty$ for negative $\eta$.

Applying the same contour deformation and change of integration variable as above, the integral $I_3$ can be expressed as   
\begin{align}
    I_3&=     \frac{i^\eta }{2\pi}\int_0^{\infty} \td tt^{\eta-1}e^{-tY^+}e^{-\frac{\bk^2}{2t}Y^-}\,.
\end{align}
Introducing a variable $\tau\equiv Y^+ t$, this becomes
\begin{align}
    I_3&=     \frac{i^\eta }{2\pi} (Y^+)^{-\eta} \int_0^{\infty} \td \tau \tau^{\eta-1}e^{-\tau}e^{-\sqrt{2\bk^2 Y^- Y^+}\frac{1}{4\tau}}=   \frac{i^\eta }{\pi} (Y^+)^{-\eta}\left(\frac{1}{2}\sqrt{2\bk^2 Y^- Y^+}\right)^\eta K_\eta \left( \sqrt{2\bk^2 Y^- Y^+}\right)\,,
\end{align}
where $K_\eta$ denotes the modified Bessel function of second kind.
The above expression vanishes in the limit $Y^\pm \to \infty$ for $\eta =0$ as well as for finite positive or negative values of $\eta$.

\vspace{0.3in}
\bibliography{mybib_New}

\end{document}